\documentclass[copyright]{eptcs}
\providecommand{\event}{MSFP 2012} 
\usepackage{breakurl}             
\usepackage{amsmath,amsthm,amssymb,amstext}
\usepackage{stmaryrd} 
\usepackage{ottlayout}
\usepackage{subfigure}
\usepackage{supertabular}
\usepackage{comment}
\usepackage{xspace}
\usepackage{xcolor}

\newif\ifcomments\commentsfalse
\definecolor{darkgreen}{RGB}{0,175,0}
\ifcomments
\newcommand{\cjc}[1]{\textcolor{darkgreen}{\textbf{[#1 ---CJC]}}}
\newcommand{\vs}[1]{\textcolor{blue}{\textbf{[#1 ---VS]}}}
\newcommand{\as}[1]{\textcolor{red}{\textbf{[#1 ---AS]}}}
\newcommand{\scw}[1]{\textcolor{red}{\textbf{[#1 ---SCW]}}}
\newcommand{\ts}[1]{\textcolor{purple}{\textbf{[#1 ---TS]}}}
\newcommand{\kya}[1]{\textcolor{magenta}{\textbf{[#1 ---KYA]}}}
\newcommand{\hde}[1]{\textcolor{magenta}{\textbf{[#1 ---HE]}}}
\newcommand{\ntc}[1]{\textcolor{teal}{\textbf{[#1 ---NTC]}}}
\newcommand{\gk}[1]{\textcolor{orange}{\textbf{[#1 ---GK]}}}
\else
\newcommand{\cjc}[1]{}
\newcommand{\vs}[1]{}
\newcommand{\as}[1]{}
\newcommand{\scw}[1]{}
\newcommand{\ts}[1]{}
\newcommand{\kya}[1]{}
\newcommand{\hde}[1]{}
\newcommand{\ntc}[1]{}
\newcommand{\gk}[1]{}
\fi

\newtheorem{thm}{Theorem}
\newtheorem{lemma}[thm]{Lemma}

\newtheorem{definition}[thm]{Definition}

\newcommand{\ottdrule}[4][]{{\displaystyle\frac{\begin{array}{l}#2\end{array}}{#3}\quad\ottdrulename{#4}}}
\newcommand{\ottusedrule}[1]{\[#1\]}
\newcommand{\ottpremise}[1]{ #1 \\}
\newenvironment{ottdefnblock}[3][]{ \framebox{\mbox{#2}} \quad #3 \\[0pt]}{}

\newcommand{\ottnt}[1]{\mathit{#1}}
\newcommand{\ottmv}[1]{\mathit{#1}}
\newcommand{\ottkw}[1]{\mathbf{#1}}
\newcommand{\ottsym}[1]{#1}
\newcommand{\ottcom}[1]{\text{#1}}
\newcommand{\ottdrulename}[1]{\textsc{#1}}
\newcommand{\ottcomplu}[5]{\overline{#1}^{\,#2\in #3 #4 #5}}

\newcommand{\ottgrammartabular}[1]{\begin{supertabular}{llcllllll}#1\end{supertabular}}

\newcommand{\ottrulehead}[3]{$#1$ & & $#2$ & & & \multicolumn{2}{l}{#3}}
\newcommand{\ottprodline}[6]{& & $#1$ & $#2$ & $#3 #4$ & $#5$ & $#6$}
\newcommand{\ottinterrule}{\\[5.0mm]}

\renewcommand{\ottkw}[1]{\mathsf{#1} }

\newcommand{\otttele}{
\ottrulehead{\ottnt{tele}  ,\ \Delta}{::=}{\ottcom{telescope}}\\ 
\ottprodline{|}{ \cdot }{}{}{}{\ottcom{empty telescope}}\\ 
\ottprodline{|}{\ottsym{(}  \ottmv{x}  \ottsym{:}  \ottnt{A}  \ottsym{)}  \Delta}{}{}{}{\ottcom{relevant binding}}\\ 
\ottprodline{|}{\ottsym{[}  \ottmv{x}  \ottsym{:}  \ottnt{A}  \ottsym{]}  \Delta}{}{}{}{\ottcom{irrelevant binding}}}

\newcommand{\ottteleplus}{
\ottrulehead{\ottnt{teleplus}  ,\ \Delta^{+}}{::=}{\ottcom{(relevant) telescope}}\\ 
\ottprodline{|}{ \cdot }{}{}{}{\ottcom{empty telescope}}\\ 
\ottprodline{|}{\ottsym{(}  \ottmv{x}  \ottsym{:}  \ottnt{A}  \ottsym{)}  \Delta^{+}}{}{}{}{\ottcom{relevant binding}}}

\newcommand{\ottenv}{
\ottrulehead{\ottnt{env}  ,\ \Gamma}{::=}{\ottcom{typing environment}}\\ 
\ottprodline{|}{ {\cdot } }{}{}{}{\ottcom{empty}}\\ 
\ottprodline{|}{\Gamma  \ottsym{,}  \ottnt{decl}}{}{}{}{}}

\newcommand{\ottdecl}{
\ottrulehead{\ottnt{decl}}{::=}{\ottcom{typing env declaration}}\\ 
\ottprodline{|}{ \ottmv{x}  :  \ottnt{A} }{}{}{}{\ottcom{variable}}\\ 
\ottprodline{|}{\ottkw{data} \, \ottmv{D} \, \Delta^{+} \, \ottkw{where} \, \ottsym{\{} \, \ottcomplu{\ottmv{d_{\ottmv{i}}}  \ottsym{:}  \Delta_{\ottmv{i}}  \to \; D \; \Delta^{+}}{\ottmv{i}}{{\mathrm{1}}}{..}{\ottmv{j}} \, \ottsym{\}}}{}{}{}{\ottcom{datatype}}\\ 
\ottprodline{|}{\ottkw{data} \, \ottmv{D} \, \Delta^{+}}{}{}{}{\ottcom{abstract datatype name}}}

\newcommand{\ottexp}{
\ottrulehead{\ottnt{exp}  ,\ \ottnt{a}  ,\ \ottnt{b}  ,\ \ottnt{A}  ,\ \ottnt{B}}{::=}{\ottcom{annotated expressions}}\\ 
\ottprodline{|}{ \star }{}{}{}{\ottcom{type}}\\ 
\ottprodline{|}{\ottmv{x}}{}{}{}{\ottcom{variable}}\\ 
\ottprodline{|}{ \ottmv{D}  \;  \overline{A_i} }{}{}{}{\ottcom{datatype}}\\ 
\ottprodline{|}{ \ottmv{d}  \; [  \overline{A_i}  ] \;  \overline{a_i} }{}{}{}{\ottcom{data}}\\ 
\ottprodline{|}{ \mathsf{rec}\; \ottmv{f} : \ottnt{A} . \ottnt{a} }{}{}{}{\ottcom{recursive definition}}\\ 
\ottprodline{|}{\lambda  \ottmv{x}  \ottsym{:}  \ottnt{A}  \ottsym{.}  \ottnt{a}}{}{}{}{\ottcom{$\lambda$-abstraction}}\\ 
\ottprodline{|}{\lambda  \ottsym{[}  \ottmv{x}  \ottsym{:}  \ottnt{A}  \ottsym{]}  \ottsym{.}  \ottnt{a}}{}{}{}{\ottcom{irrelevant $\lambda$-abstraction}}\\ 
\ottprodline{|}{ \ottnt{a}  \;  \ottnt{b} }{}{}{}{\ottcom{application}}\\ 
\ottprodline{|}{ \ottnt{a}  \; [  \ottnt{b}  ] }{}{}{}{\ottcom{implicit application}}\\ 
\ottprodline{|}{ ( \ottmv{x} \!:\! \ottnt{A} )  \to \,  \ottnt{B} }{}{}{}{\ottcom{function type}}\\ 
\ottprodline{|}{ [  \ottmv{x} \!:\! \ottnt{A}  ]  \to \,  \ottnt{B} }{}{}{}{\ottcom{irrelevant function type}}\\ 
\ottprodline{|}{\ottkw{case} \, \ottnt{a} \, \ottkw{as} \, \ottsym{[y]} \, \ottkw{of} \, \ottsym{\{} \, \ottcomplu{\ottmv{d_{\ottmv{j}}} \, \Delta_{\ottmv{j}}  \Rightarrow  \ottnt{b_{\ottmv{j}}}}{\ottmv{j}}{{\mathrm{1}}}{..}{\ottmv{k}} \, \ottsym{\}}}{}{}{}{\ottcom{pattern matching}}\\ 
\ottprodline{|}{\ottnt{a}  \ottsym{=}  \ottnt{b}}{}{}{}{\ottcom{equality proposition}}\\ 
\ottprodline{|}{ \mathsf{join}_{ \ottnt{a} = \ottnt{b} } \;  \ottmv{i}  \;  \ottmv{j} }{}{}{}{\ottcom{equality proof}}\\ 
\ottprodline{|}{\ottkw{injdom} \, \ottnt{a}}{}{}{}{\ottcom{equality proof}}\\ 
\ottprodline{|}{\ottkw{injrng} \, \ottnt{a} \, \ottnt{b}}{}{}{}{\ottcom{equality proof}}\\ 
\ottprodline{|}{ \mathsf{injtcon}_{ \ottmv{i} } \;  \ottnt{a} }{}{}{}{\ottcom{equality proof}}\\ 
\ottprodline{|}{\ottkw{conv} \, \ottnt{a} \, \ottkw{at} \, \ottsym{[}  \mathsf{\sim}  \ottnt{P_{{\mathrm{1}}}}  \ottsym{/}  \ottmv{x_{{\mathrm{1}}}}  \ottsym{]} \, ... \, \ottsym{[}  \mathsf{\sim}  \ottnt{P_{\ottmv{i}}}  \ottsym{/}  \ottmv{x_{\ottmv{i}}}  \ottsym{]}  \ottnt{A}}{}{}{}{\ottcom{type conversion}}\\ 
\ottprodline{|}{ \mathsf{abort}_{ \ottnt{A} } }{}{}{}{\ottcom{failure}}}

\newcommand{\ottP}{
\ottrulehead{\ottnt{P}}{::=}{\ottcom{Proofs used in conv rule}}\\ 
\ottprodline{|}{\ottnt{v}}{}{}{}{}\\ 
\ottprodline{|}{\ottsym{[}  \ottnt{a}  \ottsym{=}  \ottnt{b}  \ottsym{]}}{}{}{}{}}

\newcommand{\ottval}{
\ottrulehead{\ottnt{val}  ,\ \ottnt{v}}{::=}{\ottcom{Values}}\\ 
\ottprodline{|}{\ottmv{x}}{}{}{}{}\\ 
\ottprodline{|}{ \star }{}{}{}{}\\ 
\ottprodline{|}{ ( \ottmv{x} \!:\! \ottnt{A} )  \to \,  \ottnt{B} }{}{}{}{}\\ 
\ottprodline{|}{ [  \ottmv{x} \!:\! \ottnt{A}  ]  \to \,  \ottnt{B} }{}{}{}{}\\ 
\ottprodline{|}{\ottnt{a}  \ottsym{=}  \ottnt{b}}{}{}{}{}\\ 
\ottprodline{|}{\ottkw{conv} \, \ottnt{v} \, \ottkw{at} \, \ottsym{[}  \mathsf{\sim}  \ottnt{P_{{\mathrm{1}}}}  \ottsym{/}  \ottmv{x_{{\mathrm{1}}}}  \ottsym{]} \, ... \, \ottsym{[}  \mathsf{\sim}  \ottnt{P_{\ottmv{i}}}  \ottsym{/}  \ottmv{x_{\ottmv{i}}}  \ottsym{]}  \ottnt{A}}{}{}{}{}\\ 
\ottprodline{|}{ \mathsf{join}_{ \ottnt{a} = \ottnt{b} } \;  \ottmv{i}  \;  \ottmv{j} }{}{}{}{}\\ 
\ottprodline{|}{ \ottmv{D}  \;  \overline{A_i} }{}{}{}{}\\ 
\ottprodline{|}{ \ottmv{d}  \; [  \overline{A_i}  ] \;  \overline{v_i} }{}{}{}{}\\ 
\ottprodline{|}{\lambda  \ottmv{x}  \ottsym{:}  \ottnt{A}  \ottsym{.}  \ottnt{a}}{}{}{}{}\\ 
\ottprodline{|}{\lambda  \ottsym{[}  \ottmv{x}  \ottsym{:}  \ottnt{A}  \ottsym{]}  \ottsym{.}  \ottnt{a}}{}{}{}{}\\ 
\ottprodline{|}{ \mathsf{rec}\; \ottmv{f} : \ottnt{A} . \ottnt{a} }{}{}{}{}}

\newcommand{\ottexplist}{
\ottrulehead{\ottnt{explist}  ,\ \overline{a_i}  ,\ \overline{b_i}  ,\ \overline{A_i}  ,\ \overline{B_i}}{::=}{\ottcom{list of expressions}}\\ 
\ottprodline{|}{ \cdot }{}{}{}{\ottcom{empty}}\\ 
\ottprodline{|}{ \ottnt{a}  \;  \overline{a_i} }{}{}{}{\ottcom{relevant expression}}\\ 
\ottprodline{|}{ [  \ottnt{a}  ] \;  \overline{a_i} }{}{}{}{\ottcom{irrelevant expression}}}

\newcommand{\ottvallist}{
\ottrulehead{\ottnt{vallist}  ,\ \overline{v_i}}{::=}{\ottcom{list of values}}\\ 
\ottprodline{|}{ \cdot }{}{}{}{}\\ 
\ottprodline{|}{ \ottnt{v}  \;  \overline{v_i} }{}{}{}{}\\ 
\ottprodline{|}{ [  \ottnt{v}  ] \;  \overline{v_i} }{}{}{}{}}

\newcommand{\ottetele}{
\ottrulehead{\ottnt{etele}  ,\ \Xi}{::=}{\ottcom{unannotated telescope}}\\ 
\ottprodline{|}{ \cdot }{}{}{}{\ottcom{empty telescope}}\\ 
\ottprodline{|}{\ottsym{(}  \ottmv{x}  \ottsym{:}  \ottnt{M}  \ottsym{)}  \Xi}{}{}{}{\ottcom{relevant binding}}\\ 
\ottprodline{|}{\ottsym{[}  \ottmv{x}  \ottsym{:}  \ottnt{M}  \ottsym{]}  \Xi}{}{}{}{\ottcom{irrelevant binding}}}

\newcommand{\otteteleplus}{
\ottrulehead{\ottnt{eteleplus}  ,\ \Xi^{+}}{::=}{\ottcom{unannotated (relevant) telescope}}\\ 
\ottprodline{|}{ \cdot }{}{}{}{\ottcom{empty telescope}}\\ 
\ottprodline{|}{\ottsym{(}  \ottmv{x}  \ottsym{:}  \ottnt{M}  \ottsym{)}  \Xi^{+}}{}{}{}{\ottcom{relevant binding}}}

\newcommand{\otteenv}{
\ottrulehead{\ottnt{eenv}  ,\ \ottnt{H}}{::=}{\ottcom{typing environment}}\\ 
\ottprodline{|}{ \cdot }{}{}{}{}\\ 
\ottprodline{|}{\ottnt{H}  \ottsym{,}  \ottnt{edecl}}{}{}{}{}}

\newcommand{\ottedecl}{
\ottrulehead{\ottnt{edecl}}{::=}{\ottcom{typing env declaration}}\\ 
\ottprodline{|}{ \ottmv{x}  :  \ottnt{M} }{}{}{}{\ottcom{variable}}\\ 
\ottprodline{|}{\ottkw{data} \, \ottmv{D} \, \Xi^{+} \, \ottkw{where} \, \ottsym{\{} \, \ottcomplu{\ottmv{d_{\ottmv{i}}}  \ottsym{:}  \Xi_{\ottmv{i}}  \to \; D \; \Xi^{+}}{\ottmv{i}}{{\mathrm{1}}}{..}{\ottmv{j}} \, \ottsym{\}}}{}{}{}{\ottcom{datatype}}\\ 
\ottprodline{|}{\ottkw{data} \, \ottmv{D} \, \Xi^{+}}{}{}{}{\ottcom{abstract datatype name}}}

\newcommand{\otteenvD}{
\ottrulehead{\ottnt{eenvD}  ,\ H_{D}}{::=}{\ottcom{closed typing environment}}\\ 
\ottprodline{|}{ \cdot }{}{}{}{}\\ 
\ottprodline{|}{\ottnt{H}  \ottsym{,}  \ottnt{edeclD}}{}{}{}{}}

\newcommand{\ottedeclD}{
\ottrulehead{\ottnt{edeclD}}{::=}{}\\ 
\ottprodline{|}{\ottkw{data} \, \ottmv{D} \, \Xi^{+} \, \ottkw{where} \, \ottsym{\{} \, \ottcomplu{\ottmv{d_{\ottmv{i}}}  \ottsym{:}  \Xi_{\ottmv{i}}  \to \; D \; \Xi^{+}}{\ottmv{i}}{{\mathrm{1}}}{..}{\ottmv{j}} \, \ottsym{\}}}{}{}{}{}}

\newcommand{\otteexp}{
\ottrulehead{\ottnt{eexp}  ,\ \ottnt{m}  ,\ \ottnt{n}  ,\ \ottnt{M}  ,\ \ottnt{N}}{::=}{\ottcom{unannotated expressions}}\\ 
\ottprodline{|}{ \star }{}{}{}{\ottcom{type}}\\ 
\ottprodline{|}{\ottmv{x}}{}{}{}{\ottcom{variable}}\\ 
\ottprodline{|}{\ottmv{D} \, \overline{M_i}}{}{}{}{\ottcom{datatype}}\\ 
\ottprodline{|}{\ottmv{d} \, \overline{m_i}}{}{}{}{\ottcom{data}}\\ 
\ottprodline{|}{ \mathsf{rec}\; \ottmv{f} . \ottkw{u} }{}{}{}{\ottcom{recursive definition}}\\ 
\ottprodline{|}{\lambda  \ottmv{x}  \ottsym{.}  \ottnt{m}}{}{}{}{\ottcom{$\lambda$-abstraction}}\\ 
\ottprodline{|}{\lambda  \ottsym{[}  \ottsym{]}  \ottsym{.}  \ottnt{m}}{}{}{}{\ottcom{irrelevant $\lambda$-abstraction}}\\ 
\ottprodline{|}{ \ottnt{m}  \;  \ottnt{n} }{}{}{}{\ottcom{application}}\\ 
\ottprodline{|}{\ottnt{m}  \ottsym{[}  \ottsym{]}}{}{}{}{\ottcom{irrelevant application}}\\ 
\ottprodline{|}{ ( \ottmv{x} \!:\! \ottnt{M} )  \to \,  \ottnt{N} }{}{}{}{\ottcom{function type}}\\ 
\ottprodline{|}{ [  \ottmv{x} \!:\! \ottnt{M}  ]  \to \,  \ottnt{N} }{}{}{}{\ottcom{irrelevant function type}}\\ 
\ottprodline{|}{\ottkw{case} \, \ottnt{n} \, \ottkw{of} \, \ottsym{\{} \, \ottcomplu{\ottmv{d_{\ottmv{j}}} \, \overline{x_i}_{\ottmv{j}}  \Rightarrow  \ottnt{m_{\ottmv{j}}}}{\ottmv{j}}{{\mathrm{1}}}{..}{\ottmv{k}} \, \ottsym{\}}}{}{}{}{\ottcom{pattern matching}}\\ 
\ottprodline{|}{\ottnt{m}  \ottsym{=}  \ottnt{n}}{}{}{}{\ottcom{equality proposition}}\\ 
\ottprodline{|}{\ottkw{join}}{}{}{}{\ottcom{equality proof}}\\ 
\ottprodline{|}{\ottkw{abort}}{}{}{}{\ottcom{failure}}}

\newcommand{\otteval}{
\ottrulehead{\ottnt{eval}  ,\ \ottnt{u}}{::=}{\ottcom{values}}\\ 
\ottprodline{|}{\ottmv{x}}{}{}{}{}\\ 
\ottprodline{|}{ \star }{}{}{}{}\\ 
\ottprodline{|}{ ( \ottmv{x} \!:\! \ottnt{M} )  \to \,  \ottnt{N} }{}{}{}{}\\ 
\ottprodline{|}{ [  \ottmv{x} \!:\! \ottnt{M}  ]  \to \,  \ottnt{N} }{}{}{}{}\\ 
\ottprodline{|}{\ottnt{m}  \ottsym{=}  \ottnt{n}}{}{}{}{}\\ 
\ottprodline{|}{\ottkw{join}}{}{}{}{}\\ 
\ottprodline{|}{\ottmv{D} \, \overline{M_i}}{}{}{}{}\\ 
\ottprodline{|}{\ottmv{d} \, \overline{u_i}}{}{}{}{}\\ 
\ottprodline{|}{ \mathsf{rec}\; \ottmv{f} . \ottkw{u} }{}{}{}{}\\ 
\ottprodline{|}{\lambda  \ottmv{x}  \ottsym{.}  \ottnt{m}}{}{}{}{}\\ 
\ottprodline{|}{\lambda  \ottsym{[}  \ottsym{]}  \ottsym{.}  \ottnt{m}}{}{}{}{}}

\newcommand{\otteexplist}{
\ottrulehead{\ottnt{eexplist}  ,\ \overline{m_i}  ,\ \overline{n_i}  ,\ \overline{M_i}  ,\ \overline{N_i}}{::=}{\ottcom{list of expressions}}\\ 
\ottprodline{|}{ \cdot }{}{}{}{}\\ 
\ottprodline{|}{ \ottnt{m}  \;  \overline{m_i} }{}{}{}{}\\ 
\ottprodline{|}{ [] \;  \overline{m_i} }{}{}{}{}}

\newcommand{\ottevallist}{
\ottrulehead{\ottnt{evallist}  ,\ \overline{u_i}}{::=}{}\\ 
\ottprodline{|}{ \cdot }{}{}{}{}\\ 
\ottprodline{|}{ \ottnt{u}  \;  \overline{u_i} }{}{}{}{}\\ 
\ottprodline{|}{ [] \;  \overline{u_i} }{}{}{}{}}

\newcommand{\ottevalctx}{
\ottrulehead{\ottnt{evalctx}  ,\ \mathcal{E}}{::=}{\ottcom{Evaluation contexts}}\\ 
\ottprodline{|}{ \bullet }{}{}{}{}\\ 
\ottprodline{|}{\bullet  \ottnt{m}}{}{}{}{}\\ 
\ottprodline{|}{\ottnt{u}  \bullet}{}{}{}{}\\ 
\ottprodline{|}{\bullet  \ottsym{[]}}{}{}{}{}\\ 
\ottprodline{|}{\ottkw{case} \, \bullet \, \ottkw{of} \, \ottsym{\{}  \ottmv{d_{\ottmv{j}}} \, \overline{x_i}_{\ottmv{j}}  \Rightarrow  \ottnt{m_{\ottmv{j}}}  \ottsym{\}}}{}{}{}{}\\ 
\ottprodline{|}{\ottmv{d} \, \overline{u_i}  \bullet  \overline{m_i}}{}{}{}{}}

\newcommand{\ottdrulescXXappbeta}[1]{\ottdrule[#1]{%
}{
 \ottsym{(}  \lambda  \ottmv{x}  \ottsym{.}  \ottnt{m}  \ottsym{)}  \;  \ottnt{u}   \leadsto_{\mathsf{cbv} }  \ottsym{[}  \ottnt{u}  \ottsym{/}  \ottmv{x}  \ottsym{]}  \ottnt{m}}{%
{\ottdrulename{sc\_appbeta}}{}%
}}

\newcommand{\ottdrulescXXapprec}[1]{\ottdrule[#1]{%
}{
 \ottsym{(}   \mathsf{rec}\; \ottmv{f} . \ottkw{u}   \ottsym{)}  \;  \ottnt{u_{{\mathrm{2}}}}   \leadsto_{\mathsf{cbv} }   \ottsym{(}  \ottsym{[}   \mathsf{rec}\; \ottmv{f} . \ottkw{u}   \ottsym{/}  \ottmv{f}  \ottsym{]}  \ottnt{u_{{\mathrm{1}}}}  \ottsym{)}  \;  \ottnt{u_{{\mathrm{2}}}} }{%
{\ottdrulename{sc\_apprec}}{}%
}}

\newcommand{\ottdrulescXXiappbeta}[1]{\ottdrule[#1]{%
}{
\ottsym{(}  \lambda  \ottsym{[}  \ottsym{]}  \ottsym{.}  \ottnt{m}  \ottsym{)}  \ottsym{[}  \ottsym{]}  \leadsto_{\mathsf{cbv} }  \ottnt{m}}{%
{\ottdrulename{sc\_iappbeta}}{}%
}}

\newcommand{\ottdrulescXXiapprec}[1]{\ottdrule[#1]{%
}{
\ottsym{(}   \mathsf{rec}\; \ottmv{f} . \ottkw{u}   \ottsym{)}  \ottsym{[}  \ottsym{]}  \leadsto_{\mathsf{cbv} }  \ottsym{(}  \ottsym{[}   \mathsf{rec}\; \ottmv{f} . \ottkw{u}   \ottsym{/}  \ottmv{f}  \ottsym{]}  \ottnt{u_{{\mathrm{1}}}}  \ottsym{)}  \ottsym{[}  \ottsym{]}}{%
{\ottdrulename{sc\_iapprec}}{}%
}}

\newcommand{\ottdrulescXXcasebeta}[1]{\ottdrule[#1]{%
}{
\ottkw{case} \, \ottsym{(}  \ottmv{d_{\ottmv{l}}} \, \overline{u_i}  \ottsym{)} \, \ottkw{of} \, \ottsym{\{} \, \ottcomplu{\ottmv{d_{\ottmv{j}}} \, \overline{x_i}_{\ottmv{j}}  \Rightarrow  \ottnt{m_{\ottmv{j}}}}{\ottmv{j}}{{\mathrm{1}}}{..}{\ottmv{k}} \, \ottsym{\}}  \leadsto_{\mathsf{cbv} }  \ottsym{[}  \overline{u_i}  \ottsym{/}  \overline{x_i}_{\ottmv{l}}  \ottsym{]}  \ottnt{m_{\ottmv{l}}}}{%
{\ottdrulename{sc\_casebeta}}{}%
}}

\newcommand{\ottdrulescXXabort}[1]{\ottdrule[#1]{%
}{
\mathcal{E}  \ottsym{[}  \ottkw{abort}  \ottsym{]}  \leadsto_{\mathsf{cbv} }  \ottkw{abort}}{%
{\ottdrulename{sc\_abort}}{}%
}}

\newcommand{\ottdrulescXXctx}[1]{\ottdrule[#1]{%
\ottpremise{\ottnt{m}  \leadsto_{\mathsf{cbv} }  \ottnt{n}}%
}{
\mathcal{E}  \ottsym{[}  \ottnt{m}  \ottsym{]}  \leadsto_{\mathsf{cbv} }  \mathcal{E}  \ottsym{[}  \ottnt{n}  \ottsym{]}}{%
{\ottdrulename{sc\_ctx}}{}%
}}

\newcommand{\ottdefnstepc}[1]{\begin{ottdefnblock}[#1]{$\ottnt{m}  \leadsto_{\mathsf{cbv} }  \ottnt{n}$}{}
\ottusedrule{\ottdrulescXXappbeta{}}
\ottusedrule{\ottdrulescXXapprec{}}
\ottusedrule{\ottdrulescXXiappbeta{}}
\ottusedrule{\ottdrulescXXiapprec{}}
\ottusedrule{\ottdrulescXXcasebeta{}}
\ottusedrule{\ottdrulescXXabort{}}
\ottusedrule{\ottdrulescXXctx{}}
\end{ottdefnblock}}

\newcommand{\ottdefnsJStepCBV}{
\ottdefnstepc{}}

\newcommand{\ottdrulespXXrefl}[1]{\ottdrule[#1]{%
}{
\ottnt{m}  \leadsto_{\mathsf{p} }  \ottnt{m}}{%
{\ottdrulename{sp\_refl}}{}%
}}

\newcommand{\ottdrulespXXrec}[1]{\ottdrule[#1]{%
\ottpremise{\ottnt{u}  \leadsto_{\mathsf{p} }  \ottnt{u'}}%
}{
 \mathsf{rec}\; \ottmv{f} . \ottkw{u}   \leadsto_{\mathsf{p} }   \mathsf{rec}\; \ottmv{f} . \ottkw{u} }{%
{\ottdrulename{sp\_rec}}{}%
}}

\newcommand{\ottdrulespXXabs}[1]{\ottdrule[#1]{%
\ottpremise{\ottnt{m}  \leadsto_{\mathsf{p} }  \ottnt{m'}}%
}{
\lambda  \ottmv{x}  \ottsym{.}  \ottnt{m}  \leadsto_{\mathsf{p} }  \lambda  \ottmv{x}  \ottsym{.}  \ottnt{m'}}{%
{\ottdrulename{sp\_abs}}{}%
}}

\newcommand{\ottdrulespXXpi}[1]{\ottdrule[#1]{%
\ottpremise{\ottnt{M}  \leadsto_{\mathsf{p} }  \ottnt{M'}}%
\ottpremise{\ottnt{N}  \leadsto_{\mathsf{p} }  \ottnt{N'}}%
}{
 ( \ottmv{x} \!:\! \ottnt{M} )  \to \,  \ottnt{N}   \leadsto_{\mathsf{p} }   ( \ottmv{x} \!:\! \ottnt{M'} )  \to \,  \ottnt{N'} }{%
{\ottdrulename{sp\_pi}}{}%
}}

\newcommand{\ottdrulespXXipi}[1]{\ottdrule[#1]{%
\ottpremise{\ottnt{M}  \leadsto_{\mathsf{p} }  \ottnt{M'}}%
\ottpremise{\ottnt{N}  \leadsto_{\mathsf{p} }  \ottnt{N'}}%
}{
 [  \ottmv{x} \!:\! \ottnt{M}  ]  \to \,  \ottnt{N}   \leadsto_{\mathsf{p} }   [  \ottmv{x} \!:\! \ottnt{M'}  ]  \to \,  \ottnt{N'} }{%
{\ottdrulename{sp\_ipi}}{}%
}}

\newcommand{\ottdrulespXXeq}[1]{\ottdrule[#1]{%
\ottpremise{\ottnt{m}  \leadsto_{\mathsf{p} }  \ottnt{m'}}%
\ottpremise{\ottnt{n}  \leadsto_{\mathsf{p} }  \ottnt{n'}}%
}{
\ottnt{m}  \ottsym{=}  \ottnt{n}  \leadsto_{\mathsf{p} }  \ottnt{m'}  \ottsym{=}  \ottnt{n'}}{%
{\ottdrulename{sp\_eq}}{}%
}}

\newcommand{\ottdrulespXXapp}[1]{\ottdrule[#1]{%
\ottpremise{\ottnt{m}  \leadsto_{\mathsf{p} }  \ottnt{m'}}%
\ottpremise{\ottnt{n}  \leadsto_{\mathsf{p} }  \ottnt{n'}}%
}{
 \ottnt{m}  \;  \ottnt{n}   \leadsto_{\mathsf{p} }   \ottnt{m'}  \;  \ottnt{n'} }{%
{\ottdrulename{sp\_app}}{}%
}}

\newcommand{\ottdrulespXXappbeta}[1]{\ottdrule[#1]{%
\ottpremise{\ottnt{m}  \leadsto_{\mathsf{p} }  \ottnt{m'}}%
\ottpremise{\ottnt{u}  \leadsto_{\mathsf{p} }  \ottnt{u'}}%
}{
 \ottsym{(}  \lambda  \ottmv{x}  \ottsym{.}  \ottnt{m}  \ottsym{)}  \;  \ottnt{u}   \leadsto_{\mathsf{p} }  \ottsym{[}  \ottnt{u'}  \ottsym{/}  \ottmv{x}  \ottsym{]}  \ottnt{m'}}{%
{\ottdrulename{sp\_appbeta}}{}%
}}

\newcommand{\ottdrulespXXapprec}[1]{\ottdrule[#1]{%
\ottpremise{\ottnt{u_{{\mathrm{1}}}}  \leadsto_{\mathsf{p} }  \ottnt{u'_{{\mathrm{1}}}}}%
\ottpremise{\ottnt{u_{{\mathrm{2}}}}  \leadsto_{\mathsf{p} }  \ottnt{u'_{{\mathrm{2}}}}}%
}{
 \ottsym{(}   \mathsf{rec}\; \ottmv{f} . \ottkw{u}   \ottsym{)}  \;  \ottnt{u_{{\mathrm{2}}}}   \leadsto_{\mathsf{p} }   \ottsym{(}  \ottsym{[}   \mathsf{rec}\; \ottmv{f} . \ottkw{u}   \ottsym{/}  \ottmv{f}  \ottsym{]}  \ottnt{u'_{{\mathrm{1}}}}  \ottsym{)}  \;  \ottnt{u'_{{\mathrm{2}}}} }{%
{\ottdrulename{sp\_apprec}}{}%
}}

\newcommand{\ottdrulespXXiapp}[1]{\ottdrule[#1]{%
\ottpremise{\ottnt{m}  \leadsto_{\mathsf{p} }  \ottnt{m'}}%
}{
\ottnt{m}  \ottsym{[}  \ottsym{]}  \leadsto_{\mathsf{p} }  \ottnt{m'}  \ottsym{[}  \ottsym{]}}{%
{\ottdrulename{sp\_iapp}}{}%
}}

\newcommand{\ottdrulespXXiappbeta}[1]{\ottdrule[#1]{%
\ottpremise{\ottnt{m}  \leadsto_{\mathsf{p} }  \ottnt{m'}}%
}{
\ottsym{(}  \lambda  \ottsym{[}  \ottsym{]}  \ottsym{.}  \ottnt{m}  \ottsym{)}  \ottsym{[}  \ottsym{]}  \leadsto_{\mathsf{p} }  \ottnt{m'}}{%
{\ottdrulename{sp\_iappbeta}}{}%
}}

\newcommand{\ottdrulespXXiapprec}[1]{\ottdrule[#1]{%
\ottpremise{\ottnt{u_{{\mathrm{1}}}}  \leadsto_{\mathsf{p} }  \ottnt{u'_{{\mathrm{1}}}}}%
}{
\ottsym{(}   \mathsf{rec}\; \ottmv{f} . \ottkw{u}   \ottsym{)}  \ottsym{[}  \ottsym{]}  \leadsto_{\mathsf{p} }  \ottsym{(}  \ottsym{[}   \mathsf{rec}\; \ottmv{f} . \ottkw{u}   \ottsym{/}  \ottmv{f}  \ottsym{]}  \ottnt{u'_{{\mathrm{1}}}}  \ottsym{)}  \ottsym{[}  \ottsym{]}}{%
{\ottdrulename{sp\_iapprec}}{}%
}}

\newcommand{\ottdrulespXXtcon}[1]{\ottdrule[#1]{%
\ottpremise{ \forall  \ottmv{i}  .\ \  \ottnt{M_{\ottmv{i}}}  \leadsto_{\mathsf{p} }  \ottnt{M'_{\ottmv{i}}} }%
}{
\ottmv{D} \, \overline{M_i}  \leadsto_{\mathsf{p} }  \ottmv{D} \, \overline{M_i}'}{%
{\ottdrulename{sp\_tcon}}{}%
}}

\newcommand{\ottdrulespXXdcon}[1]{\ottdrule[#1]{%
\ottpremise{ \forall  \ottmv{i}  .\ \  \ottnt{m_{\ottmv{i}}}  \leadsto_{\mathsf{p} }  \ottnt{m'_{\ottmv{i}}} }%
}{
\ottmv{d} \, \overline{m_i}  \leadsto_{\mathsf{p} }  \ottmv{d} \, \overline{m_i}}{%
{\ottdrulename{sp\_dcon}}{}%
}}

\newcommand{\ottdrulespXXcase}[1]{\ottdrule[#1]{%
\ottpremise{\ottnt{m}  \leadsto_{\mathsf{p} }  \ottnt{m'}}%
\ottpremise{ \forall  \ottmv{j}  .\ \  \ottnt{m_{\ottmv{j}}}  \leadsto_{\mathsf{p} }  \ottnt{m'_{\ottmv{j}}} }%
}{
\ottkw{case} \, \ottnt{m} \, \ottkw{of} \, \ottsym{\{} \, \ottcomplu{\ottmv{d_{\ottmv{j}}} \, \overline{x_i}_{\ottmv{j}}  \Rightarrow  \ottnt{m_{\ottmv{j}}}}{\ottmv{j}}{{\mathrm{1}}}{..}{\ottmv{k}} \, \ottsym{\}}  \leadsto_{\mathsf{p} }  \ottkw{case} \, \ottnt{m'} \, \ottkw{of} \, \ottsym{\{} \, \ottcomplu{\ottmv{d_{\ottmv{j}}} \, \overline{x_i}_{\ottmv{j}}  \Rightarrow  \ottnt{m'_{\ottmv{j}}}}{\ottmv{j}}{{\mathrm{1}}}{..}{\ottmv{k}} \, \ottsym{\}}}{%
{\ottdrulename{sp\_case}}{}%
}}

\newcommand{\ottdrulespXXcasebeta}[1]{\ottdrule[#1]{%
\ottpremise{ \forall  \ottmv{i}  .\ \  \ottnt{u_{\ottmv{i}}}  \leadsto_{\mathsf{p} }  \ottnt{u'_{\ottmv{i}}} }%
\ottpremise{\ottnt{m_{\ottmv{l}}}  \leadsto_{\mathsf{p} }  \ottnt{m'_{\ottmv{l}}}}%
}{
\ottkw{case} \, \ottsym{(}  \ottmv{d_{\ottmv{l}}} \, \overline{u_i}  \ottsym{)} \, \ottkw{of} \, \ottsym{\{} \, \ottcomplu{\ottmv{d_{\ottmv{j}}} \, \overline{x_i}_{\ottmv{j}}  \Rightarrow  \ottnt{m_{\ottmv{j}}}}{\ottmv{j}}{{\mathrm{1}}}{..}{\ottmv{k}} \, \ottsym{\}}  \leadsto_{\mathsf{p} }  \ottsym{[}  \overline{u_i}'  \ottsym{/}  \overline{x_i}_{\ottmv{l}}  \ottsym{]}  \ottnt{m'_{\ottmv{l}}}}{%
{\ottdrulename{sp\_casebeta}}{}%
}}

\newcommand{\ottdrulespXXabort}[1]{\ottdrule[#1]{%
}{
\mathcal{E}  \ottsym{[}  \ottkw{abort}  \ottsym{]}  \leadsto_{\mathsf{p} }  \ottkw{abort}}{%
{\ottdrulename{sp\_abort}}{}%
}}

\newcommand{\ottdefnstepp}[1]{\begin{ottdefnblock}[#1]{$\ottnt{m}  \leadsto_{\mathsf{p} }  \ottnt{n}$}{}
\ottusedrule{\ottdrulespXXrefl{}}
\ottusedrule{\ottdrulespXXrec{}}
\ottusedrule{\ottdrulespXXabs{}}
\ottusedrule{\ottdrulespXXpi{}}
\ottusedrule{\ottdrulespXXipi{}}
\ottusedrule{\ottdrulespXXeq{}}
\ottusedrule{\ottdrulespXXapp{}}
\ottusedrule{\ottdrulespXXappbeta{}}
\ottusedrule{\ottdrulespXXapprec{}}
\ottusedrule{\ottdrulespXXiapp{}}
\ottusedrule{\ottdrulespXXiappbeta{}}
\ottusedrule{\ottdrulespXXiapprec{}}
\ottusedrule{\ottdrulespXXtcon{}}
\ottusedrule{\ottdrulespXXdcon{}}
\ottusedrule{\ottdrulespXXcase{}}
\ottusedrule{\ottdrulespXXcasebeta{}}
\ottusedrule{\ottdrulespXXabort{}}
\end{ottdefnblock}}

\newcommand{\ottdefnsJStepP}{
\ottdefnstepp{}}

\newcommand{\ottdrulejXXjoin}[1]{\ottdrule[#1]{%
\ottpremise{\ottnt{m_{{\mathrm{1}}}}  \leadsto^{*}_{\mathsf{p} }  \ottnt{n}}%
\ottpremise{\ottnt{m_{{\mathrm{2}}}}  \leadsto^{*}_{\mathsf{p} }  \ottnt{n}}%
}{
 \ottnt{m_{{\mathrm{2}}}} \, \curlyveedownarrow \, \ottnt{m_{{\mathrm{2}}}} }{%
{\ottdrulename{j\_join}}{}%
}}

\newcommand{\ottdefnjoin}[1]{\begin{ottdefnblock}[#1]{$ \ottnt{n} \, \curlyveedownarrow \, \ottnt{m} $}{}
\ottusedrule{\ottdrulejXXjoin{}}
\end{ottdefnblock}}

\newcommand{\ottdefnsJJoin}{
\ottdefnjoin{}}

\newcommand{\ottdruleetXXtype}[1]{\ottdrule[#1]{%
\ottpremise{ \vdash   \ottnt{H} }%
}{
 \ottnt{H} \vdash  \star  :  \star  }{%
{\ottdrulename{et\_type}}{}%
}}

\newcommand{\ottdruleetXXcase}[1]{\ottdrule[#1]{%
\ottpremise{ \ottnt{H} \vdash \ottnt{n} : \ottmv{D} \, \overline{n_i} }%
\ottpremise{ \ottnt{H} \vdash \ottnt{M} :  \star  }%
\ottpremise{ \ottkw{data} \, \ottmv{D} \, \Xi^{+} \, \ottkw{where} \, \ottsym{\{} \, \ottcomplu{\ottmv{d_{\ottmv{i}}}  \ottsym{:}  \Xi_{\ottmv{i}}  \to \; D \; \Xi^{+}}{\ottmv{i}}{{\mathrm{1}}}{..}{\ottmv{l}} \, \ottsym{\}} \in \ottnt{H} }%
\ottpremise{ \forall  \ottmv{i}  .\ \   \ottnt{H}  \ottsym{,}  \ottsym{[}  \overline{n_i}  \ottsym{/}  \Xi^{+}  \ottsym{]}  \Xi_{\ottmv{i}}  \ottsym{,}   \ottmv{y}  :  \ottnt{n}  \ottsym{=}  \ottmv{d_{\ottmv{i}}} \, \Xi_{\ottmv{i}}  \vdash \ottnt{m_{\ottmv{i}}} : \ottnt{M}  }%
\ottpremise{ \forall  \ottmv{i}  .\ \   \ottsym{\{}  \ottmv{y}  \ottsym{\}}  \cup   \mathsf{dom}^-( \Xi_{\ottmv{i}} )  \ \#\  \ottkw{FV} \, \ottsym{(}  \ottnt{m_{\ottmv{i}}}  \ottsym{)}  }%
\ottpremise{ \text{$ \overline{x_i}_{\ottmv{i}} $ is $  \mathsf{dom}^+( \Xi_{\ottmv{i}} )  $} }%
}{
 \ottnt{H} \vdash \ottkw{case} \, \ottnt{n} \, \ottkw{of} \, \ottsym{\{} \, \ottcomplu{\ottmv{d_{\ottmv{i}}} \, \overline{x_i}_{\ottmv{i}}  \Rightarrow  \ottnt{m_{\ottmv{i}}}}{\ottmv{i}}{{\mathrm{1}}}{..}{\ottmv{l}} \, \ottsym{\}} : \ottnt{M} }{%
{\ottdrulename{et\_case}}{}%
}}

\newcommand{\ottdruleetXXvar}[1]{\ottdrule[#1]{%
\ottpremise{  \ottmv{x}  :  \ottnt{M}  \in \ottnt{H} }%
\ottpremise{ \vdash   \ottnt{H} }%
}{
 \ottnt{H} \vdash \ottmv{x} : \ottnt{M} }{%
{\ottdrulename{et\_var}}{}%
}}

\newcommand{\ottdruleetXXpi}[1]{\ottdrule[#1]{%
\ottpremise{ \ottnt{H} \vdash \ottnt{M} :  \star   \quad  \ottnt{H}  \ottsym{,}   \ottmv{x}  :  \ottnt{M}  \vdash \ottnt{N} :  \star  }%
}{
 \ottnt{H} \vdash  ( \ottmv{x} \!:\! \ottnt{M} )  \to \,  \ottnt{N}  :  \star  }{%
{\ottdrulename{et\_pi}}{}%
}}

\newcommand{\ottdruleetXXipi}[1]{\ottdrule[#1]{%
\ottpremise{ \ottnt{H} \vdash \ottnt{M} :  \star   \quad  \ottnt{H}  \ottsym{,}   \ottmv{x}  :  \ottnt{M}  \vdash \ottnt{N} :  \star  }%
}{
 \ottnt{H} \vdash  [  \ottmv{x} \!:\! \ottnt{M}  ]  \to \,  \ottnt{N}  :  \star  }{%
{\ottdrulename{et\_ipi}}{}%
}}

\newcommand{\ottdruleetXXtcon}[1]{\ottdrule[#1]{%
\ottpremise{ \ottkw{data} \, \ottmv{D} \, \Xi^{+} \, \ottkw{where} \, \ottsym{\{} \, \ottcomplu{\ottmv{d_{\ottmv{i}}}  \ottsym{:}  \Xi_{\ottmv{i}}  \to \; D \; \Xi^{+}}{\ottmv{i}}{{\mathrm{1}}}{..}{\ottmv{j}} \, \ottsym{\}} \in \ottnt{H} }%
\ottpremise{\ottnt{H}  \vdash  \overline{M_i}  \ottsym{:}  \Xi^{+}}%
}{
 \ottnt{H} \vdash \ottmv{D} \, \overline{M_i} :  \star  }{%
{\ottdrulename{et\_tcon}}{}%
}}

\newcommand{\ottdruleetXXabstcon}[1]{\ottdrule[#1]{%
\ottpremise{ \ottkw{data} \, \ottmv{D} \, \Xi^{+} \in \ottnt{H} }%
\ottpremise{\ottnt{H}  \vdash  \overline{M_i}  \ottsym{:}  \Xi^{+}}%
}{
 \ottnt{H} \vdash \ottmv{D} \, \overline{M_i} :  \star  }{%
{\ottdrulename{et\_abstcon}}{}%
}}

\newcommand{\ottdruleetXXdcon}[1]{\ottdrule[#1]{%
\ottpremise{ \ottkw{data} \, \ottmv{D} \, \Xi^{+} \, \ottkw{where} \, \ottsym{\{} \, \ottcomplu{\ottmv{d_{\ottmv{i}}}  \ottsym{:}  \Xi_{\ottmv{i}}  \to \; D \; \Xi^{+}}{\ottmv{i}}{{\mathrm{1}}}{..}{\ottmv{j}} \, \ottsym{\}} \in \ottnt{H} }%
\ottpremise{\ottnt{H}  \vdash  \overline{M_i}  \ottsym{:}  \Xi^{+}}%
\ottpremise{\ottnt{H}  \vdash  \overline{m_i}  \ottsym{:}  \ottsym{[}  \overline{M_i}  \ottsym{/}  \Xi  \ottsym{]}  \Xi_{\ottmv{i}}}%
}{
 \ottnt{H} \vdash \ottmv{d_{\ottmv{k}}} \, \overline{m_i} : \ottmv{D} \, \overline{M_i} }{%
{\ottdrulename{et\_dcon}}{}%
}}

\newcommand{\ottdruleetXXabs}[1]{\ottdrule[#1]{%
\ottpremise{ \ottnt{H}  \ottsym{,}   \ottmv{x}  :  \ottnt{M}  \vdash \ottnt{n} : \ottnt{N} }%
}{
 \ottnt{H} \vdash \lambda  \ottmv{x}  \ottsym{.}  \ottnt{n} :  ( \ottmv{x} \!:\! \ottnt{M} )  \to \,  \ottnt{N}  }{%
{\ottdrulename{et\_abs}}{}%
}}

\newcommand{\ottdruleetXXiabs}[1]{\ottdrule[#1]{%
\ottpremise{ \ottnt{H}  \ottsym{,}   \ottmv{x}  :  \ottnt{M}  \vdash \ottnt{n} : \ottnt{N} }%
\ottpremise{ \ottmv{x} \ \notin \ottkw{FV} \, \ottsym{(}  \ottnt{n}  \ottsym{)} \ }%
}{
 \ottnt{H} \vdash \lambda  \ottsym{[}  \ottsym{]}  \ottsym{.}  \ottnt{n} :  [  \ottmv{x} \!:\! \ottnt{M}  ]  \to \,  \ottnt{N}  }{%
{\ottdrulename{et\_iabs}}{}%
}}

\newcommand{\ottdruleetXXrec}[1]{\ottdrule[#1]{%
\ottpremise{ \ottnt{H}  \ottsym{,}   \ottmv{f}  :  \ottnt{M}  \vdash \ottnt{u} : \ottnt{M} }%
\ottpremise{ \ottnt{H} \vdash \ottnt{M} :  \star  }%
\ottpremise{ \text{$ \ottnt{M} $ is $  ( \ottmv{x} \!:\! \ottnt{M_{{\mathrm{1}}}} )  \to \,  \ottnt{M_{{\mathrm{2}}}}  $ or $  [  \ottmv{x} \!:\! \ottnt{M_{{\mathrm{1}}}}  ]  \to \,  \ottnt{M_{{\mathrm{2}}}}  $} }%
}{
 \ottnt{H} \vdash  \mathsf{rec}\; \ottmv{f} . \ottkw{u}  : \ottnt{M} }{%
{\ottdrulename{et\_rec}}{}%
}}

\newcommand{\ottdruleetXXapp}[1]{\ottdrule[#1]{%
\ottpremise{ \ottnt{H} \vdash \ottnt{m} :  ( \ottmv{x} \!:\! \ottnt{M} )  \to \,  \ottnt{N}  }%
\ottpremise{ \ottnt{H} \vdash \ottnt{n} : \ottnt{M} }%
\ottpremise{ \ottnt{H} \vdash \ottsym{[}  \ottnt{n}  \ottsym{/}  \ottmv{x}  \ottsym{]}  \ottnt{N} :  \star  }%
}{
 \ottnt{H} \vdash  \ottnt{m}  \;  \ottnt{n}  : \ottsym{[}  \ottnt{n}  \ottsym{/}  \ottmv{x}  \ottsym{]}  \ottnt{N} }{%
{\ottdrulename{et\_app}}{}%
}}

\newcommand{\ottdruleetXXiapp}[1]{\ottdrule[#1]{%
\ottpremise{ \ottnt{H} \vdash \ottnt{m} :  [  \ottmv{x} \!:\! \ottnt{M}  ]  \to \,  \ottnt{N}  }%
\ottpremise{ \ottnt{H} \vdash \ottnt{u} : \ottnt{M} }%
}{
 \ottnt{H} \vdash \ottnt{m}  \ottsym{[}  \ottsym{]} : \ottsym{[}  \ottnt{u}  \ottsym{/}  \ottmv{x}  \ottsym{]}  \ottnt{N} }{%
{\ottdrulename{et\_iapp}}{}%
}}

\newcommand{\ottdruleetXXabort}[1]{\ottdrule[#1]{%
\ottpremise{ \ottnt{H} \vdash \ottnt{M} :  \star  }%
}{
 \ottnt{H} \vdash \ottkw{abort} : \ottnt{M} }{%
{\ottdrulename{et\_abort}}{}%
}}

\newcommand{\ottdruleetXXeq}[1]{\ottdrule[#1]{%
\ottpremise{ \ottnt{H} \vdash \ottnt{m} : \ottnt{M}  \quad  \ottnt{H} \vdash \ottnt{n} : \ottnt{N} }%
}{
 \ottnt{H} \vdash \ottnt{m}  \ottsym{=}  \ottnt{n} :  \star  }{%
{\ottdrulename{et\_eq}}{}%
}}

\newcommand{\ottdruleetXXjoin}[1]{\ottdrule[#1]{%
\ottpremise{ \ottnt{m} \, \curlyveedownarrow \, \ottnt{n} }%
\ottpremise{ \ottnt{H} \vdash \ottnt{m}  \ottsym{=}  \ottnt{n} :  \star  }%
}{
 \ottnt{H} \vdash \ottkw{join} : \ottnt{m}  \ottsym{=}  \ottnt{n} }{%
{\ottdrulename{et\_join}}{}%
}}

\newcommand{\ottdruleetXXconv}[1]{\ottdrule[#1]{%
\ottpremise{ \ottnt{H} \vdash \ottnt{u_{{\mathrm{1}}}} : \ottnt{M_{{\mathrm{1}}}}  \ottsym{=}  \ottnt{N_{{\mathrm{1}}}}  \quad ... \quad  \ottnt{H} \vdash \ottnt{u_{\ottmv{i}}} : \ottnt{M_{\ottmv{i}}}  \ottsym{=}  \ottnt{N_{\ottmv{i}}} }%
\ottpremise{ \ottnt{H} \vdash \ottnt{m} : \ottsym{[}  \ottnt{M_{{\mathrm{1}}}}  \ottsym{/}  \ottmv{x_{{\mathrm{1}}}}  \ottsym{]} \, ... \, \ottsym{[}  \ottnt{M_{\ottmv{i}}}  \ottsym{/}  \ottmv{x_{\ottmv{i}}}  \ottsym{]}  \ottnt{M} }%
\ottpremise{ \ottnt{H} \vdash \ottsym{[}  \ottnt{N_{{\mathrm{1}}}}  \ottsym{/}  \ottmv{x_{{\mathrm{1}}}}  \ottsym{]} \, ... \, \ottsym{[}  \ottnt{N_{\ottmv{i}}}  \ottsym{/}  \ottmv{x_{\ottmv{i}}}  \ottsym{]}  \ottnt{M} :  \star  }%
}{
 \ottnt{H} \vdash \ottnt{m} : \ottsym{[}  \ottnt{N_{{\mathrm{1}}}}  \ottsym{/}  \ottmv{x_{{\mathrm{1}}}}  \ottsym{]} \, ... \, \ottsym{[}  \ottnt{N_{\ottmv{i}}}  \ottsym{/}  \ottmv{x_{\ottmv{i}}}  \ottsym{]}  \ottnt{M} }{%
{\ottdrulename{et\_conv}}{}%
}}

\newcommand{\ottdruleetXXinjdom}[1]{\ottdrule[#1]{%
\ottpremise{ \ottnt{H} \vdash \ottnt{u_{{\mathrm{1}}}} :  ( \ottmv{x} \!:\! \ottnt{M_{{\mathrm{1}}}} )  \to \,  \ottnt{N_{{\mathrm{1}}}}  \ottsym{=}   ( \ottmv{x} \!:\! \ottnt{M_{{\mathrm{2}}}} )  \to \,  \ottnt{N_{{\mathrm{2}}}}   }%
}{
 \ottnt{H} \vdash \ottkw{join} : \ottnt{M_{{\mathrm{1}}}}  \ottsym{=}  \ottnt{M_{{\mathrm{2}}}} }{%
{\ottdrulename{et\_injdom}}{}%
}}

\newcommand{\ottdruleetXXinjrng}[1]{\ottdrule[#1]{%
\ottpremise{ \ottnt{H} \vdash \ottnt{u_{{\mathrm{1}}}} :  ( \ottmv{x} \!:\! \ottnt{M} )  \to \,  \ottnt{N_{{\mathrm{1}}}}  \ottsym{=}   ( \ottmv{x} \!:\! \ottnt{M} )  \to \,  \ottnt{N_{{\mathrm{2}}}}   }%
\ottpremise{ \ottnt{H} \vdash \ottnt{u} : \ottnt{M} }%
}{
 \ottnt{H} \vdash \ottkw{join} : \ottsym{[}  \ottnt{u}  \ottsym{/}  \ottmv{x}  \ottsym{]}  \ottnt{N_{{\mathrm{1}}}}  \ottsym{=}  \ottsym{[}  \ottnt{u}  \ottsym{/}  \ottmv{x}  \ottsym{]}  \ottnt{N_{{\mathrm{2}}}} }{%
{\ottdrulename{et\_injrng}}{}%
}}

\newcommand{\ottdruleetXXiinjdom}[1]{\ottdrule[#1]{%
\ottpremise{ \ottnt{H} \vdash \ottnt{u_{{\mathrm{1}}}} :  [  \ottmv{x} \!:\! \ottnt{M_{{\mathrm{1}}}}  ]  \to \,  \ottnt{N_{{\mathrm{1}}}}  \ottsym{=}   [  \ottmv{x} \!:\! \ottnt{M_{{\mathrm{2}}}}  ]  \to \,  \ottnt{N_{{\mathrm{2}}}}   }%
}{
 \ottnt{H} \vdash \ottkw{join} : \ottnt{M_{{\mathrm{1}}}}  \ottsym{=}  \ottnt{M_{{\mathrm{2}}}} }{%
{\ottdrulename{et\_iinjdom}}{}%
}}

\newcommand{\ottdruleetXXiinjrng}[1]{\ottdrule[#1]{%
\ottpremise{ \ottnt{H} \vdash \ottnt{u_{{\mathrm{1}}}} :  [  \ottmv{x} \!:\! \ottnt{M}  ]  \to \,  \ottnt{N_{{\mathrm{1}}}}  \ottsym{=}   [  \ottmv{x} \!:\! \ottnt{M}  ]  \to \,  \ottnt{N_{{\mathrm{2}}}}   }%
\ottpremise{ \ottnt{H} \vdash \ottnt{u} : \ottnt{M} }%
}{
 \ottnt{H} \vdash \ottkw{join} : \ottsym{[}  \ottnt{u}  \ottsym{/}  \ottmv{x}  \ottsym{]}  \ottnt{N_{{\mathrm{1}}}}  \ottsym{=}  \ottsym{[}  \ottnt{u}  \ottsym{/}  \ottmv{x}  \ottsym{]}  \ottnt{N_{{\mathrm{2}}}} }{%
{\ottdrulename{et\_iinjrng}}{}%
}}

\newcommand{\ottdruleetXXinjtcon}[1]{\ottdrule[#1]{%
\ottpremise{ \ottnt{H} \vdash \ottnt{u_{{\mathrm{1}}}} : \ottmv{D} \, \overline{n_i}  \ottsym{=}  \ottmv{D} \, \overline{n_i}' }%
}{
 \ottnt{H} \vdash \ottkw{join} : \ottnt{n_{\ottmv{k}}}  \ottsym{=}  \ottnt{n'_{\ottmv{k}}} }{%
{\ottdrulename{et\_injtcon}}{}%
}}

\newcommand{\ottdefnetyping}[1]{\begin{ottdefnblock}[#1]{$ \ottnt{H} \vdash \ottnt{m} : \ottnt{M} $}{}
\ottusedrule{\ottdruleetXXtype{}}
\ottusedrule{\ottdruleetXXcase{}}
\ottusedrule{\ottdruleetXXvar{}}
\ottusedrule{\ottdruleetXXpi{}}
\ottusedrule{\ottdruleetXXipi{}}
\ottusedrule{\ottdruleetXXtcon{}}
\ottusedrule{\ottdruleetXXabstcon{}}
\ottusedrule{\ottdruleetXXdcon{}}
\ottusedrule{\ottdruleetXXabs{}}
\ottusedrule{\ottdruleetXXiabs{}}
\ottusedrule{\ottdruleetXXrec{}}
\ottusedrule{\ottdruleetXXapp{}}
\ottusedrule{\ottdruleetXXiapp{}}
\ottusedrule{\ottdruleetXXabort{}}
\ottusedrule{\ottdruleetXXeq{}}
\ottusedrule{\ottdruleetXXjoin{}}
\ottusedrule{\ottdruleetXXconv{}}
\ottusedrule{\ottdruleetXXinjdom{}}
\ottusedrule{\ottdruleetXXinjrng{}}
\ottusedrule{\ottdruleetXXiinjdom{}}
\ottusedrule{\ottdruleetXXiinjrng{}}
\ottusedrule{\ottdruleetXXinjtcon{}}
\end{ottdefnblock}}

\newcommand{\ottdruleeenvXXwfXXempty}[1]{\ottdrule[#1]{%
}{
 \vdash    \cdot  }{%
{\ottdrulename{eenv\_wf\_empty}}{}%
}}

\newcommand{\ottdruleeenvXXwfXXvar}[1]{\ottdrule[#1]{%
\ottpremise{ \vdash   \ottnt{H}  \quad  \ottmv{x} \ \notin \ottkw{dom} \, \ottsym{(}  \ottnt{H}  \ottsym{)} \ }%
\ottpremise{ \ottnt{H} \vdash \ottnt{M} :  \star  }%
}{
 \vdash   \ottnt{H}  \ottsym{,}   \ottmv{x}  :  \ottnt{M}  }{%
{\ottdrulename{eenv\_wf\_var}}{}%
}}

\newcommand{\ottdruleeenvXXwfXXdtype}[1]{\ottdrule[#1]{%
\ottpremise{ \vdash   \ottnt{H}  \ottsym{,}  \Xi }%
\ottpremise{ \ottmv{D} \ \notin  \ottkw{dom} \, \ottsym{(}  \ottnt{H}  \ottsym{)} \ }%
\ottpremise{ \forall  \ottmv{i}  .\ \   \ottmv{d_{\ottmv{i}}} \ \notin  \ottkw{dom} \, \ottsym{(}  \ottnt{H}  \ottsym{)} \  }%
\ottpremise{ \forall  \ottmv{i}  .\ \   \ottnt{H}  \ottsym{,}  \ottkw{data} \, \ottmv{D} \, \Xi^{+}  \ottsym{,}  \Xi \vdash \Xi_{\ottmv{i}}  \to  \ottmv{D} \, \Xi^{+} :  \star   }%
}{
 \vdash   \ottnt{H}  \ottsym{,}  \ottkw{data} \, \ottmv{D} \, \Xi^{+} \, \ottkw{where} \, \ottsym{\{} \, \ottcomplu{\ottmv{d_{\ottmv{i}}}  \ottsym{:}  \Xi_{\ottmv{i}}  \to \; D \; \Xi^{+}}{\ottmv{i}}{{\mathrm{1}}}{..}{\ottmv{j}} \, \ottsym{\}} }{%
{\ottdrulename{eenv\_wf\_dtype}}{}%
}}

\newcommand{\ottdruleeenvXXwfXXabsdtype}[1]{\ottdrule[#1]{%
\ottpremise{ \vdash   \ottnt{H}  \ottsym{,}  \Xi  \quad  \ottmv{D} \ \notin  \ottkw{dom} \, \ottsym{(}  \ottnt{H}  \ottsym{)} \ }%
}{
 \vdash   \ottnt{H}  \ottsym{,}  \ottkw{data} \, \ottmv{D} \, \Xi^{+} }{%
{\ottdrulename{eenv\_wf\_absdtype}}{}%
}}

\newcommand{\ottdefneenvXXwf}[1]{\begin{ottdefnblock}[#1]{$ \vdash   \ottnt{H} $}{\ottcom{$\ottnt{H}$ is a well-formed environment}}
\ottusedrule{\ottdruleeenvXXwfXXempty{}}
\ottusedrule{\ottdruleeenvXXwfXXvar{}}
\ottusedrule{\ottdruleeenvXXwfXXdtype{}}
\ottusedrule{\ottdruleeenvXXwfXXabsdtype{}}
\end{ottdefnblock}}

\newcommand{\ottdruleetlXXempty}[1]{\ottdrule[#1]{%
}{
\ottnt{H}  \vdash   \cdot   \ottsym{:}   \cdot }{%
{\ottdrulename{etl\_empty}}{}%
}}

\newcommand{\ottdruleetlXXcons}[1]{\ottdrule[#1]{%
\ottpremise{ \ottnt{H} \vdash \ottnt{m} : \ottnt{M} }%
\ottpremise{ \ottnt{H} \vdash \ottnt{M} :  \star  }%
\ottpremise{\ottnt{H}  \vdash  \overline{m_i}  \ottsym{:}  \ottsym{[}  \ottnt{m}  \ottsym{/}  \ottmv{x}  \ottsym{]}  \Xi}%
}{
\ottnt{H}  \vdash   \ottnt{m}  \;  \overline{m_i}   \ottsym{:}  \ottsym{(}  \ottmv{x}  \ottsym{:}  \ottnt{M}  \ottsym{)}  \Xi}{%
{\ottdrulename{etl\_cons}}{}%
}}

\newcommand{\ottdruleetlXXicons}[1]{\ottdrule[#1]{%
\ottpremise{ \ottnt{H} \vdash \ottnt{u} : \ottnt{M} }%
\ottpremise{ \ottnt{H} \vdash \ottnt{M} :  \star  }%
\ottpremise{\ottnt{H}  \vdash  \overline{m_i}  \ottsym{:}  \ottsym{[}  \ottnt{u}  \ottsym{/}  \ottmv{x}  \ottsym{]}  \Xi}%
}{
\ottnt{H}  \vdash   [] \;  \overline{m_i}   \ottsym{:}  \ottsym{[}  \ottmv{x}  \ottsym{:}  \ottnt{M}  \ottsym{]}  \Xi}{%
{\ottdrulename{etl\_icons}}{}%
}}

\newcommand{\ottdefnetypingl}[1]{\begin{ottdefnblock}[#1]{$\ottnt{H}  \vdash  \overline{m_i}  \ottsym{:}  \Xi$}{}
\ottusedrule{\ottdruleetlXXempty{}}
\ottusedrule{\ottdruleetlXXcons{}}
\ottusedrule{\ottdruleetlXXicons{}}
\end{ottdefnblock}}

\newcommand{\ottdefnsJetyp}{
\ottdefnetyping{}
\ottdefneenvXXwf{}
\ottdefnetypingl{}}

\newcommand{\ottdruletXXtype}[1]{\ottdrule[#1]{%
\ottpremise{ \vdash   \Gamma }%
}{
 \Gamma \vdash  \star  :  \star  }{%
{\ottdrulename{t\_type}}{}%
}}

\newcommand{\ottdruletXXcase}[1]{\ottdrule[#1]{%
\ottpremise{ \Gamma \vdash \ottnt{b} :  \ottmv{D}  \;  \overline{B_i}  }%
\ottpremise{ \Gamma \vdash \ottnt{A} :  \star  }%
\ottpremise{ \ottkw{data} \, \ottmv{D} \, \Delta^{+} \, \ottkw{where} \, \ottsym{\{} \, \ottcomplu{\ottmv{d_{\ottmv{i}}}  \ottsym{:}  \Delta_{\ottmv{i}}  \to \; D \; \Delta^{+}}{\ottmv{i}}{{\mathrm{1}}}{..}{\ottmv{l}} \, \ottsym{\}}  \in  \Gamma }%
\ottpremise{ \forall  \ottmv{i}  .\ \   \Gamma  \ottsym{,}  \ottsym{[}  \overline{B_i}  \ottsym{/}  \Delta^{+}  \ottsym{]}  \Delta_{\ottmv{i}}  \ottsym{,}   \ottmv{y}  :  \ottnt{b}  \ottsym{=}  \ottmv{d_{\ottmv{i}}} \, \Delta_{\ottmv{i}}  \vdash \ottnt{a_{\ottmv{i}}} : \ottnt{A}  }%
\ottpremise{ \forall  \ottmv{i}  .\ \   \ottsym{\{}  \ottmv{y}  \ottsym{\}}  \cup  \ottsym{dom-}  \ottsym{(}  \Delta_{\ottmv{i}}  \ottsym{)} \ \#\  \ottkw{FV} \, \ottsym{(}  \ottsym{\mbox{$\mid$}}  \ottnt{a_{\ottmv{i}}}  \ottsym{\mbox{$\mid$}}  \ottsym{)}  }%
}{
 \Gamma \vdash \ottkw{case} \, \ottnt{b} \, \ottkw{as} \, \ottsym{[y]} \, \ottkw{of} \, \ottsym{\{} \, \ottcomplu{\ottmv{d_{\ottmv{i}}} \, \Delta_{\ottmv{i}}  \Rightarrow  \ottnt{a_{\ottmv{i}}}}{\ottmv{i}}{{\mathrm{1}}}{..}{\ottmv{l}} \, \ottsym{\}} : \ottnt{A} }{%
{\ottdrulename{t\_case}}{}%
}}

\newcommand{\ottdruletXXvar}[1]{\ottdrule[#1]{%
\ottpremise{  \ottmv{x}  :  \ottnt{A}   \in  \Gamma }%
\ottpremise{ \vdash   \Gamma }%
}{
 \Gamma \vdash \ottmv{x} : \ottnt{A} }{%
{\ottdrulename{t\_var}}{}%
}}

\newcommand{\ottdruletXXpi}[1]{\ottdrule[#1]{%
\ottpremise{ \Gamma \vdash \ottnt{A} :  \star   \quad  \Gamma  \ottsym{,}   \ottmv{x}  :  \ottnt{A}  \vdash \ottnt{B} :  \star  }%
}{
 \Gamma \vdash  ( \ottmv{x} \!:\! \ottnt{A} )  \to \,  \ottnt{B}  :  \star  }{%
{\ottdrulename{t\_pi}}{}%
}}

\newcommand{\ottdruletXXipi}[1]{\ottdrule[#1]{%
\ottpremise{ \Gamma \vdash \ottnt{A} :  \star   \quad  \Gamma  \ottsym{,}   \ottmv{x}  :  \ottnt{A}  \vdash \ottnt{B} :  \star  }%
}{
 \Gamma \vdash  [  \ottmv{x} \!:\! \ottnt{A}  ]  \to \,  \ottnt{B}  :  \star  }{%
{\ottdrulename{t\_ipi}}{}%
}}

\newcommand{\ottdruletXXtcon}[1]{\ottdrule[#1]{%
\ottpremise{ \ottkw{data} \, \ottmv{D} \, \Delta^{+} \, \ottkw{where} \, \ottsym{\{} \, \ottcomplu{\ottmv{d_{\ottmv{i}}}  \ottsym{:}  \Delta_{\ottmv{i}}  \to \; D \; \Delta^{+}}{\ottmv{i}}{{\mathrm{1}}}{..}{\ottmv{j}} \, \ottsym{\}}  \in  \Gamma }%
\ottpremise{\Gamma  \vdash  \overline{A_i}  \ottsym{:}  \Delta^{+}}%
}{
 \Gamma \vdash  \ottmv{D}  \;  \overline{A_i}  :  \star  }{%
{\ottdrulename{t\_tcon}}{}%
}}

\newcommand{\ottdruletXXabstcon}[1]{\ottdrule[#1]{%
\ottpremise{ \ottkw{data} \, \ottmv{D} \, \Delta^{+}  \in  \Gamma }%
\ottpremise{\Gamma  \vdash  \overline{A_i}  \ottsym{:}  \Delta^{+}}%
}{
 \Gamma \vdash  \ottmv{D}  \;  \overline{A_i}  :  \star  }{%
{\ottdrulename{t\_abstcon}}{}%
}}

\newcommand{\ottdruletXXdcon}[1]{\ottdrule[#1]{%
\ottpremise{ \ottkw{data} \, \ottmv{D} \, \Delta^{+} \, \ottkw{where} \, \ottsym{\{} \, \ottcomplu{\ottmv{d_{\ottmv{i}}}  \ottsym{:}  \Delta_{\ottmv{i}}  \to \; D \; \Delta^{+}}{\ottmv{i}}{{\mathrm{1}}}{..}{\ottmv{j}} \, \ottsym{\}}  \in  \Gamma }%
\ottpremise{\Gamma  \vdash  \overline{A_i}  \ottsym{:}  \Delta^{+}}%
\ottpremise{\Gamma  \vdash  \overline{a_i}  \ottsym{:}  \ottsym{[}  \overline{A_i}  \ottsym{/}  \Delta  \ottsym{]}  \Delta_{\ottmv{i}}}%
}{
 \Gamma \vdash  \ottmv{d_{\ottmv{k}}}  \; [  \overline{A_i}  ] \;  \overline{a_i}  :  \ottmv{D}  \;  \overline{A_i}  }{%
{\ottdrulename{t\_dcon}}{}%
}}

\newcommand{\ottdruletXXabs}[1]{\ottdrule[#1]{%
\ottpremise{ \Gamma  \ottsym{,}   \ottmv{x}  :  \ottnt{A}  \vdash \ottnt{b} : \ottnt{B} }%
}{
 \Gamma \vdash \lambda  \ottmv{x}  \ottsym{:}  \ottnt{A}  \ottsym{.}  \ottnt{b} :  ( \ottmv{x} \!:\! \ottnt{A} )  \to \,  \ottnt{B}  }{%
{\ottdrulename{t\_abs}}{}%
}}

\newcommand{\ottdruletXXiabs}[1]{\ottdrule[#1]{%
\ottpremise{ \Gamma  \ottsym{,}   \ottmv{x}  :  \ottnt{A}  \vdash \ottnt{b} : \ottnt{B} }%
\ottpremise{ \ottmv{x} \ \notin \ottkw{FV} \, \ottsym{(}  \ottsym{\mbox{$\mid$}}  \ottnt{b}  \ottsym{\mbox{$\mid$}}  \ottsym{)} \ }%
}{
 \Gamma \vdash \lambda  \ottsym{[}  \ottmv{x}  \ottsym{:}  \ottnt{A}  \ottsym{]}  \ottsym{.}  \ottnt{b} :  [  \ottmv{x} \!:\! \ottnt{A}  ]  \to \,  \ottnt{B}  }{%
{\ottdrulename{t\_iabs}}{}%
}}

\newcommand{\ottdruletXXrec}[1]{\ottdrule[#1]{%
\ottpremise{ \Gamma  \ottsym{,}   \ottmv{f}  :  \ottnt{A}  \vdash \ottnt{v} : \ottnt{A}  \quad  \Gamma \vdash \ottnt{A} :  \star  }%
\ottpremise{ \text{$ \ottnt{A} $ is $  ( \ottmv{x} \!:\! \ottnt{A_{{\mathrm{1}}}} )  \to \,  \ottnt{A_{{\mathrm{2}}}}  $ or $  [  \ottmv{x} \!:\! \ottnt{A_{{\mathrm{1}}}}  ]  \to \,  \ottnt{A_{{\mathrm{2}}}}  $} }%
}{
 \Gamma \vdash  \mathsf{rec}\; \ottmv{f} : \ottnt{A} . \ottnt{v}  : \ottnt{A} }{%
{\ottdrulename{t\_rec}}{}%
}}

\newcommand{\ottdruletXXapp}[1]{\ottdrule[#1]{%
\ottpremise{ \Gamma \vdash \ottnt{a} :  ( \ottmv{x} \!:\! \ottnt{A} )  \to \,  \ottnt{B}  }%
\ottpremise{ \Gamma \vdash \ottnt{b} : \ottnt{A} }%
\ottpremise{ \Gamma \vdash \ottsym{[}  \ottnt{b}  \ottsym{/}  \ottmv{x}  \ottsym{]}  \ottnt{B} :  \star  }%
}{
 \Gamma \vdash  \ottnt{a}  \;  \ottnt{b}  : \ottsym{[}  \ottnt{b}  \ottsym{/}  \ottmv{x}  \ottsym{]}  \ottnt{B} }{%
{\ottdrulename{t\_app}}{}%
}}

\newcommand{\ottdruletXXiapp}[1]{\ottdrule[#1]{%
\ottpremise{ \Gamma \vdash \ottnt{a} :  [  \ottmv{x} \!:\! \ottnt{A}  ]  \to \,  \ottnt{B}  }%
\ottpremise{ \Gamma \vdash \ottnt{v} : \ottnt{A} }%
}{
 \Gamma \vdash  \ottnt{a}  \; [  \ottnt{v}  ]  : \ottsym{[}  \ottnt{v}  \ottsym{/}  \ottmv{x}  \ottsym{]}  \ottnt{B} }{%
{\ottdrulename{t\_iapp}}{}%
}}

\newcommand{\ottdruletXXabort}[1]{\ottdrule[#1]{%
\ottpremise{ \Gamma \vdash \ottnt{A} :  \star  }%
}{
 \Gamma \vdash  \mathsf{abort}_{ \ottnt{A} }  : \ottnt{A} }{%
{\ottdrulename{t\_abort}}{}%
}}

\newcommand{\ottdruletXXeq}[1]{\ottdrule[#1]{%
\ottpremise{ \Gamma \vdash \ottnt{a} : \ottnt{A}  \quad  \Gamma \vdash \ottnt{b} : \ottnt{B} }%
}{
 \Gamma \vdash \ottnt{a}  \ottsym{=}  \ottnt{b} :  \star  }{%
{\ottdrulename{t\_eq}}{}%
}}

\newcommand{\ottdruletXXjoin}[1]{\ottdrule[#1]{%
\ottpremise{ \ottsym{\mbox{$\mid$}}  \ottnt{a}  \ottsym{\mbox{$\mid$}}   \leadsto_{\mathsf{cbv} }  ^{ \ottmv{i} }  \ottnt{n}  \quad  \ottsym{\mbox{$\mid$}}  \ottnt{b}  \ottsym{\mbox{$\mid$}}   \leadsto_{\mathsf{cbv} }  ^{ \ottmv{j} }  \ottnt{n} }%
\ottpremise{ \Gamma \vdash \ottnt{a}  \ottsym{=}  \ottnt{b} :  \star  }%
}{
 \Gamma \vdash  \mathsf{join}_{ \ottnt{a} = \ottnt{b} } \;  \ottmv{i}  \;  \ottmv{j}  : \ottnt{a}  \ottsym{=}  \ottnt{b} }{%
{\ottdrulename{t\_join}}{}%
}}

\newcommand{\ottdruletXXconv}[1]{\ottdrule[#1]{%
\ottpremise{ \forall  \ottmv{j}  .\ \  \ottsym{(}   \ottsym{(}    \text{$ \ottnt{P_{\ottmv{j}}} $ is $ \ottnt{v_{\ottmv{j}}} $ }   \text{ and }   \Gamma \vdash \ottnt{v_{\ottmv{j}}} : \ottnt{A_{\ottmv{j}}}  \ottsym{=}  \ottnt{B_{\ottmv{j}}}    \ottsym{)}  \text{ or }  \ottsym{(}    \text{$ \ottnt{P_{\ottmv{j}}} $ is $ \ottsym{[}  \ottnt{A_{\ottmv{j}}}  \ottsym{=}  \ottnt{B_{\ottmv{j}}}  \ottsym{]} $ }   \text{ and }   \ottmv{x_{\ottmv{j}}} \ \notin \ottkw{FV} \, \ottsym{(}  \ottsym{\mbox{$\mid$}}  \ottnt{A}  \ottsym{\mbox{$\mid$}}  \ottsym{)} \    \ottsym{)}   \ottsym{)} }%
\ottpremise{ \Gamma \vdash \ottnt{a} : \ottsym{[}  \ottnt{A_{{\mathrm{1}}}}  \ottsym{/}  \ottmv{x_{{\mathrm{1}}}}  \ottsym{]} \, ... \, \ottsym{[}  \ottnt{A_{\ottmv{i}}}  \ottsym{/}  \ottmv{x_{\ottmv{i}}}  \ottsym{]}  \ottnt{A} }%
\ottpremise{ \Gamma \vdash \ottsym{[}  \ottnt{B_{{\mathrm{1}}}}  \ottsym{/}  \ottmv{x_{{\mathrm{1}}}}  \ottsym{]} \, ... \, \ottsym{[}  \ottnt{B_{\ottmv{i}}}  \ottsym{/}  \ottmv{x_{\ottmv{i}}}  \ottsym{]}  \ottnt{A} :  \star  }%
}{
 \Gamma \vdash \ottkw{conv} \, \ottnt{a} \, \ottkw{at} \, \ottsym{[}  \mathsf{\sim}  \ottnt{P_{{\mathrm{1}}}}  \ottsym{/}  \ottmv{x_{{\mathrm{1}}}}  \ottsym{]} \, ... \, \ottsym{[}  \mathsf{\sim}  \ottnt{P_{\ottmv{i}}}  \ottsym{/}  \ottmv{x_{\ottmv{i}}}  \ottsym{]}  \ottnt{A} : \ottsym{[}  \ottnt{B_{{\mathrm{1}}}}  \ottsym{/}  \ottmv{x_{{\mathrm{1}}}}  \ottsym{]} \, ... \, \ottsym{[}  \ottnt{B_{\ottmv{i}}}  \ottsym{/}  \ottmv{x_{\ottmv{i}}}  \ottsym{]}  \ottnt{A} }{%
{\ottdrulename{t\_conv}}{}%
}}

\newcommand{\ottdruletXXinjdom}[1]{\ottdrule[#1]{%
\ottpremise{ \Gamma \vdash \ottnt{v_{{\mathrm{1}}}} : \ottsym{(}   ( \ottmv{x} \!:\! \ottnt{A_{{\mathrm{1}}}} )  \to \,  \ottnt{B_{{\mathrm{1}}}}   \ottsym{)}  \ottsym{=}  \ottsym{(}   ( \ottmv{x} \!:\! \ottnt{A_{{\mathrm{2}}}} )  \to \,  \ottnt{B_{{\mathrm{2}}}}   \ottsym{)} }%
}{
 \Gamma \vdash \ottkw{injdom} \, \ottnt{v_{{\mathrm{1}}}} : \ottnt{A_{{\mathrm{1}}}}  \ottsym{=}  \ottnt{A_{{\mathrm{2}}}} }{%
{\ottdrulename{t\_injdom}}{}%
}}

\newcommand{\ottdruletXXinjrng}[1]{\ottdrule[#1]{%
\ottpremise{ \Gamma \vdash \ottnt{v_{{\mathrm{1}}}} : \ottsym{(}   ( \ottmv{x} \!:\! \ottnt{A} )  \to \,  \ottnt{B_{{\mathrm{1}}}}   \ottsym{)}  \ottsym{=}  \ottsym{(}   ( \ottmv{x} \!:\! \ottnt{A} )  \to \,  \ottnt{B_{{\mathrm{2}}}}   \ottsym{)}  \quad  \Gamma \vdash \ottnt{v} : \ottnt{A} }%
}{
 \Gamma \vdash \ottkw{injrng} \, \ottnt{v_{{\mathrm{1}}}} \, \ottnt{v} : \ottsym{[}  \ottnt{v}  \ottsym{/}  \ottmv{x}  \ottsym{]}  \ottnt{B_{{\mathrm{1}}}}  \ottsym{=}  \ottsym{[}  \ottnt{v}  \ottsym{/}  \ottmv{x}  \ottsym{]}  \ottnt{B_{{\mathrm{2}}}} }{%
{\ottdrulename{t\_injrng}}{}%
}}

\newcommand{\ottdruletXXiinjdom}[1]{\ottdrule[#1]{%
\ottpremise{ \Gamma \vdash \ottnt{v_{{\mathrm{1}}}} : \ottsym{(}   [  \ottmv{x} \!:\! \ottnt{A_{{\mathrm{1}}}}  ]  \to \,  \ottnt{B_{{\mathrm{1}}}}   \ottsym{)}  \ottsym{=}  \ottsym{(}   [  \ottmv{x} \!:\! \ottnt{A_{{\mathrm{2}}}}  ]  \to \,  \ottnt{B_{{\mathrm{2}}}}   \ottsym{)} }%
}{
 \Gamma \vdash \ottkw{injdom} \, \ottnt{v_{{\mathrm{1}}}} : \ottnt{A_{{\mathrm{1}}}}  \ottsym{=}  \ottnt{A_{{\mathrm{2}}}} }{%
{\ottdrulename{t\_iinjdom}}{}%
}}

\newcommand{\ottdruletXXiinjrng}[1]{\ottdrule[#1]{%
\ottpremise{ \Gamma \vdash \ottnt{v_{{\mathrm{1}}}} : \ottsym{(}   [  \ottmv{x} \!:\! \ottnt{A}  ]  \to \,  \ottnt{B_{{\mathrm{1}}}}   \ottsym{)}  \ottsym{=}  \ottsym{(}   [  \ottmv{x} \!:\! \ottnt{A}  ]  \to \,  \ottnt{B_{{\mathrm{2}}}}   \ottsym{)}  \quad  \Gamma \vdash \ottnt{v} : \ottnt{A} }%
}{
 \Gamma \vdash \ottkw{injrng} \, \ottnt{v_{{\mathrm{1}}}} \, \ottnt{v} : \ottsym{[}  \ottnt{v}  \ottsym{/}  \ottmv{x}  \ottsym{]}  \ottnt{B_{{\mathrm{1}}}}  \ottsym{=}  \ottsym{[}  \ottnt{v}  \ottsym{/}  \ottmv{x}  \ottsym{]}  \ottnt{B_{{\mathrm{2}}}} }{%
{\ottdrulename{t\_iinjrng}}{}%
}}

\newcommand{\ottdruletXXinjtcon}[1]{\ottdrule[#1]{%
\ottpremise{ \Gamma \vdash \ottnt{v_{{\mathrm{1}}}} :  \ottmv{D}  \;  \overline{A_i}   \ottsym{=}   \ottmv{D}  \;  \overline{A_i}'  }%
}{
 \Gamma \vdash  \mathsf{injtcon}_{ \ottmv{k} } \;  \ottnt{v_{{\mathrm{1}}}}  : \ottnt{A_{\ottmv{k}}}  \ottsym{=}  \ottnt{A'_{\ottmv{k}}} }{%
{\ottdrulename{t\_injtcon}}{}%
}}

\newcommand{\ottdefntyping}[1]{\begin{ottdefnblock}[#1]{$ \Gamma \vdash \ottnt{a} : \ottnt{A} $}{}
\ottusedrule{\ottdruletXXtype{}}
\ottusedrule{\ottdruletXXcase{}}
\ottusedrule{\ottdruletXXvar{}}
\ottusedrule{\ottdruletXXpi{}}
\ottusedrule{\ottdruletXXipi{}}
\ottusedrule{\ottdruletXXtcon{}}
\ottusedrule{\ottdruletXXabstcon{}}
\ottusedrule{\ottdruletXXdcon{}}
\ottusedrule{\ottdruletXXabs{}}
\ottusedrule{\ottdruletXXiabs{}}
\ottusedrule{\ottdruletXXrec{}}
\ottusedrule{\ottdruletXXapp{}}
\ottusedrule{\ottdruletXXiapp{}}
\ottusedrule{\ottdruletXXabort{}}
\ottusedrule{\ottdruletXXeq{}}
\ottusedrule{\ottdruletXXjoin{}}
\ottusedrule{\ottdruletXXconv{}}
\ottusedrule{\ottdruletXXinjdom{}}
\ottusedrule{\ottdruletXXinjrng{}}
\ottusedrule{\ottdruletXXiinjdom{}}
\ottusedrule{\ottdruletXXiinjrng{}}
\ottusedrule{\ottdruletXXinjtcon{}}
\end{ottdefnblock}}

\newcommand{\ottdruleenvXXwfXXempty}[1]{\ottdrule[#1]{%
}{
 \vdash    {\cdot }  }{%
{\ottdrulename{env\_wf\_empty}}{}%
}}

\newcommand{\ottdruleenvXXwfXXvar}[1]{\ottdrule[#1]{%
\ottpremise{ \vdash   \Gamma  \quad  \ottmv{x} \ \notin \ottkw{dom} \, \ottsym{(}  \Gamma  \ottsym{)} \ }%
\ottpremise{ \Gamma \vdash \ottnt{A} :  \star  }%
}{
 \vdash   \Gamma  \ottsym{,}   \ottmv{x}  :  \ottnt{A}  }{%
{\ottdrulename{env\_wf\_var}}{}%
}}

\newcommand{\ottdruleenvXXwfXXdtype}[1]{\ottdrule[#1]{%
\ottpremise{ \vdash   \Gamma  \ottsym{,}  \Delta  \quad  \ottmv{D} \ \notin  \ottkw{dom} \, \ottsym{(}  \ottnt{H}  \ottsym{)} \  \quad \ottcomplu{ \Gamma  \ottsym{,}  \ottkw{data} \, \ottmv{D} \, \Delta^{+}  \ottsym{,}  \Delta \vdash \Delta_{\ottmv{i}}  \to  \ottmv{D} \, \Delta^{+} :  \star  }{\ottmv{i}}{{\mathrm{1}}}{..}{\ottmv{j}}}%
}{
 \vdash   \Gamma  \ottsym{,}  \ottkw{data} \, \ottmv{D} \, \Delta^{+} \, \ottkw{where} \, \ottsym{\{} \, \ottcomplu{\ottmv{d_{\ottmv{i}}}  \ottsym{:}  \Delta_{\ottmv{i}}  \to \; D \; \Delta^{+}}{\ottmv{i}}{{\mathrm{1}}}{..}{\ottmv{j}} \, \ottsym{\}} }{%
{\ottdrulename{env\_wf\_dtype}}{}%
}}

\newcommand{\ottdruleenvXXwfXXabsdtype}[1]{\ottdrule[#1]{%
\ottpremise{ \vdash   \Gamma  \ottsym{,}  \Delta  \quad  \ottmv{D} \ \notin  \ottkw{dom} \, \ottsym{(}  \ottnt{H}  \ottsym{)} \ }%
}{
 \vdash   \Gamma  \ottsym{,}  \ottkw{data} \, \ottmv{D} \, \Delta^{+} }{%
{\ottdrulename{env\_wf\_absdtype}}{}%
}}

\newcommand{\ottdefnenvXXwf}[1]{\begin{ottdefnblock}[#1]{$ \vdash   \Gamma $}{\ottcom{$\Gamma$ is a well-formed environment}}
\ottusedrule{\ottdruleenvXXwfXXempty{}}
\ottusedrule{\ottdruleenvXXwfXXvar{}}
\ottusedrule{\ottdruleenvXXwfXXdtype{}}
\ottusedrule{\ottdruleenvXXwfXXabsdtype{}}
\end{ottdefnblock}}

\newcommand{\ottdruletlXXempty}[1]{\ottdrule[#1]{%
}{
\Gamma  \vdash   \cdot   \ottsym{:}   \cdot }{%
{\ottdrulename{tl\_empty}}{}%
}}

\newcommand{\ottdruletlXXcons}[1]{\ottdrule[#1]{%
\ottpremise{ \Gamma \vdash \ottnt{a} : \ottnt{A} }%
\ottpremise{ \Gamma \vdash \ottnt{A} :  \star  }%
\ottpremise{\Gamma  \vdash  \overline{a_i}  \ottsym{:}  \ottsym{[}  \ottnt{a}  \ottsym{/}  \ottmv{x}  \ottsym{]}  \Delta}%
}{
\Gamma  \vdash   \ottnt{a}  \;  \overline{a_i}   \ottsym{:}  \ottsym{(}  \ottmv{x}  \ottsym{:}  \ottnt{A}  \ottsym{)}  \Delta}{%
{\ottdrulename{tl\_cons}}{}%
}}

\newcommand{\ottdruletlXXicons}[1]{\ottdrule[#1]{%
\ottpremise{ \Gamma \vdash \ottnt{a} : \ottnt{A} }%
\ottpremise{ \Gamma \vdash \ottnt{A} :  \star  }%
\ottpremise{\Gamma  \vdash  \overline{a_i}  \ottsym{:}  \ottsym{[}  \ottnt{a}  \ottsym{/}  \ottmv{x}  \ottsym{]}  \Delta}%
}{
\Gamma  \vdash   [  \ottnt{a}  ] \;  \overline{a_i}   \ottsym{:}  \ottsym{[}  \ottmv{x}  \ottsym{:}  \ottnt{A}  \ottsym{]}  \Delta}{%
{\ottdrulename{tl\_icons}}{}%
}}

\newcommand{\ottdefntypingl}[1]{\begin{ottdefnblock}[#1]{$\Gamma  \vdash  \overline{a_i}  \ottsym{:}  \Delta$}{}
\ottusedrule{\ottdruletlXXempty{}}
\ottusedrule{\ottdruletlXXcons{}}
\ottusedrule{\ottdruletlXXicons{}}
\end{ottdefnblock}}

\newcommand{\ottdefnsJtyp}{
\ottdefntyping{}
\ottdefnenvXXwf{}
\ottdefntypingl{}}

\newcommand{\ottdrulejoinXXnoXXannot}[1]{\ottdrule[#1]{%
\ottpremise{\ottsym{\mbox{$\mid$}}  \ottnt{a}  \ottsym{\mbox{$\mid$}}  \leadsto^{*}_{\mathsf{p} }  \ottnt{n} \quad \ottsym{\mbox{$\mid$}}  \ottnt{b}  \ottsym{\mbox{$\mid$}}  \leadsto^{*}_{\mathsf{p} }  \ottnt{n}}%
\ottpremise{ \Gamma \vdash \ottnt{a}  \ottsym{=}  \ottnt{b} :  \star  }%
}{
 \Gamma \vdash \ottkw{join} : \ottnt{a}  \ottsym{=}  \ottnt{b} }{%
{\ottdrulename{join\_no\_annot}}{}%
}}

\newcommand{\ottdruleappXXval}[1]{\ottdrule[#1]{%
\ottpremise{ \Gamma \vdash \ottnt{a} :  ( \ottmv{x} \!:\! \ottnt{A} )  \to \,  \ottnt{B}  }%
\ottpremise{ \Gamma \vdash \ottnt{v} : \ottnt{A} }%
}{
 \Gamma \vdash  \ottnt{a}  \;  \ottnt{v}  : \ottsym{[}  \ottnt{v}  \ottsym{/}  \ottmv{x}  \ottsym{]}  \ottnt{B} }{%
{\ottdrulename{app\_val}}{}%
}}

\newcommand{\ottdruleappXXnondep}[1]{\ottdrule[#1]{%
\ottpremise{ \Gamma \vdash \ottnt{a} : \ottnt{A}  \to  \ottnt{B} }%
\ottpremise{ \Gamma \vdash \ottnt{b} : \ottnt{A} }%
}{
 \Gamma \vdash  \ottnt{a}  \;  \ottnt{b}  : \ottnt{B} }{%
{\ottdrulename{app\_nondep}}{}%
}}

\newcommand{\ottdrulevconv}[1]{\ottdrule[#1]{%
\ottpremise{ \Gamma \vdash \ottnt{a} : \ottnt{A}  \quad  \Gamma \vdash \ottnt{v} : \ottnt{A}  \ottsym{=}  \ottnt{B} }%
\ottpremise{ \Gamma \vdash \ottnt{B} :  \star  }%
}{
 \Gamma \vdash \ottkw{conv} \, \ottnt{a} \, \ottkw{at} \, \mathsf{\sim}  \ottnt{v} : \ottnt{B} }{%
{\ottdrulename{vconv}}{}%
}}

\newcommand{\ottdruleconvXXsubst}[1]{\ottdrule[#1]{%
\ottpremise{ \Gamma \vdash \ottnt{a} : \ottsym{[}  \ottnt{B_{{\mathrm{1}}}}  \ottsym{/}  \ottmv{x}  \ottsym{]}  \ottnt{A}  \quad  \Gamma \vdash \ottnt{v} : \ottnt{B_{{\mathrm{1}}}}  \ottsym{=}  \ottnt{B_{{\mathrm{2}}}} }%
\ottpremise{ \Gamma \vdash \ottsym{[}  \ottnt{B_{{\mathrm{2}}}}  \ottsym{/}  \ottmv{x}  \ottsym{]}  \ottnt{A} :  \star  }%
}{
 \Gamma \vdash \ottkw{conv} \, \ottnt{a} \, \ottkw{at} \, \ottsym{[}  \mathsf{\sim}  \ottnt{v}  \ottsym{/}  \ottmv{x}  \ottsym{]}  \ottnt{A} : \ottsym{[}  \ottnt{B_{{\mathrm{2}}}}  \ottsym{/}  \ottmv{x}  \ottsym{]}  \ottnt{A} }{%
{\ottdrulename{conv\_subst}}{}%
}}

\newcommand{\ottdruleconvXXmultisubst}[1]{\ottdrule[#1]{%
\ottpremise{ \Gamma \vdash \ottnt{v_{{\mathrm{1}}}} : \ottnt{A_{{\mathrm{1}}}}  \ottsym{=}  \ottnt{B_{{\mathrm{1}}}}  \quad ... \quad  \Gamma \vdash \ottnt{v_{\ottmv{i}}} : \ottnt{A_{\ottmv{i}}}  \ottsym{=}  \ottnt{B_{\ottmv{i}}} }%
\ottpremise{ \Gamma \vdash \ottnt{a} : \ottsym{[}  \ottnt{A_{{\mathrm{1}}}}  \ottsym{/}  \ottmv{x_{{\mathrm{1}}}}  \ottsym{]} \, ... \, \ottsym{[}  \ottnt{A_{\ottmv{i}}}  \ottsym{/}  \ottmv{x_{\ottmv{i}}}  \ottsym{]}  \ottnt{A} }%
\ottpremise{ \Gamma \vdash \ottsym{[}  \ottnt{B_{{\mathrm{1}}}}  \ottsym{/}  \ottmv{x_{{\mathrm{1}}}}  \ottsym{]} \, ... \, \ottsym{[}  \ottnt{B_{\ottmv{i}}}  \ottsym{/}  \ottmv{x_{\ottmv{i}}}  \ottsym{]}  \ottnt{A} :  \star  }%
}{
 \Gamma \vdash \ottkw{conv} \, \ottnt{a} \, \ottkw{at} \, \ottsym{[}  \mathsf{\sim}  \ottnt{v_{{\mathrm{1}}}}  \ottsym{/}  \ottmv{x_{{\mathrm{1}}}}  \ottsym{]} \, ... \, \ottsym{[}  \mathsf{\sim}  \ottnt{v_{\ottmv{i}}}  \ottsym{/}  \ottmv{x_{\ottmv{i}}}  \ottsym{]}  \ottnt{A} : \ottsym{[}  \ottnt{B_{{\mathrm{1}}}}  \ottsym{/}  \ottmv{x_{{\mathrm{1}}}}  \ottsym{]} \, ... \, \ottsym{[}  \ottnt{B_{\ottmv{i}}}  \ottsym{/}  \ottmv{x_{\ottmv{i}}}  \ottsym{]}  \ottnt{A} }{%
{\ottdrulename{conv\_multisubst}}{}%
}}

\usepackage{listings}
\lstnewenvironment{code}
    {\lstset{}%
      \csname lst@SetFirstLabel\endcsname}
    {\csname lst@SaveFirstLabel\endcsname}
\renewcommand{\ottdrule}[4][]{{\displaystyle\frac{\begin{array}{l}#2\end{array}}{#3}\hspace{0.0cm}\ottdrulename{#4}}}

\newcommand{\figspace}{\vspace{9pt}}
\newcommand{\ruleline}[1]{
    \begin{center} \(
    #1
    \) \end{center}
    \figspace
}

\title{Irrelevance, Heterogeneous Equality, and Call-by-value Dependent Type Systems}

\author{
     Vilhelm Sj\"oberg
        \institute{University of Pennsylvania}
        \email{vilhelm@cis.upenn.edu}
\and Chris Casinghino
        \institute{University of Pennsylvania}
        \email{ccasin@cis.upenn.edu}
\and Ki Yung Ahn     
        \institute{Portland State University}
        \email{kya@cs.pdx.edu}
\and Nathan Collins
        \institute{Portland State University}
        \email{nathan.collins@gmail.com}
\and Harley D. Eades III
        \institute{University of Iowa}
        \email{harley-eades@uiowa.edu}
\and Peng Fu
        \institute{University of Iowa}
        \email{peng-fu@uiowa.edu}
\and Garrin Kimmell
        \institute{University of Iowa}
        \email{garrin-kimmell@uiowa.edu}
\and Tim Sheard
        \institute{Portland State University}
        \email{sheard@cis.pdx.edu}
\and Aaron Stump
        \institute{University of Iowa}
        \email{astump@acm.org}
\and Stephanie Weirich
        \institute{University of Pennsylvania}
        \email{sweirich@cis.upenn.edu}
}

\def\titlerunning{Irrelevance, Heterogeneous Equality, and CBV}
\def\authorrunning{Sj\"{o}berg et al.}
\begin{document}
\maketitle

\begin{abstract}
We present a full-spectrum dependently typed core language which
includes both \emph{nontermination} and \emph{computational
irrelevance} (a.k.a. erasure), a combination which has not been studied
before. The two features interact: to protect type safety we must be
careful to only erase terminating expressions.
Our language design is strongly influenced by the choice of CBV
evaluation, and by our novel treatment of propositional equality which
has a heterogeneous, completely erased elimination form.
\end{abstract}


\section{Introduction}

The Trellys project is a collaborative effort to design a new
dependently typed programming language. Our goal is to bridge the gap
between ordinary functional programming and program verification with
dependent types. Programmers should be able to port their existing
functional programs to Trellys with minor modifications, and then
gradually add more expressive types as appropriate.

This goal has implications for our design. First, and most
importantly, we must consider {\bf nontermination} and other
effects. Unlike Coq and Agda, functional programming languages like
OCaml and Haskell allow general recursive functions, so to accept
functional programs `as-is' we must be able to turn off the
termination checker.  We also want to use dependent types even with
programs that may diverge.
 
Second, the fact that our language includes effects 
means that order of evaluation matters. We choose {\bf
  call-by-value} order, both because it has a simple cost model (enabling
programmers to understand the running time and space usage of their
programs), and also because CBV seems to to work particularly well for
nonterminating dependent languages (as we explain in
section~\ref{sec:CBV-partial-correctness}).

Finally, to be able to add precise types to a program without slowing
it down, we believe it is essential to support {\bf computational
  irrelevance}---expressions in the program which are only needed for
type-checking should be erased during compilation and require no
run-time representation. We also want to reflect irrelevance in the
type system, where it can also help reason about a program.

These three features interact in nontrivial ways. Nontermination makes
irrelevance more complicated, because we must be careful to only erase
terminating expressions. \scw{What causes this unsafety? abort? CBV?
  nontermination?}\ntc{Could forward reference the safediv example
in section \ref{sec:CBV-partial-correctness} here.} \cjc{I think an
example would ruin the flow, better to say ``As we explain in section
such-and-such, nontermination makes irrelevence...''}
 On the other hand CBV helps, since it lets us treat
variables in the typing context as terminating.

To study this interaction, we have designed a full-spectrum\vs{define
  this term somewhere?}\ntc{yes}, dependently-typed core language with a
small-step call-by-value operational semantics.  This language is
inconsistent as a logic, but very expressive as a programming
language: it includes general recursion, datatypes, abort,
large eliminations and ``Type-in-Type''.

The subtleties of adding irrelevance to a dependent type system all
have to do with equality of expressions.  Therefore many language
design decisions are influenced by our {\bf novel treatment of
  propositional equality}. This primitive equality has two unusual
features: it is computationally irrelevant (equality proofs do not
need to be examined during computation), and it is ``very
heterogenous'' (we can state \emph{and} use equations between terms of
different types).

This paper discusses some of the key insights that we have gained in
the process of this design. In particular, the contributions of this
paper include:

\begin{enumerate}
\item The presence of nontermination means that the application rule
  must be restricted. This paper presents the most generous
  application rule to date (section~\ref{sec:cbv-apprule}).

\item Our language includes a primitive equality type, which may be
  eliminated in an irrelevant manner
  (section~\ref{sec:irrelevant-conv}).

\item The equality type is also ``very heterogenous'' (section~\ref{sec:heterogenous-eq}), and we design a
  new variation of the elimination rule, ``$n$-ary conv'', to better
  exploit this feature (section~\ref{sec:n-ary-conv}). We also discuss how
  to add type annotations to this rule (section~\ref{sec:annotations}).

\item We support irrelevant arguments and data structure components.
  We show by example that in the presence of nontermination/abort the usual
  rule for irrelevant function application must be restricted, and
  propose a new rule with a value restriction (section~\ref{sec:irrarg}).

\item We prove that our language is type safe (section~\ref{sec:metatheory}). 
\end{enumerate}

The design choices for each language feature affects the others. By
combining concrete proposals for evaluation-order, erasure, and
equality in a single language, we have found interactions that are not
apparent in isolation.

\subsection{CBV, nontermination, and ``partial correctness''}
\label{sec:CBV-partial-correctness}

There is a particularly nice fit between nonterminating dependent
languages and CBV evaluation, because the strictness of evaluation
partially compensates for the fact that all types are inhabited.

For example, consider integer division. Suppose the standard library
provides a function
\begin{code}
div : Nat -> Nat -> Nat
\end{code}
which performs truncating division and throws an error if the divisor
is zero. If we are concerned about runtime errors, we might want to be
more careful. One way to proceed is to define a wrapper around \lstinline$div$,
which requires a proof of div's {\bf precondition} that the
denominator be non-zero:
\begin{code}
safediv : Nat -> (y:Nat) -> (p: isZero y = false) -> Nat
safediv = \x:Nat.\y:Nat.\p:(isZero y = false).div x y
\end{code}
Programs written using \lstinline$safediv$ are guaranteed to not divide by
zero, even though our language is inconsistent as a logic. This works
because $\lambda$-abstractions are strict in their arguments, so if we
provide an infinite loop as the proof in \lstinline$safediv 1 0 loop$ the
entire expression diverges and never reaches the division.  
In the \lstinline$safediv$ example, strictness was a matter of
expressivity, since it allowed us to maintain a strong invariant. But
when type conversion is involved, strictness
is required for type safety. For example, if a purported proof
of $ \mathsf{Bool}   \ottsym{=}   \mathsf{Nat} $ were not evaluated strictly, we
could use an infinite loop as a proof and try to add two booleans.
This is recognized by, e.g. GHC Core,\gk{How about ``A similar
  technique is used in GHC Core,''} which does most evaluation lazily
but is strict when computing type-equality
proofs~\cite{vytiniotis-practical11}.

While strict $\lambda$-abstractions give preconditions, strict data
constructors can be used to express {\bf postconditions}. For example, we
might define a datatype characterizing what it means for a string
(represented as a list of characters) to match a regular expression
\begin{code}
data Match : String -> Regexp -> * where
  MChar : (x:Char) -> Match (x::nil) (RCh x)
  MStar0 : (r:Regexp) -> Match (nil) (RStar r)
  MStar1 : (r:Regexp) -> (s s':String) ->  
   Match s r -> Match s' (RStar r) -> Match (s ++ s') (RStar r)
  ...
\end{code}
and then define a regexp matching function to return a proof of the
match
\begin{code}
match : (s:String) -> (r:Regexp) -> Maybe (Matches s r)
\end{code}
Such a type can be read as a partial correctness assertion: we have no
guarantee that the function will terminate, but if it does and says
that there was a match, then there really was. Even though we are
working in an inconsistent logic, if the function returns at all we
know that the constructors of \lstinline$Match$ were not given bogus
looping terms.

Compared to normalizing languages, the properties
our types can express are limited in two ways. First, of course,
there is no way to state total correctness. Second, we are
limited to predicates that can be witnessed by a first-order type like
\lstinline$Match$. In Coq or Agda we could give \lstinline$match$ the
more informative type
\begin{code}
match : (s:String) -> (r:Regexp) -> Either (Matches s r) (Matches s r -> False)
\end{code}
which says that it is a decision-procedure. But in our language a
value of type \lstinline$Matches s r -> False$ is not necessarily a valid
proof, since it could be a function that always diverges when called.

\section{Language Design}

\newcommand{\alt}{\ensuremath{\ |\ }}
\begin{figure}
\[
\begin{array}{lrcl}
&x,y,f,p    & \in & \mbox { variables } \\           
&D          & \in & \mbox { data types, including }  \mathsf{Nat}  \\
&d          & \in & \mbox { constructors, including } 
                     \ottsym{0} \mbox { and }  \mathsf{S}  \\
&i,j        & \in & \mbox { natural numbers } \\
\\
\text{expressions}
& a, b, A, B & ::= &  
   \star  \alt \ottmv{x} \alt  \mathsf{rec}\; \ottmv{f} : \ottnt{A} . \ottnt{a}  \alt  \mathsf{abort}_{ \ottnt{A} }  \\
&  &\alt&  ( \ottmv{x} \!:\! \ottnt{A} )  \to \,  \ottnt{B}   \alt \lambda  \ottmv{x}  \ottsym{:}  \ottnt{A}  \ottsym{.}  \ottnt{a} \alt  \ottnt{a}  \;  \ottnt{b}    \\
&  &\alt&   [  \ottmv{x} \!:\! \ottnt{A}  ]  \to \,  \ottnt{B}    \alt \lambda  \ottsym{[}  \ottmv{x}  \ottsym{:}  \ottnt{A}  \ottsym{]}  \ottsym{.}  \ottnt{a} \alt  \ottnt{a}  \; [  \ottnt{b}  ]   \\
&  &\alt&  \ottmv{D}  \;  \overline{A_i}   \alt  \ottmv{d}  \; [  \overline{A_i}  ] \;  \overline{a_i}  \alt 
    \ottkw{case} \, \ottnt{a} \, \ottkw{as} \, \ottsym{[y]} \, \ottkw{of} \, \ottsym{\{} \, \ottcomplu{\ottmv{d_{\ottmv{j}}} \, \Delta_{\ottmv{j}}  \Rightarrow  \ottnt{b_{\ottmv{j}}}}{\ottmv{j}}{{\mathrm{1}}}{..}{\ottmv{k}} \, \ottsym{\}} \\
&  &\alt& \ottnt{a}  \ottsym{=}  \ottnt{b} \alt  \mathsf{join}_{ \ottnt{a} = \ottnt{b} } \;  \ottmv{i}  \;  \ottmv{j}  \alt \ottkw{injdom} \, \ottnt{a} \alt
    \ottkw{injrng} \, \ottnt{a} \, \ottnt{b} \alt  \mathsf{injtcon}_{ \ottmv{i} } \;  \ottnt{a}  \\
&  &\alt&  \ottkw{conv} \, \ottnt{a} \, \ottkw{at} \, \ottsym{[}  \mathsf{\sim}  \ottnt{P_{{\mathrm{1}}}}  \ottsym{/}  \ottmv{x_{{\mathrm{1}}}}  \ottsym{]} \, ... \, \ottsym{[}  \mathsf{\sim}  \ottnt{P_{\ottmv{i}}}  \ottsym{/}  \ottmv{x_{\ottmv{i}}}  \ottsym{]}  \ottnt{A} \\
\text{telescopes}
& \Delta & ::= & \cdot \alt  \ottsym{(}  \ottmv{x}  \ottsym{:}  \ottnt{A}  \ottsym{)}  \Delta \alt\ottsym{[}  \ottmv{x}  \ottsym{:}  \ottnt{A}  \ottsym{]}  \Delta \\
\text{expression lists}
& \overline{a_i} & ::= & \cdot \alt   \ottnt{a}  \;  \overline{a_i}  \alt [  \ottnt{a}  ] \;  \overline{a_i}  \\
\text{conv proofs}
& \ottnt{P} & ::= & \ottnt{v} \alt \ottsym{[}  \ottnt{a}  \ottsym{=}  \ottnt{b}  \ottsym{]} \\
\\
\text{values}
& v & ::= &  \star  \alt \ottmv{x} \alt  \mathsf{rec}\; \ottmv{f} : \ottnt{A} . \ottnt{v}  \\
&   &\alt &  ( \ottmv{x} \!:\! \ottnt{A} )  \to \,  \ottnt{B}  \alt \lambda  \ottmv{x}  \ottsym{:}  \ottnt{A}  \ottsym{.}  \ottnt{a} \\
&   &\alt &  [  \ottmv{x} \!:\! \ottnt{A}  ]  \to \,  \ottnt{B}  \alt \lambda  \ottsym{[}  \ottmv{x}  \ottsym{:}  \ottnt{A}  \ottsym{]}  \ottsym{.}  \ottnt{a} \\
&   &\alt &  \ottmv{D}  \;  \overline{A_i}  \alt  \ottmv{d}  \; [  \overline{A_i}  ] \;  \overline{v_i}  \\
&   &\alt & \ottnt{a}  \ottsym{=}  \ottnt{b} \alt  \mathsf{join}_{ \ottnt{a} = \ottnt{b} } \;  \ottmv{i}  \;  \ottmv{j}  \alt \ottkw{injdom} \, \ottnt{a} \alt
     \ottkw{injrng} \, \ottnt{a} \, \ottnt{b} \alt  \mathsf{injtcon}_{ \ottmv{i} } \;  \ottnt{a}  \\
&   &\alt & \ottkw{conv} \, \ottnt{v} \, \ottkw{at} \, \ottsym{[}  \mathsf{\sim}  \ottnt{P_{{\mathrm{1}}}}  \ottsym{/}  \ottmv{x_{{\mathrm{1}}}}  \ottsym{]} \, ... \, \ottsym{[}  \mathsf{\sim}  \ottnt{P_{\ottmv{i}}}  \ottsym{/}  \ottmv{x_{\ottmv{i}}}  \ottsym{]}  \ottnt{A} 
\end{array}
\]
\caption{Syntax of the annotated language}
\label{fig:syntax}
\end{figure}

\begin{figure} 
\[
\begin{array}{lrcl}
\text{expressions}  
&m,n,M,N & ::= &    \star  \alt \ottmv{x} \alt  \mathsf{rec}\; \ottmv{f} . \ottkw{u}  \alt \ottkw{abort} \\
&        &\alt &  ( \ottmv{x} \!:\! \ottnt{M} )  \to \,  \ottnt{N}  \alt \lambda  \ottmv{x}  \ottsym{.}  \ottnt{m} \alt  \ottnt{m}  \;  \ottnt{n}  \\
&        &\alt &  [  \ottmv{x} \!:\! \ottnt{M}  ]  \to \,  \ottnt{N}  \alt \lambda  \ottsym{[}  \ottsym{]}  \ottsym{.}  \ottnt{m} \alt \ottnt{m}  \ottsym{[}  \ottsym{]} \\
&        &\alt & \ottmv{D} \, \overline{M_i} \alt \ottmv{d} \, \overline{m_i} \alt
   \ottkw{case} \, \ottnt{n} \, \ottkw{of} \, \ottsym{\{} \, \ottcomplu{\ottmv{d_{\ottmv{j}}} \, \overline{x_i}_{\ottmv{j}}  \Rightarrow  \ottnt{m_{\ottmv{j}}}}{\ottmv{j}}{{\mathrm{1}}}{..}{\ottmv{k}} \, \ottsym{\}} \\
&        &\alt & \ottnt{m}  \ottsym{=}  \ottnt{n} \alt \ottkw{join} \\
\text{telescopes}
& \Xi & ::= & \cdot \alt  \ottsym{(}  \ottmv{x}  \ottsym{:}  \ottnt{M}  \ottsym{)}  \Xi \alt \ottsym{[}  \ottmv{x}  \ottsym{:}  \ottnt{M}  \ottsym{]}  \Xi \\
\text{expression lists}
& \overline{m_i} & ::= & \cdot \alt   \ottnt{m}  \;  \overline{m_i}  \alt  [] \;  \overline{m_i}  \\
\\
\text{values}  
& u      & ::= &  \star  \alt \ottmv{x} \alt  \mathsf{rec}\; \ottmv{f} . \ottkw{u}  \\
&        &\alt &  ( \ottmv{x} \!:\! \ottnt{M} )  \to \,  \ottnt{N}  \alt \lambda  \ottmv{x}  \ottsym{.}  \ottnt{m} \\
&        &\alt &  [  \ottmv{x} \!:\! \ottnt{M}  ]  \to \,  \ottnt{N}  \alt \lambda  \ottsym{[}  \ottsym{]}  \ottsym{.}  \ottnt{m} \\
&        &\alt & \ottmv{D} \, \overline{M_i} \alt \ottmv{d} \, \overline{u_i} \\
&        &\alt & \ottnt{m}  \ottsym{=}  \ottnt{n} \alt \ottkw{join}
\\
\text{evaluation contexts}
& \mathcal{E} & ::= &  \bullet  \alt \bullet  \ottnt{m} \alt \ottnt{u}  \bullet \alt \bullet  \ottsym{[]}  \alt
\ottmv{d} \, \overline{u_i}  \bullet  \overline{m_i} \alt \ottkw{case} \, \bullet \, \ottkw{of} \, \ottsym{\{}  \ottmv{d_{\ottmv{j}}} \, \overline{x_i}_{\ottmv{j}}  \Rightarrow  \ottnt{m_{\ottmv{j}}}  \ottsym{\}}
\end{array}  
\]
\framebox{\mbox{$\ottnt{m}  \leadsto_{\mathsf{cbv} }  \ottnt{m'}$}}
\ruleline{
  \ottdrulescXXappbeta{} \qquad \ottdrulescXXapprec{} 
}
\ruleline{
  \ottdrulescXXiappbeta{} \qquad \ottdrulescXXiapprec{}
} 
\ruleline{
  \ottdrulescXXcasebeta{}
}
\ruleline{
  \ottdrulescXXctx{} \qquad \ottdrulescXXabort{}
}
\caption{Syntax and operational semantics of the unannotated language}
\label{fig:unannotated}
\end{figure}

We now go on to describe the syntax and type system of our language,
focusing on its novel contributions.

The syntax of the language is shown in figure~\ref{fig:syntax}.
Terms, types, and sorts are collapsed into one syntactic category as in
the presentation of the lambda cube~\cite{Barendregt92lambdacalculi},
but by convention we use uppercase metavariables $\ottnt{A}, \ottnt{B}$
for expressions that are types. Some of the expressions are standard:
the type of types $ \star $~\cite{Cardelli86apolymorphic}, variables, recursive definitions, error,
the usual dependent function type, function definition, and function
application. The language also includes expressions dealing with irrelevance,
datatypes, and propositional equality; these will be explained in
detail in the following subsections.

The typing judgment is written $ \Gamma \vdash \ottnt{a} : \ottnt{A} $. The full
definition can be found in appendix~\ref{sec:annotated-type-system}. In
the rest of the paper we will highlight the interesting rules
when we describe the corresponding language features. The typing
contexts $\Gamma$ are lists containing variable declarations and datatype
declarations (discussed in section~\ref{sec:irrarg}):
\[
\Gamma ::= \cdot \alt \Gamma  \ottsym{,}   \ottmv{x}  :  \ottnt{A}  \alt
\Gamma  \ottsym{,}  \ottkw{data} \, \ottmv{D} \, \Delta^{+} \, \ottkw{where} \, \ottsym{\{} \, \ottcomplu{\ottmv{d_{\ottmv{i}}}  \ottsym{:}  \Delta_{\ottmv{i}}  \to \; D \; \Delta^{+}}{\ottmv{i}}{{\mathrm{1}}}{..}{\ottmv{j}} \, \ottsym{\}}
\]

In order to study computational irrelevance and erasure, we define a
separate language of \emph{unannotated expressions} ranged over by
metavariables $\ottnt{m}, \ottnt{M}$. The unannotated language captures runtime behavior; its definition is similar to the annotated
language but with computationally irrelevant subexpressions (e.g. type
annotations) removed. This is the language for which we define the
operational semantics (the step relation $ \leadsto_{\mathsf{cbv} } $ in
figure~\ref{fig:unannotated}). The annotated and unannotated languages
are related by an \emph{erasure operation} $| \cdot |$, which takes an
expression $\ottnt{a}$ and produces an unannotated expression $\ottsym{\mbox{$\mid$}}  \ottnt{a}  \ottsym{\mbox{$\mid$}}$
by deleting all the computationally irrelevant parts (figure~\ref{fig:erasure}). 
To show type safety we define an unannotated typing relation $ \ottnt{H} \vdash \ottnt{m} : \ottnt{M} $ and
prove preservation and progress theorems for unannotated terms. 

The relation $ \leadsto_{\mathsf{cbv} } $ models runtime evaluation. However, in the
specification of the type system we use a more
liberal notion of \emph{parallel reduction}, denoted $ \leadsto_{\mathsf{p} } $. The
difference is that $ \leadsto_{\mathsf{p} } $ allows reducing under binders,
e.g. $\ottsym{(}  \lambda  \ottmv{x}  \ottsym{.}  \ottsym{1}  \ottsym{+}  \ottsym{1}  \ottsym{)}  \leadsto_{\mathsf{p} }  \ottsym{(}  \lambda  \ottmv{x}  \ottsym{.}  \ottsym{2}  \ottsym{)}$ even though $\ottsym{(}  \lambda  \ottmv{x}  \ottsym{.}  \ottsym{1}  \ottsym{+}  \ottsym{1}  \ottsym{)}$ is
already a CBV value. The main reason for introducing $ \leadsto_{\mathsf{p} } $ in addition to $ \leadsto_{\mathsf{cbv} } $
is for the metatheory: in order to characterize when two expressions
are provably equal (lemma~\ref{lemma:equality-soundness}) we need a
notion of reduction that satisfies the substitution properties in
section~\ref{sec:parred-props}, and we defined $ \leadsto_{\mathsf{p} } $
accordingly. But because $ \leadsto_{\mathsf{p} } $ allows strictly more reductions
than $ \leadsto_{\mathsf{cbv} } $, defining the type system in terms of $ \leadsto_{\mathsf{p} } $ lets
the programmer write more programs. Since the type safety proof does not become
harder, we pick the more expressive type system.

In summary, we use the following judgments:
\begin{center}
\begin{tabular}{ll}
$ \Gamma \vdash \ottnt{a} : \ottnt{A} $  & Typing of annotated expressions \\
$ \ottnt{H} \vdash \ottnt{m} : \ottnt{M} $  & Typing of unannotated expressions \\
$\ottnt{m}  \leadsto_{\mathsf{cbv} }  \ottnt{m'}$  & (Runtime, deterministic CBV) evaluation \\
$\ottnt{m}  \leadsto_{\mathsf{p} }  \ottnt{m'}$  & (Typechecking-time, nondeterministic) parallel reduction \\
\end{tabular}
\end{center}

\paragraph{Nontermination and Error}
Before moving on to the more novel parts of the language we mention
how recursive definitions and error terms are formalized. Recursive
definitions are made using the $ \mathsf{rec}\; \ottmv{f} : \ottnt{A} . \ottnt{a} $ form, with the
typing rule 
\[
\ottdruletXXrec{}
\]
With this rule the body of a well-typed rec-expression is always a
value, but we leave it a general expression
$\ottnt{a}$ in the syntax so that substitution $\ottsym{[}  \ottnt{a}  \ottsym{/}  \ottmv{x}  \ottsym{]}  \ottnt{b}$ is always
defined. For simplicity the rule restricts $A$ so that a \lstinline$rec$ can
only have a function type, disallowing (for example) recursive types
or pairs of mutually recursive functions. A typical use of the form
will look like $ \mathsf{rec}\; \ottmv{f} :  ( \ottmv{x} \!:\! \ottnt{A} )  \to \,  \ottnt{B}  . \lambda  \ottmv{x}  \ottsym{:}  \ottnt{A}  \ottsym{.}  \ottnt{b} $. Rec-expressions are
values, and a rec-expression in an evaluation context steps by the
rule $ \ottsym{(}   \mathsf{rec}\; \ottmv{f} . \ottkw{u}   \ottsym{)}  \;  \ottnt{u_{{\mathrm{2}}}}   \leadsto_{\mathsf{cbv} }   \ottsym{(}  \ottsym{[}   \mathsf{rec}\; \ottmv{f} . \ottkw{u}   \ottsym{/}  \ottmv{f}  \ottsym{]}  \ottnt{u_{{\mathrm{1}}}}  \ottsym{)}  \;  \ottnt{u_{{\mathrm{2}}}} $. This maintains the
invariant that CBV evaluation only substitutes values for variables.

In addition to nonterminating expressions, we include explicit error
terms $ \mathsf{abort}_{ \ottnt{A} } $, which can be given any well-formed type.
\[
\ottdruletXXabort{}
\]
An abort expression propagates past any evaluation context by the rule
$\mathcal{E}  \ottsym{[}  \ottkw{abort}  \ottsym{]}  \leadsto_{\mathsf{cbv} }  \ottkw{abort}$. This is a standard treatment of errors. General
recursion already lets us give a looping expression any type in any
context, so it is not surprising that this is type safe. However, note
that the stepping rule for $\ottkw{abort}$ could be considered an
extremely simple control effect. We will see that this is
already enough to influence the language design.

\subsection{CBV Program Equivalence meets the Application rule}
\label{sec:cbv-apprule}
Adding more effects to a dependently typed language requires being
more restrictive about what expressions the type system equates.
Pure, strongly normalizing languages can allow arbitrary
$\beta$-reductions when comparing types, for example reducing
$ \ottsym{(}  \lambda  \ottmv{x}  \ottsym{.}  \ottnt{m}  \ottsym{)}  \;  \ottnt{n} $ either to $\ottsym{[}  \ottnt{n}  \ottsym{/}  \ottmv{x}  \ottsym{]}  \ottnt{m}$ or by reducing $\ottnt{n}$.
This works because any order of evaluation gives the same answer.
In our language that is not the case, e.g. $ \ottsym{(}  \lambda  \ottmv{x}  \ottsym{.}  \ottsym{3}  \ottsym{)}  \;  \ottkw{abort} $ evaluates to $\ottkw{abort}$ under
CBV but to $\ottsym{3}$ under CBN. We can not have both equations
$\ottsym{(}   \ottsym{(}  \lambda  \ottmv{x}  \ottsym{.}  \ottsym{3}  \ottsym{)}  \;  \ottkw{abort}   \ottsym{)}  \ottsym{=}  \ottkw{abort}$ and $\ottsym{(}   \ottsym{(}  \lambda  \ottmv{x}  \ottsym{.}  \ottsym{3}  \ottsym{)}  \;  \ottkw{abort}   \ottsym{)}  \ottsym{=}  \ottsym{3}$ at the
same time, since by transitivity all terms would be equal. Our type
system must commit to a particular order of evaluation.

Therefore, as in previous work~\cite{jia:lambdaeek}, our type system
uses a notion of equality that respects CBV contextual
equivalence. Two terms can by proven equal if they have a common
reduct under {\bf CBV parallel reduction} $ \leadsto_{\mathsf{p} } $.  This relation
is similar to $ \leadsto_{\mathsf{cbv} } $, except that it permits evaluation under
binders and subexpressions can be evaluated in parallel.  The rules
for $\lambda$-abstractions and applications are shown in
figure~\ref{fig:parred} (the remaining rules are in the appendix, section~\ref{sec:parallel-reduction}).  In
particular, the typechecker can only carry out a $\beta$-reduction of
an application or case expression if the argument or scrutinee is a
value. Note, however, that values include variables. Treating
variables as values is safe due to the CBV semantics, and it is
crucial when reasoning about open terms. For example, to typecheck the
usual $\mathsf{append}$ function we want $  \mathsf{Vec}   \;    \mathsf{Nat}   \;  \ottsym{(}  \ottsym{0}  \ottsym{+}  \ottmv{x}  \ottsym{)}  $ and $  \mathsf{Vec}   \;    \mathsf{Nat}   \;  \ottmv{x}  $ to be equal types.

\begin{figure}
\ruleline{
\ottdrulespXXrefl{} \qquad
\ottdrulespXXabs{}  \qquad
\ottdrulespXXapp{}  
}
\ruleline{
\ottdrulespXXappbeta{} \qquad
\ottdrulespXXapprec{}
}
\caption{Parallel reduction $ \leadsto_{\mathsf{p} } $ (Excerpt).}
\label{fig:parred}
\end{figure}

The possibility that expressions may have effects restricts the
application rule of a dependent type system. The typical rule for
typing applications in pure languages is
\[
\ottdrule{
  \ottpremise{ \Gamma \vdash \ottnt{a} :  ( \ottmv{x} \!:\! \ottnt{A} )  \to \,  \ottnt{B}  }
  \ottpremise{ \Gamma \vdash \ottnt{b} : \ottnt{A} }
}{
   \Gamma \vdash  \ottnt{a}  \;  \ottnt{b}  : \ottsym{[}  \ottnt{b}  \ottsym{/}  \ottmv{x}  \ottsym{]}  \ottnt{B} 
}{
}
\]
However, this rule does not work if $\ottnt{b}$ may have effects,
because then the type $\ottsym{[}  \ottnt{b}  \ottsym{/}  \ottmv{x}  \ottsym{]}  \ottnt{B}$ may not be well-formed. 
Although we know by regularity (lemma~\ref{lemma:regularity}) that $ ( \ottmv{x} \!:\! \ottnt{A} )  \to \,  \ottnt{B} $ is
well-formed, the derivation of $ \Gamma \vdash  ( \ottmv{x} \!:\! \ottnt{A} )  \to \,  \ottnt{B}  :  \star  $ may involve
reductions, and substituting a non-value $\ottnt{b}$ for $\ottmv{x}$ may block
a $\beta$-reduction that used to have $\ottmv{x}$ as an argument.
Intuitively this makes sense: under CBV-semantics, $\ottnt{a}$ is really
called on the \emph{value} of $\ottnt{b}$, so the type $\ottnt{B}$
should be able to assume that $\ottmv{x}$ is an (effect-free) value.
Our fix is to add a premise that the result type is well-formed.
This additional premise is exactly what is required to prove type safety.
\[
\ottdruletXXapp{}
\]
This rule is simple, yet expressive. Previous work
~\cite{vytiniotis:pie,jia+08,swamy-fstar}\scw{Some CMU language?}
uses a more restrictive typing for applications, splitting it into two rules: one which
permits only \emph{value dependency}, and requires the argument to 
be a value, and one which allows an application to an arbitrary
argument when there is no dependency. 

Because our annotated type system satisfies substitution of values,
both of these rules are special cases of our rule above (proofs are in
appendix~\ref{sec:ann-theory}):

\begin{lemma}[Substitution for the annotated language]
 If $ \Gamma_{{\mathrm{1}}}  \ottsym{,}   \ottmv{x_{{\mathrm{1}}}}  :  \ottnt{A_{{\mathrm{1}}}}   \ottsym{,}  \Gamma_{{\mathrm{2}}} \vdash \ottnt{a} : \ottnt{A} $, then
 $ \Gamma_{{\mathrm{1}}}  \ottsym{,}  \ottsym{[}  \ottnt{v_{{\mathrm{1}}}}  \ottsym{/}  \ottmv{x_{{\mathrm{1}}}}  \ottsym{]}  \Gamma_{{\mathrm{2}}} \vdash \ottsym{[}  \ottnt{v_{{\mathrm{1}}}}  \ottsym{/}  \ottmv{x_{{\mathrm{1}}}}  \ottsym{]}  \ottnt{a} : \ottsym{[}  \ottnt{v_{{\mathrm{1}}}}  \ottsym{/}  \ottmv{x_{{\mathrm{1}}}}  \ottsym{]}  \ottnt{A} $.
\end{lemma}

\begin{lemma}[Regularity for the annotated language]
\label{lemma:regularity}
If $ \Gamma \vdash \ottnt{a} : \ottnt{A} $, then $ \Gamma \vdash \ottnt{A} :  \star  $.
\end{lemma}

\begin{lemma}The following rules are admissible.
\[
\begin{array}{lll}
\ottdruleappXXval{}
& \qquad &
\ottdruleappXXnondep{}
\end{array}
\]
\end{lemma}

\subsection{Equality and irrelevant type conversions}
\label{sec:irrelevant-conv}

One crucial point in the design of a dependently typed language is the
elimination form for propositional equality, \emph{conversion}.
\footnote{Some authors reserve the word ``conversion'' for
  \emph{definitional} equality. Our type system does not have a
  definitional equality judgment, so we hope our use of the word
  does not cause confusion.} 
 Given an expression $ \Gamma \vdash \ottnt{a} : \ottnt{A} $ and a proof $ \Gamma \vdash \ottnt{b} : \ottsym{(}  \ottnt{A}  \ottsym{=}  \ottnt{A'}  \ottsym{)} $,
we should be able to convert the type of $\ottnt{a}$ to $\ottnt{A'}$. We write
this operation as $\ottkw{conv} \, \ottnt{a} \, \ottkw{at} \, \mathsf{\sim}  \ottnt{b}$.

In most languages, the proof $\ottnt{b}$ in such a conversion
affects the operational semantics of the expression; we say that it is
\emph{computationally relevant}. For example, in Coq the operational
behavior of \lstinline$conv$ is to first evaluate $\ottnt{b}$ until it reaches
$ \mathsf{refl\_eq} $, the only constructor of the equality type, and
then step by $\ottkw{conv} \, \ottnt{a} \, \ottkw{at} \, \mathsf{\sim}   \mathsf{refl\_eq} $ $\leadsto$ $\ottnt{a}$.

However, relevance can get in the way of reasoning about programs.
Equations involving \lstinline$conv$ such as $\ottsym{(}  \ottkw{conv} \, \ottnt{a} \, \ottkw{at} \, \mathsf{\sim}  \ottnt{b}  \ottsym{)}  \ottsym{=}  \ottnt{a}$
are not easily provable in Coq unless $\ottnt{b}$ is $ \mathsf{refl\_eq} $.
Indeed, because Coq's built-in equality is homogeneous, such
equalities are often difficult even to state.  This issue can be a
practical problem when reasoning about programs operating on indexed
data.  One workaround is to assert additional axioms about equality
and conversion, such as Streicher's Axiom K~\cite{streicher:iiitt}.
\vs{This is not the full story. As Chris points out there is
  additional problems with that equation in Coq due to homogenous
  typing. But isn't there some similar axiom for JMeq, which is maybe
  related to K somehow? Also, OTT asserts exactly that equation as an
  axiom ``coherence'', should maybe mention that.}  The situation is
frustrating because the computationally relevant behavior of
conversion does not actually correspond to the compiled code. Coq's
extraction mechanism will erase $\ottnt{b}$ and turn $\ottkw{conv} \, \ottnt{a} \, \ottkw{at} \, \mathsf{\sim}  \ottnt{b}$
into just $\ottnt{a}$. But the Coq typechecker does not know about
extraction.

\begin{figure}
\[
  \begin{array}{lll}
  \ottsym{\mbox{$\mid$}}   \star   \ottsym{\mbox{$\mid$}} =   \star  &\quad \ottsym{\mbox{$\mid$}}  \ottmv{x}  \ottsym{\mbox{$\mid$}} = \ottmv{x} &\quad\ottsym{\mbox{$\mid$}}   \mathsf{rec}\; \ottmv{f} : \ottnt{A} . \ottnt{v}   \ottsym{\mbox{$\mid$}}  =  \mathsf{rec}\; \ottmv{f} . \ottkw{u}  \\
  \ottsym{\mbox{$\mid$}}   \mathsf{abort}_{ \ottnt{A} }   \ottsym{\mbox{$\mid$}} = \ottkw{abort} \\
  \ottsym{\mbox{$\mid$}}   ( \ottmv{x} \!:\! \ottnt{A} )  \to \,  \ottnt{B}   \ottsym{\mbox{$\mid$}} =  ( \ottmv{x} \!:\! \ottsym{\mbox{$\mid$}}  \ottnt{A}  \ottsym{\mbox{$\mid$}} )  \to \,  \ottsym{\mbox{$\mid$}}  \ottnt{B}  \ottsym{\mbox{$\mid$}}  &\quad \ottsym{\mbox{$\mid$}}  \lambda  \ottmv{x}  \ottsym{:}  \ottnt{A}  \ottsym{.}  \ottnt{a}  \ottsym{\mbox{$\mid$}} = \lambda  \ottmv{x}  \ottsym{.}  \ottsym{\mbox{$\mid$}}  \ottnt{a}  \ottsym{\mbox{$\mid$}} &\quad\ottsym{\mbox{$\mid$}}   \ottnt{a}  \;  \ottnt{b}   \ottsym{\mbox{$\mid$}} = \ottsym{\mbox{$\mid$}}  \ottnt{a}  \ottsym{\mbox{$\mid$}}\ \ottsym{\mbox{$\mid$}}  \ottnt{b}  \ottsym{\mbox{$\mid$}} \\
  \ottsym{\mbox{$\mid$}}   [  \ottmv{x} \!:\! \ottnt{A}  ]  \to \,  \ottnt{B}   \ottsym{\mbox{$\mid$}} =  [  \ottmv{x} \!:\! \ottsym{\mbox{$\mid$}}  \ottnt{A}  \ottsym{\mbox{$\mid$}}  ]  \to \,  \ottsym{\mbox{$\mid$}}  \ottnt{B}  \ottsym{\mbox{$\mid$}}  &\quad \ottsym{\mbox{$\mid$}}  \lambda  \ottsym{[}  \ottmv{x}  \ottsym{:}  \ottnt{A}  \ottsym{]}  \ottsym{.}  \ottnt{a}  \ottsym{\mbox{$\mid$}}  = \lambda  \ottsym{[}  \ottsym{]}  \ottsym{.}  \ottsym{\mbox{$\mid$}}  \ottnt{a}  \ottsym{\mbox{$\mid$}} &\quad\ottsym{\mbox{$\mid$}}   \ottnt{a}  \; [  \ottnt{b}  ]   \ottsym{\mbox{$\mid$}} = \ottsym{\mbox{$\mid$}}  \ottnt{a}  \ottsym{\mbox{$\mid$}}  \ottsym{[}  \ottsym{]} \\
  \ottsym{\mbox{$\mid$}}   \ottmv{D}  \;  \overline{A_i}   \ottsym{\mbox{$\mid$}} = \ottmv{D} \, \ottsym{\mbox{$\mid$}}  \overline{A_i}  \ottsym{\mbox{$\mid$}}  &\quad \ottsym{\mbox{$\mid$}}   \ottmv{d}  \; [  \overline{A_i}  ] \;  \overline{a_i}   \ottsym{\mbox{$\mid$}}= \ottmv{d} \, \ottsym{\mbox{$\mid$}}  \overline{a_i}  \ottsym{\mbox{$\mid$}} \\
  \multicolumn{3}{l}{
  \hspace{-1.75mm} \begin{array}{ll}
  \ottsym{\mbox{$\mid$}}  \ottkw{case} \, \ottnt{a} \, \ottkw{as} \, \ottsym{[y]} \, \ottkw{of} \, \ottsym{\{} \, \ottcomplu{\ottmv{d_{\ottmv{j}}} \, \Delta_{\ottmv{j}}  \Rightarrow  \ottnt{b_{\ottmv{j}}}}{\ottmv{j}}{{\mathrm{1}}}{..}{\ottmv{k}} \, \ottsym{\}}  \ottsym{\mbox{$\mid$}} 
  &=
  \ottkw{case} \, \ottsym{\mbox{$\mid$}}  \ottnt{a}  \ottsym{\mbox{$\mid$}} \, \ottkw{of} \, \ottsym{\{} \, \ottcomplu{\ottmv{d_{\ottmv{j}}} \, \overline{x_i}_{\ottmv{j}}  \Rightarrow  \ottsym{\mbox{$\mid$}}  \ottnt{b_{\ottmv{j}}}  \ottsym{\mbox{$\mid$}}}{\ottmv{j}}{{\mathrm{1}}}{..}{\ottmv{k}} \, \ottsym{\}} \\
  & \quad\text{where $\overline{x_i}_{\ottmv{j}}$ are the relevant variables of  $\Delta_{\ottmv{j}}$} \\
  \end{array}
  }\\
  \ottsym{\mbox{$\mid$}}  \ottnt{a}  \ottsym{=}  \ottnt{b}  \ottsym{\mbox{$\mid$}} = \ottsym{(}  \ottsym{\mbox{$\mid$}}  \ottnt{a}  \ottsym{\mbox{$\mid$}}  \ottsym{=}  \ottsym{\mbox{$\mid$}}  \ottnt{b}  \ottsym{\mbox{$\mid$}}  \ottsym{)} & \multicolumn{2}{l}{
  \quad\ottsym{\mbox{$\mid$}}   \mathsf{join}_{ \ottnt{a} = \ottnt{b} } \;  \ottmv{i}  \;  \ottmv{j}   \ottsym{\mbox{$\mid$}} = \ottsym{\mbox{$\mid$}}  \ottkw{injdom} \, \ottnt{a}  \ottsym{\mbox{$\mid$}} = \ottsym{\mbox{$\mid$}}  \ottkw{injrng} \, \ottnt{a} \, \ottnt{b}  \ottsym{\mbox{$\mid$}} =
       \ottsym{\mbox{$\mid$}}   \mathsf{injtcon}_{ \ottmv{i} } \;  \ottnt{a}   \ottsym{\mbox{$\mid$}} =   \ottkw{join}  
  } \\ 
  \ottsym{\mbox{$\mid$}}  \ottkw{conv} \, \ottnt{a} \, \ottkw{at} \, \ottsym{[}  \mathsf{\sim}  \ottnt{P_{{\mathrm{1}}}}  \ottsym{/}  \ottmv{x_{{\mathrm{1}}}}  \ottsym{]} \, ... \, \ottsym{[}  \mathsf{\sim}  \ottnt{P_{\ottmv{i}}}  \ottsym{/}  \ottmv{x_{\ottmv{i}}}  \ottsym{]}  \ottnt{A}  \ottsym{\mbox{$\mid$}} = \ottsym{\mbox{$\mid$}}  \ottnt{a}  \ottsym{\mbox{$\mid$}} \\
  \\
  | \cdot | = \cdot &\quad \ottsym{\mbox{$\mid$}}   \ottnt{a}  \;  \overline{a_i}   \ottsym{\mbox{$\mid$}} =  \ottsym{\mbox{$\mid$}}  \ottnt{a}  \ottsym{\mbox{$\mid$}}  \;  \ottsym{\mbox{$\mid$}}  \overline{a_i}  \ottsym{\mbox{$\mid$}}  &\quad\ottsym{\mbox{$\mid$}}   [  \ottnt{a}  ] \;  \overline{a_i}   \ottsym{\mbox{$\mid$}} =  [] \;  \ottsym{\mbox{$\mid$}}  \overline{a_i}  \ottsym{\mbox{$\mid$}} 
  \end{array}
\]
\caption{The erasure function $|\cdot|$}
\label{fig:erasure}
\end{figure}

\gk{The following paragraph is stated, in part, in section 3. Can it
  be refactored?}  Our language integrates extraction into the
type-system, similarly to ICC*~\cite{barras+08}. Specifically, we
define an {\bf erasure function} $|\cdot|$ which takes an
annotated expression $a$ and produces an unannotated expression $m
\equiv \ottsym{\mbox{$\mid$}}  \ottnt{a}  \ottsym{\mbox{$\mid$}}$. The definition of $|\cdot|$ is given in
Figure~\ref{fig:erasure}. In most cases it just traverses
$\ottnt{a}$, but it erases type annotations from abstractions, it deletes
irrelevant arguments (see section~\ref{sec:irrarg}), and it completely
deletes conversions leaving just the subject of the cast.

The unannotated system is used to determine when expressions are equal.
\[
\ottdrulejoinXXnoXXannot{}
\]
The rule says that the term $ \ottkw{join} $ is a proof of an
equality $\ottnt{a}  \ottsym{=}  \ottnt{b}$ if the \emph{erasures} of the expressions
$\ottnt{a}$ and $\ottnt{b}$ parallel-reduce to a common reduct. Therefore,
when reasoning about a program we can completely ignore the parts of
it that will not remain at runtime. (The rule presented above is
somewhat simplified from our actual system---it is type safe, but as
we discuss in section~\ref{sec:annotations} it needs additional
annotations to make type checking algorithmic.)

Erasing conversions requires a corresponding restriction on the \lstinline$conv$
typing rule. As we noted before, conversion must evaluate equality
proofs strictly in order to not be fooled by infinite loops, but if
the proofs are erased there is nothing left at runtime to evaluate.
The fix is to restrict the proof term to be a syntactic value: 
\[
\ottdrulevconv{}
\]
\label{vconvrule}
(We will discuss the third premise $ \Gamma \vdash \ottnt{B} :  \star  $ in
section~\ref{sec:heterogenous-eq}).
In the case where the proof is a variable (for instance, the
equalities that come out of a dependent pattern match), the
value restriction is already satisfied. Otherwise (for example, when the proof is
the application of a lemma) we can satisfy the requirement by
rewriting ($\ottkw{conv} \, \ottnt{a} \, \ottkw{at} \, \mathsf{\sim}  \ottnt{b}$) to ($\ottkw{let} \, \ottmv{x}  \ottsym{=}  \ottnt{b} \, \ottkw{in} \, \ottkw{conv} \, \ottnt{a} \, \ottkw{at} \, \mathsf{\sim}  \ottmv{x}$),
making explicit the sequencing that Coq integrates into the evaluation rule.

Most languages make conversion computationally relevant in order to
ensure strong normalization for open terms. If conversion is irrelevant, then
in a context containing the assumption $ \mathsf{Nat}   \ottsym{=}  \ottsym{(}   \mathsf{Nat}   \to   \mathsf{Nat}   \ottsym{)}$ it is
possible to type the unannotated looping term $ \ottsym{(}   \lambda  \ottmv{x}  \ottsym{.}  \ottmv{x}  \;  \ottmv{x}   \ottsym{)}  \;  \ottsym{(}   \lambda  \ottmv{x}  \ottsym{.}  \ottmv{x}  \;  \ottmv{x}   \ottsym{)} $ since
evaluation does not get stuck on the assumption. Of course, in
our language expressions are not normalizing in the first place.

Making conversions completely erased blurs the usual distinction
between \emph{definitional} and \emph{propositional} equality.
Typically, definitional equality is a decidable comparison
which is automatically
applied everywhere, while propositional equality uses arbitrary proofs
but has to be marked with an explicit elimination form.

There are two main reasons languages use a distinguished definitional
equality in addition to the propositional one, but neither of them
applies to our language. First, if there exists a straightforward algorithm
for testing definitional equality (e.g., just reduce both sides to
normal form, as in PTSs~\cite{Barendregt92lambdacalculi}),
then it is convenient for the programmer to have it
applied automatically. However, our language has non-terminating expressions,
and we don't want the type checker to loop trying to normalize them.

Second, languages where the use of propositional equalities is
computationally relevant and marked need automatic conversion for a
technical reason in the preservation proof. As an application steps,
$ \ottnt{m}  \;  \ottnt{n}   \leadsto_{\mathsf{cbv} }   \ottnt{m}  \;  \ottnt{n'} $, its type changes from $\ottsym{[}  \ottnt{n}  \ottsym{/}  \ottmv{x}  \ottsym{]}  \ottnt{N}$ to
$\ottsym{[}  \ottnt{n'}  \ottsym{/}  \ottmv{x}  \ottsym{]}  \ottnt{N}$ and has to be converted back to $\ottsym{[}  \ottnt{n}  \ottsym{/}  \ottmv{x}  \ottsym{]}  \ottnt{N}$.
Because the operational semantics does not introduce any explicit
conversion into the term, this conversion needs to be automatic. 
However, in our unannotated language uses of propositional equations are never
marked, so we can use the propositional equality at this point in the proof.

\subsection{Irrelevant arguments, and reasoning about  indexed data}
\label{sec:irrarg}

Above, we discussed how conversions get erased. Our language also
includes a more general feature where arguments to functions and data
constructors can be marked as irrelevant so that they are erased as
well.

To motivate this feature, we consider
\emph{vectors} (i.e. length-indexed lists). Suppose we
have defined the usual vector data type and append function, with
types

\begin{code} 
Vec (a:*) :  Nat -> * where
  nil  : Vec a 0
  cons : (n:Nat) -> Vec a n -> Vec a (S n)

app : (n1 n2 : Nat) -> (a : *) -> Vec a n1 -> Vec a n2 -> Vec a (n1+n2)
app n1 n2 a xs ys =
  case xs of
    nil => ys
    (cons n x xs) => cons a (n+n2) x (app n n2 a xs ys)
\end{code}

Having defined this operation, we might wish to prove that the append operation
is associative. This amounts to defining a recursive function of type
\begin{code}
  app-assoc : (n1 n2 n3:Nat) -> 
              (v1 : Vec a n1) -> (v2 : Vec a n2) -> (v3 : Vec a n3) ->
              (app a n1 (n2+n3) v1 (app a n2 n3 v2 v3)) 
                =  (app a (n1+n2) n3 (app a n1 n2 v1 v2) v3)
\end{code}
If we proceed by pattern-matching on \lstinline$v1$, 
then when \lstinline$v1 = cons n x v$ we have to show, after reducing the RHS, that
\begin{code}
    (cons a (n + (n2 + n3)) x (app a    n     (n2 + n3)         v       (app a n2 n3 v2 v3)))
  = (cons a ((n + n2) + n3) x (app a (n + n2)    n3     (app a n n2 v v2)       v3))
\end{code}
By a recursive call/induction hypothesis, we have that the tails of the vectors are equal,
so we are \emph{almost} done\dots except we also need to show 
\begin{code}
  n + (n2 + n3) = (n + n2) + n3
\end{code}
which requires a separate lemma about associativity of addition.
In other words, when reasoning about indexed data, we are also 
forced to reason about their indices. In this case it is
particularly frustrating because these indices are completely
determined by the shape of the data---a Sufficiently Smart Compiler
would not even need to keep them around at runtime~\cite{Brady04inductivefamilies}.
Unfortunately, nobody told our typechecker that. 

The solution is to make the length argument to cons an {\bf irrelevant
  argument}.  We change the definition of \lstinline$Vec$ to
syntactically indicate that \lstinline$n$ is irrelevant by surrounding
it with square brackets.
\begin{code}
data Vec' (a:*) : Nat -> * where
  nil' : Vec' a 0
  cons' : [n:Nat] -> a -> Vec' a n -> Vec' a (S n)
\end{code}
Irrelevant constructor arguments are not represented in memory at
run-time, and equations between irrelevant arguments are trivially true since our
\ottdrulename{t\_join} rule is stated using erasure.


The basic building block of irrelevance is irrelevant function types
$ [  \ottmv{x} \!:\! \ottnt{M}  ]  \to \,  \ottnt{N} $, which are inhabited by irrelevant $\lambda$-abstractions
$\lambda  \ottsym{[}  \ottmv{x}  \ottsym{:}  \ottnt{A}  \ottsym{]}  \ottsym{.}  \ottnt{b}$ and eliminated by irrelevant applications $ \ottnt{a}  \; [  \ottnt{b}  ] $. The
introduction rule for irrelevant $\lambda$s is similar to the rule for normal $\lambda$s, 
with one restriction:
\[
\ottdruletXXiabs{}
\]
The free variable condition ensures that the argument $\ottmv{x}$ is not used
at runtime, since it does not remain in the erasure of the body
$\ottnt{b}$. So $\ottmv{x}$ can only  appear in irrelevant positions in $\ottnt{b}$,
such as type annotations and proofs for conversions. On the other
hand, $\ottmv{x}$ is available at type-checking time, so it can occur
freely in the type $\ottnt{B}$.

Since the bound variable is not used at runtime, we can erase it,
leaving only a placeholder for the abstraction or application:
$\ottsym{\mbox{$\mid$}}  \lambda  \ottsym{[}  \ottmv{x}  \ottsym{:}  \ottnt{A}  \ottsym{]}  \ottsym{.}  \ottnt{a}  \ottsym{\mbox{$\mid$}}$ goes to $\lambda  \ottsym{[}  \ottsym{]}  \ottsym{.}  \ottsym{\mbox{$\mid$}}  \ottnt{a}  \ottsym{\mbox{$\mid$}}$ and $\ottsym{\mbox{$\mid$}}   \ottnt{a}  \; [  \ottnt{b}  ]   \ottsym{\mbox{$\mid$}}$ goes to
$\ottsym{\mbox{$\mid$}}  \ottnt{a}  \ottsym{\mbox{$\mid$}}  \ottsym{[}  \ottsym{]}$. As a result, the term $\ottnt{b}$ is not present in memory
and does not get in the way of equational reasoning.

The reason we leave placeholders is to ensure that
syntactic values get erased to syntactic values.
Since we make conversion irrelevant this invariant is needed for type-safety~\cite{sjoberg:quasi}.
For example, using a hypothetical equality we can type the term
\[
\lambda  \ottsym{[}  \ottmv{p}  \ottsym{:}   \mathsf{Bool}   \ottsym{=}   \mathsf{Nat}   \ottsym{]}  \ottsym{.}  \ottsym{1}  \ottsym{+}  \ottkw{conv} \, \ottkw{true} \, \ottkw{at} \, \mathsf{\sim}  \ottmv{p} :  [  \ottmv{p} \!:\!  \mathsf{Bool}   \ottsym{=}   \mathsf{Nat}   ]  \to \,   \mathsf{Nat}  .
\]
In our language this term erases to the value $\lambda [].\ottsym{1}  \ottsym{+}  \ottkw{true}$. On the
other hand, if it erased to the stuck expression $\ottsym{1}  \ottsym{+}  \ottkw{true}$
then progress would fail.

Irrelevant arguments are very useful in dependently typed
programming. In addition to datatype indices, they can be used for
type arguments of polymorphic functions (we could make the argument
\lstinline$a$ of the \lstinline$app$ function irrelevant), and for
proofs of preconditions (we could make the argument \lstinline$p$ of
the \lstinline$safediv$ function in
section~\ref{sec:CBV-partial-correctness} irrelevant).

\paragraph{Value restriction}
The treatment of erasure as discussed so far is closely inspired by
ICC*~\cite{barras+08} and EPTS~\cite{mishra-linger+08}, while a
related system is described by Abel~\cite{abel:fossacs11} and
implemented in recent versions of Agda.

However, the presence of nontermination adds a twist
because normal\gk{``computationally relevant''?} and irrelevant arguments have different
evaluation behavior. In a CBV language, normal arguments are
evaluated to values, but irrelevant arguments just get erased. So
similarly to erased conversions we need to be careful---while we
argued earlier that $  \ottsym{(}  \lambda  \ottmv{x}  \ottsym{:}  \ottsym{(}   \mathsf{Bool}   \ottsym{=}   \mathsf{Nat}   \ottsym{)}  \ottsym{.}  \ottnt{a}  \ottsym{)}  \;  \ottsym{(}  \ottkw{loop}  \ottsym{)}  $ will not lead to
type error thanks to our CBV semantics, the same reasoning clearly
does not work for $ \ottsym{(}  \lambda  \ottsym{[}  \ottmv{x}  \ottsym{:}   \mathsf{Bool}   \ottsym{=}   \mathsf{Nat}   \ottsym{]}  \ottsym{.}  \ottnt{a}  \ottsym{)}  \; [  \ottkw{loop}  ] $. To maintain the
invariant that variables always stand for values, we restrict the
irrelevant application rule to only allow values in the argument
position:
\[
\ottdruletXXiapp{}
\]

This restriction is necessary because allowing
nonterminating expressions to be erased would break type safety for
our language. The problem is not only infinite loops directly
inhabiting bogus equalities like $ \mathsf{Bool}   \ottsym{=}   \mathsf{Nat} $ (as above). The following counter-example
shows that we can get in trouble even by erasing an $\ottkw{abort}$ of
type $ \mathsf{Nat} $. First, note that since the reduction relation treats
variables as values, $ \ottsym{(}  \lambda  \ottmv{x}  \ottsym{.}   \mathsf{Bool}   \ottsym{)}  \;  \ottmv{x}   \leadsto_{\mathsf{p} }   \mathsf{Bool} $. So we have $ \ottkw{join}  : \ottsym{(}   \ottsym{(}  \lambda  \ottmv{x}  \ottsym{:}   \mathsf{Nat}   \ottsym{.}   \mathsf{Bool}   \ottsym{)}  \;   \ottmv{x}  \ottsym{=}  \ottsym{(}  \lambda  \ottmv{x}  \ottsym{:}   \mathsf{Nat}   \ottsym{.}   \mathsf{Nat}   \ottsym{)}  \;  \ottmv{x}    \ottsym{)}  \ottsym{=}  \ottsym{(}   \mathsf{Bool}   \ottsym{=}   \mathsf{Nat}   \ottsym{)}$.
Then the following term typechecks:
\[
\begin{array}{l}
\lambda  \ottsym{[}  \ottmv{x}  \ottsym{:}   \mathsf{Nat}   \ottsym{]}  \ottsym{.}  \lambda  \ottmv{p}  \ottsym{:}  \ottsym{(}   \ottsym{(}  \lambda  \ottmv{x}  \ottsym{:}   \mathsf{Nat}   \ottsym{.}   \mathsf{Bool}   \ottsym{)}  \;  \ottmv{x}   \ottsym{)}  \ottsym{=}  \ottsym{(}   \ottsym{(}  \lambda  \ottmv{x}  \ottsym{:}   \mathsf{Nat}   \ottsym{.}   \mathsf{Nat}   \ottsym{)}  \;  \ottmv{x}   \ottsym{)}  \ottsym{.}  \ottkw{conv} \, \ottmv{p} \, \ottkw{at} \, \mathsf{\sim}  \ottkw{join}
\\
\qquad\qquad :
  [  \ottmv{x} \!:\!  \mathsf{Nat}   ]  \to \,   ( \ottmv{p} \!:\!  \ottsym{(}  \lambda  \ottmv{x}  \ottsym{:}   \mathsf{Nat}   \ottsym{.}   \mathsf{Bool}   \ottsym{)}  \;   \ottmv{x}  \ottsym{=}  \ottsym{(}  \lambda  \ottmv{x}  \ottsym{:}   \mathsf{Nat}   \ottsym{.}   \mathsf{Nat}   \ottsym{)}  \;  \ottmv{x}   )  \to \,  \ottsym{(}   \mathsf{Bool}   \ottsym{=}   \mathsf{Nat}   \ottsym{)}   
\end{array}
\]
On the other hand, by our reduction rule for error terms,
$ \ottsym{(}  \lambda  \ottmv{x}  \ottsym{.}   \mathsf{Bool}   \ottsym{)}  \;  \ottkw{abort}   \leadsto_{\mathsf{p} }  \ottkw{abort}$, so
\[
\ottkw{join} : \ottsym{(}   \ottsym{(}  \lambda  \ottmv{x}  \ottsym{:}   \mathsf{Nat}   \ottsym{.}   \mathsf{Bool}   \ottsym{)}  \;   \mathsf{abort}_{  \mathsf{Nat}  }    \ottsym{)}  \ottsym{=}  \ottsym{(}   \ottsym{(}  \lambda  \ottmv{x}  \ottsym{:}   \mathsf{Nat}   \ottsym{.}   \mathsf{Nat}   \ottsym{)}  \;   \mathsf{abort}_{  \mathsf{Nat}  }    \ottsym{)}.
\]
So if we allowed $  \mathsf{abort}_{  \mathsf{Nat}  }  $ to be given as an irrelevant
argument, then we could write a terminating proof of $ \mathsf{Bool}   \ottsym{=}   \mathsf{Nat} $. Note that all the
equality proofs involved are just $\ottkw{join}$, so this
example does not depend on conversions being computationally
irrelevant. This illustrates a general issue when combining
effects and irrelevance.

\paragraph{The need for termination checking}
The value restriction is a severe limitation on the practical use of irrelevant
arguments. For example, even if we make the length argument to
\lstinline$cons'$ irrelevant, we cannot make the length arguments to
\lstinline$app$ irrelevant.  The problem is that in the recursive case
we would want to return
\begin{code}
cons' a [n+n2] x' (app a [n] xs' [n2] ys)
\end{code}
but \lstinline$n+n2$ is not a value. To make the function typecheck we
must work around the restriction by computing the value of the sum at
runtime. A first attempt would look like
\begin{code}
app : (n1 n2 : Nat) -> (a : *) -> Vec' n1 a -> Vec' n2 a -> Vec' (n1+n2) a
app n1 n2 a xs ys =
  case xs of
    nil' => ys    
    (cons' [n] x xs) => let m = n1-1+n2 in
                          cons' a [m] x (app (n1-1) n2 a xs ys)
\end{code}
This carries out the addition at runtime, so the application of
\lstinline$cons'$ is accepted. But the program still does not
typecheck, due to the mismatch between $m+1$ and $n_1 + n_2$. To make
it check, we need to insert type conversions. Even worse, the
conversions rely on the fact that $n_1 - 1 + n_2 + 1 = n_1 + n_2$. The
proof of this uses induction on $n_2$, i.e. a call to a recursive
function, so the proof also can not be erased and has to be evaluated
at runtime. The rewritten \lstinline$app$ function is more complicated, and
because of the proof even asymptotically slower, which is quite
unsatisfying.

However, to ensure type safety we believe it is enough to ensure that
erased expressions have normal forms.  In this paper we use a
syntactic value check as the very simplest example of a termination
analysis. For a full language, we would mark certain expressions as
belonging to a terminating sublanguage (with, perhaps, the full power
of Type Theory available for termination proofs). To allow the desired
definition of \lstinline$app$ the termination analysis only has to
prove that addition terminates, which is not hard.

Value-restricted irrelevance already has uses: for example,
except for type-level computation all types are values, so we could
compile ML into our language erasing all types. But to support
precisely-typed programs without a performance penalty it is
essential to also be able to erase proofs, and as we demonstrated
above this is not possible without some form of termination
analysis. Therefore, we  consider this language as only a
step towards a practical design.

\paragraph{Datatypes}
In addition to irrelevant $\lambda$-abstractions we also allow
irrelevant arguments in data types, like \lstinline$Vec'$.
Datatype declarations have the form
\[
\ottkw{data} \, \ottmv{D} \, \Delta^{+} \, \ottkw{where} \, \ottsym{\{} \, \ottcomplu{\ottmv{d_{\ottmv{i}}}  \ottsym{:}  \Delta_{\ottmv{i}}  \to \; D \; \Delta^{+}}{\ottmv{i}}{{\mathrm{1}}}{..}{\ottmv{j}} \, \ottsym{\}}
\]
The rules for datatypes in dependently-typed languages often look
intimidating. We tried to make ours as simple as we could.
First, to reduce clutter we write the rules using {\bf telescope
  notation}.  Metavariables $\Delta$ range over lists of
relevance-annotated variable declarations like
$\ottsym{(}  \ottmv{x}  \ottsym{:}  \ottnt{A}  \ottsym{)}  \ottsym{[}  \ottmv{y}  \ottsym{:}  \ottnt{B}  \ottsym{]}  \ottsym{(}  \ottmv{z}  \ottsym{:}  C  \ottsym{)}   \cdot $, also known
as telescopes, while overlined metavariables $\overline{a_i}$ range over
lists of terms. Metavariables $\Delta^{+}$ range over telescopes that have
only relevant declarations.
 Depending on where in an expression they occur,
telescope metavariables stand in for either declarations or lists of
variables, according to the following scheme: if $\Delta$ is
 $(x:A) \; [y:B] \; (z:C)$ and $\overline{a_i}$ is $\ottnt{a}\;[ \ottnt{b} ]\;c$,
then\dots

\begin{tabular}{l l}
\dots this:   &     is shorthand for this: \\
$\ottnt{a_{{\mathrm{1}}}} \, \Delta$     &     $      \ottnt{a_{{\mathrm{1}}}}  \;  \ottmv{x}    \; [  \ottmv{y}  ]    \;  \ottmv{z}  $ \\ 
$\Delta  \to  \ottnt{A_{{\mathrm{1}}}}$    &     $ ( \ottmv{x} \!:\! \ottnt{A} )  \to \,   [  \ottmv{y} \!:\! \ottnt{B}  ]  \to \,   ( \ottmv{z} \!:\! C )  \to \,  \ottnt{A_{{\mathrm{1}}}}   $ \\
$\ottsym{[}  \overline{a_i}  \ottsym{/}  \Delta  \ottsym{]}  \ottnt{a_{{\mathrm{1}}}}$ &  $\ottsym{[}  \ottnt{a}  \ottsym{/}  \ottmv{x}  \ottsym{]}  \ottsym{[}  \ottnt{b}  \ottsym{/}  \ottmv{y}  \ottsym{]}  \ottsym{[}  c  \ottsym{/}  \ottmv{z}  \ottsym{]}  \ottnt{a_{{\mathrm{1}}}}$ \\
$\Gamma  \ottsym{,}  \Delta$  &   $\Gamma  \ottsym{,}   \ottmv{x}  :  \ottnt{A}   \ottsym{,}   \ottmv{y}  :  \ottnt{B}   \ottsym{,}   \ottmv{z}  :  C $ \\
$\Gamma  \vdash  \overline{a_i}  \ottsym{:}  \Delta$ & $ \Gamma \vdash \ottnt{a} : \ottnt{A}  \land  \Gamma \vdash \ottnt{b} : \ottsym{[}  \ottnt{a}  \ottsym{/}  \ottmv{x}  \ottsym{]}  \ottnt{B}  \land  \Gamma \vdash c : \ottsym{[}  \ottnt{b}  \ottsym{/}  \ottmv{y}  \ottsym{]}  \ottsym{[}  \ottnt{a}  \ottsym{/}  \ottmv{x}  \ottsym{]}  C $
\end{tabular}

We also simplify the rules by having {\bf parameters but not
indices}. Each datatype has a list of parameters $\Delta$, and
these are instantiated uniformly (i.e. the type of each data
constructor $\ottmv{d_{\ottmv{i}}}$ ends with $\ottmv{D} \, \Delta^{+}$, the type constructor
$\ottmv{D}$ applied to a list of variables). This restriction does not limit
expressivity, because we can elaborate non-uniform indexes into a
combination of parameters and equality proofs (this is how Haskell
GADTs are elaborated into GHC Core~\cite{sulzmann-fc}). For example, the declaration of
\lstinline{Vec'} above can be reformulated without indices as
\begin{code}
data Vec' (a:*) (n:Nat) where
  nil'  : [p:n=0] -> Vec' a n
  cons' : [m:Nat] -> [p:n=S m] -> a -> Vec' a m -> Vec' a n
\end{code}

To make the statement of the  canonical forms lemma simpler (see lemma
\ref{lemma:canonical-forms} below) we require constructors to be fully applied, so they
do not pollute the function space. In other words, $\ottmv{d}$ by itself
is not a well-formed expression, it must be applied to a list of
parameters and a list of arguments $ \ottmv{d}  \; [  \overline{A_i}  ] \;  \overline{a_i} $.

In the corresponding elimination form (the case expression $\ottkw{case} \, \ottnt{b} \, \ottkw{as} \, \ottsym{[y]} \, \ottkw{of} \, \ottsym{\{} \, \ottcomplu{\ottmv{d_{\ottmv{i}}} \, \Delta_{\ottmv{i}}  \Rightarrow  \ottnt{a_{\ottmv{i}}}}{\ottmv{i}}{{\mathrm{1}}}{..}{\ottmv{l}} \, \ottsym{\}}$) the programmer
must write one branch $\ottmv{d_{\ottmv{i}}} \, \Delta_{\ottmv{i}}  \Rightarrow  \ottnt{a_{\ottmv{i}}}$ for each
constructor of the datatype $\ottmv{D}$. The branch only introduces
pattern variables for the constructor arguments, as the parameters are
fixed throughout the case. However, the parameters are used to refine
the context that the match is checked in: if $ \Gamma \vdash \ottnt{b} :  \ottmv{D}  \;  \overline{B_i}  $,
then for each case we check
\[
  \Gamma  \ottsym{,}  \ottsym{[}  \overline{B_i}  \ottsym{/}  \Delta^{+}  \ottsym{]}  \Delta_{\ottmv{i}}  \ottsym{,}   \ottmv{y}  :  \ottnt{b}  \ottsym{=}  \ottmv{d_{\ottmv{i}}} \, \Delta_{\ottmv{i}}  \vdash \ottnt{a_{\ottmv{i}}} : \ottnt{A} 
\]
The context also introduces an equality proof $y$ which can by used (in irrelevant
positions) to exploit the new information about which constructor matched.

So far, this is a fairly standard treatment of datatypes. However, we
want to point out how {\bf irrelevant parameters and constructor
arguments} work. 

First, parameters to data constructors are always irrelevant, since
they are completely fixed by the types. The erasure operation
simply deletes them: $\ottsym{\mbox{$\mid$}}   \ottmv{d}  \; [  \overline{A_i}  ] \;  \overline{a_i}   \ottsym{\mbox{$\mid$}}$ is $\ottmv{d} \, \ottsym{\mbox{$\mid$}}  \overline{a_i}  \ottsym{\mbox{$\mid$}}$.
On the other hand, it never makes sense for parameters to data\emph{type}
constructors to be irrelevant. For example, if the parameters to
\lstinline$Vec$ were made irrelevant, the join rule would validate
$ \Gamma \vdash \ottkw{join} : \ottsym{(}     \mathsf{Vec}   \; [   \mathsf{Nat}   ]   \; [  \ottsym{1}  ]   \ottsym{)}  \ottsym{=}  \ottsym{(}     \mathsf{Vec}   \; [   \mathsf{Bool}   ]   \; [  \ottsym{2}  ]   \ottsym{)} $, which would defeat the
point of having the parameters in the first place. This is reflected
in our syntax for datatypes, which requires that the list of
parameters is a $\Delta^{+}$ (i.e. contains no irrelevant declarations).
In order to typecheck a datatype constructor, we look up the
corresponding datatype declaration in the context and check that the
provided parameters have the right type.
\[
\ottdruletXXtcon{}
\]

Finally, arguments to data constructors $\ottmv{d_{\ottmv{i}}}$ can be marked as
relevant or not in the telescope $\Delta_{\ottmv{i}}$, and this is
automatically reflected in the typing rule for constructor application
and erasure. For example, given the above declaration of \lstinline$Vec'$,
the annotated expression 
\begin{code}
cons' [Bool] [1] [0] [join] true (nil' [Bool] [0] [join])
\end{code}
is well-typed and erases to \lstinline$cons' [] [] true (nil' [])$. 
However, making a constructor argument erased carries a corresponding
restriction in the case statement: since the argument has no run-time
representation it may only be used in irrelevant positions. For
example, in a case branch
\begin{code}
  cons' [m:Nat] [p:n=S m] (x:a) (xs:Vec' a m) => ...body... 
\end{code}
the code in the body can use \lstinline$x$ without restrictions but
can only use \lstinline$m$ in irrelevant positions such as type
annotations and conversions.  With the original \lstinline$Vec$ type we
could write a constant-time length function by projecting out \lstinline$m$,
but that is not possible with \lstinline$Vec'$.

\subsection{Very heterogenous equality}
\label{sec:heterogenous-eq}
The \lstinline$app-assoc$ example also illustrates a
different problem with indexed data: the two sides of the 
equation have different types (namely \lstinline$Vec a (n1+(n2+n3))$
versus \lstinline$Vec a ((n1+n2)+n3)$)---so, e.g., the usual
equality in Coq does not even allow writing down the 
equation!\gk{Can this be phrased more precisely?} We need some form of {\bf heterogenous equality}.
The most popular choice for this is JMeq~\cite{mcbride:motive},
which allows you to \emph{state} any equality, but only \emph{use}
them if both sides have (definitionally) the same type. Massaging goals into a
form where the equalities are usable often requires certain tricks
and idioms (see e.g. \cite{chlipala:cpdt}, chapter 10).

For this language, we wanted something simpler. Like JMeq, we allow
stating any equation as long as the two sides are
well-typed.  Our formation rule for the equality type is
\[
\ottdruletXXeq{}
\]
Unlike JMeq, however, conversion can use an equality even if the two
sides have different types. This is similar to the way equality is handled in
Guru~\cite{guru09}, although the details differ.

We showed a simplified version (\ottdrulename{vconv}) of our conversion rule on
page~\pageref{vconvrule}; we present the full rule (\ottdrulename{t\_conv}) in
section~\ref{sec:annotations}.  We now build-up the full rule from the simplified
rule, step-by-step, motivating each addition along the way.
First, in order to be able to change only parts of a type, we phrase
the rule in terms of substituting into a template $\ottnt{A}$.
\[
\ottdruleconvXXsubst{}
\]
For example, given a proof $ \Gamma \vdash \ottnt{v} : \ottmv{y}  \ottsym{=}  \ottsym{0} $, we can convert the type
$  \mathsf{Vec}   \;    \mathsf{Nat}   \;  \ottsym{(}  \ottmv{y}  \ottsym{+}  \ottmv{y}  \ottsym{)}  $ to $  \mathsf{Vec}   \;    \mathsf{Nat}   \;  \ottsym{(}  \ottmv{y}  \ottsym{+}  \ottsym{0}  \ottsym{)}  $ using the template substitution
$\ottkw{Vec} \, \ottkw{Nat} \, \ottsym{(y}  \ottsym{+}  \mathsf{\sim}  \ottsym{v)}$.

We need the premise $ \Gamma \vdash \ottsym{[}  \ottnt{B_{{\mathrm{2}}}}  \ottsym{/}  \ottmv{x}  \ottsym{]}  \ottnt{A} :  \star  $ for two reasons. First,
since our equality is heterogenous, we do not know that $\ottnt{B_{{\mathrm{2}}}}$ is a
type even if $\ottnt{B_{{\mathrm{1}}}}$ is. It is possible to write a function
that takes a proof of $ \mathsf{Nat}   \ottsym{=}  \ottsym{3}$ as an argument (although it will
never be possible to actually call it). But even if equality were
homogenous we would still need the wellformedness premise for the
same reason we need it in the application rule. If $\ottnt{B_{{\mathrm{1}}}}$ is a
value and $\ottnt{B_{{\mathrm{2}}}}$ is not, then $\ottsym{[}  \ottnt{B_{{\mathrm{2}}}}  \ottsym{/}  \ottmv{x}  \ottsym{]}  \ottnt{A}$ is not guaranteed to
be well-formed.

\subsection{Multiple simultaneous conversions}
\label{sec:n-ary-conv}
Next, to achieve the full potential of our flexible
elimination rule we find it is not sufficient to eliminate one
equality at a time. For a simple example, consider trying to prove
$  \ottmv{f}  \;  \ottmv{x}   =   \ottmv{f'}  \;  \ottmv{x'}  $ in the context
\[
 \ottmv{f}  :  \ottnt{A}  \to  \ottnt{B} ,  \ottmv{f'}  :  \ottnt{A'}  \to  \ottnt{B} ,  \ottmv{x}  :  \ottnt{A} , \ottmv{x'}  :  \ottnt{A'} , 
 \ottmv{p}  :  \ottmv{f}  \ottsym{=}  \ottmv{f'} , \ottmv{q}  :  \ottmv{x}  \ottsym{=}  \ottmv{x'} 
\]
Note that there is no equation relating $\ottnt{A}$ and $\ottnt{A'}$.  Using
one equality at a time, the only way to make progress is by
transitivity, that is by trying to prove $  \ottmv{f}  \;  \ottmv{x}   =   \ottmv{f}  \;  \ottmv{x'}  $ and
$  \ottmv{f}  \;  \ottmv{x'}   =   \ottmv{f'}  \;  \ottmv{x'}  $. However, the intermediate expression $  \ottmv{f}  \;  \ottmv{x'}  $ is
not well-typed so the attempt fails. To make propositional
equality a congruence with respect to application, we are led to a
conversion rule that allows eliminating several equations at once.
\[ 
\ottdruleconvXXmultisubst{}
\]

Of course, the above example is artificial: we don't really expect
that a programmer would often want to prove equations between terms of
unrelated types. A more practical motivation comes from proofs about
indexed data like vectors, where $\ottnt{A}$ might be \lstinline$Vec a (n+(n2+n3))$
and $\ottnt{A'}$ be \lstinline$Vec a ((n+n2)+n3)$. In such an example, 
$\ottnt{A}$ and $\ottnt{A'}$ are indeed provably equal, but 
our $n$-ary conversion rule obviates the need to provide that proof.

The fact that our conversion can use heterogenous equations also has a
downside: we are unable to support certain type-directed
equality rules. In particular, adding \emph{functional extensionality}
would be unsound. Extensionality implies $\ottsym{(}  \lambda  \ottmv{x}  \ottsym{:}  \ottsym{(}  \ottsym{1}  \ottsym{=}  \ottsym{0}  \ottsym{)}  \ottsym{.}  \ottsym{1}  \ottsym{)} =
\ottsym{(}  \lambda  \ottmv{x}  \ottsym{:}  \ottsym{(}  \ottsym{1}  \ottsym{=}  \ottsym{0}  \ottsym{)}  \ottsym{.}  \ottsym{0}  \ottsym{)}$ since the two functions agree on all arguments (vacuously). But 
our annotation-ignoring equality shows $\ottsym{(}  \lambda  \ottmv{x}  \ottsym{:}  \ottsym{(}  \ottsym{1}  \ottsym{=}  \ottsym{0}  \ottsym{)}  \ottsym{.}  \ottsym{1}  \ottsym{)} =
\ottsym{(}  \lambda  \ottmv{x}  \ottsym{:}   \mathsf{Nat}   \ottsym{.}  \ottsym{1}  \ottsym{)}$, so by transitivity we would get
$\ottsym{(}  \lambda  \ottmv{x}  \ottsym{:}   \mathsf{Nat}   \ottsym{.}  \ottsym{1}  \ottsym{)}  \ottsym{=}  \ottsym{(}  \lambda  \ottmv{x}  \ottsym{:}   \mathsf{Nat}   \ottsym{.}  \ottsym{0}  \ottsym{)}$, and from there to $\ottsym{1}  \ottsym{=}  \ottsym{0}$.

\subsection{Annotating equality and conversion}
\label{sec:annotations}
Ultimately, the unannotated language is the most interesting artifact,
since that is what actually gets executed. The point of defining an
annotated language is to make it convenient to write down typings of
unannotated terms. (We could consider the annotated terms as reified
typing derivations). We designed the annotated language by starting with the unannotated
language and adding just enough annotations that a typechecker
traversing an annotated term will always know what to do.  For most
language constructs this was straightforward, e.g. adding a type annotation to
$\lambda$-abstractions. The annotated programs get quite verbose, so
for a full language more sophisticated methods like bidirectional type
checking, local type inference, or unification-based inference would
be helpful, but these techniques are beyond the scope of this paper.

The last step is to understand how nontermination and irrelevance
affect the final annotated conv and join rules, 
\ottdrulename{t\_conv} and \ottdrulename{t\_join} below.
The conv rule in the erased language, including $n$-ary
substitution, looks like
\[
\ottdruleetXXconv{}
\]
To guide the typechecker, in addition to the annotated version of
$\ottnt{m}$ we need to supply the (annotated
versions of) the proof values $\ottnt{u_{\ottmv{i}}}$ and the (annotated version
of) the ``template'' type $\ottnt{M}$ that we are substituting into.
A first attempt at a corresponding annotated rule would look like
the \ottdrulename{conv\_multisubst} rule we showed above.\vs{change to ``on the previous page'', if appropriate}

However, \ottdrulename{conv\_multisubst} needs one more modification. 
In order to correspond exactly
to the unannotated conv rule it should ignore expressions in
irrelevant positions. For example, consider proving the equation
$  \ottmv{f}  \; [  \ottnt{A}  ]   \;  \ottnt{a}  =   \ottmv{f}  \; [  \ottnt{B}  ]   \;  \ottnt{b} $, which erases to $ \ottmv{f}  \ottsym{[}  \ottsym{]}  \;  \ottsym{\mbox{$\mid$}}  \ottnt{a}  \ottsym{\mbox{$\mid$}}  =
 \ottmv{f}  \ottsym{[}  \ottsym{]}  \;  \ottsym{\mbox{$\mid$}}  \ottnt{b}  \ottsym{\mbox{$\mid$}} $. The unannotated conv rule only requires a proof of
$\ottsym{\mbox{$\mid$}}  \ottnt{a}  \ottsym{\mbox{$\mid$}}  \ottsym{=}  \ottsym{\mbox{$\mid$}}  \ottnt{b}  \ottsym{\mbox{$\mid$}}$, so in the annotated language we should not have to
provide a proof involving $\ottnt{A}$ and $\ottnt{B}$.
Therefore, in the annotated rule we allow two kinds of
evidence $\ottnt{P}$: either a value $\ottnt{v}$ which is a proof of an
equation, or just an annotation $\ottsym{[}  \ottnt{a}  \ottsym{=}  \ottnt{b}  \ottsym{]}$ stating how an irrelevant
subexpression should be changed. We also need to specify the template
that the substitution is applied to. As a matter of concrete
syntax, we prefer writing the evidence $\ottnt{P_{\ottmv{j}}}$ interleaved with the
template, marking it with a tilde. So our final annotated rule looks
like this:
\[
\begin{array}{l}
  \begin{array}{lcl}
  \ottnt{P} & ::= & \ottnt{v} \alt \ottsym{[}  \ottnt{a}  \ottsym{=}  \ottnt{b}  \ottsym{]}
  \end{array} 
\\
\\
  \ottdruletXXconv{}
\end{array}
\]
For example, if \lstinline$a : Vec A x$ and \lstinline$y : x = 3$,
then \lstinline$conv a at Vec A ~y$ has type \lstinline$Vec A 3$.

Next, consider the equality introduction rule. In the unannotated
language it is simply
\[
\ottdrule{
  \ottpremise{ \ottnt{m_{{\mathrm{1}}}}  \leadsto^{*}_{\mathsf{p} }  \ottnt{n} \quad \ottnt{m_{{\mathrm{2}}}}  \leadsto^{*}_{\mathsf{p} }  \ottnt{n} }
  \ottpremise{  \ottnt{H} \vdash \ottnt{m_{{\mathrm{1}}}}  \ottsym{=}  \ottnt{m_{{\mathrm{2}}}} :  \star   }
}{
   \ottnt{H} \vdash \ottkw{join} : \ottnt{m_{{\mathrm{1}}}}  \ottsym{=}  \ottnt{m_{{\mathrm{2}}}} 
}{
\ottdrulename{et\_join}
}
\]
This is very similar to what other dependent languages, such as PTSs,
offer. In those languages, this rule may be implemented by evaluating
both sides to normal forms and comparing. Unfortunately, in the
presence of nontermination there is no similarly simple
algorithm---the parallel reduction relation is nondeterministic, and
since we are not guaranteed to hit a normal form we would have to
search through all possible evaluation orders.

One possibility would be to write down the expression to be reduced,
and tag sub-expressions of it with how many steps to take, perhaps
marked with tildes. In our experiments with a prototype type-checker
for our language, we have adopted a simpler scheme. The join rule only
does deterministic CBV evaluation for at most a specified number of
steps.  So, our final annotated join rule looks like
\[
\ottdruletXXjoin{}
\]
where $\ottmv{i}$ and $\ottmv{j}$ are integer literals.
In the common case when both $\ottnt{a}$ and $\ottnt{b}$ quickly reach
normal forms, the programmer can simply pick high numbers for the step
counts, and in the concrete syntax we treat \lstinline$join$ as an
abbreviation for \lstinline$join 100 100$. When we want to prove
equations between terms that are already values, we can use
\lstinline$conv$ to change subterms of them. For example, to prove the
equality $  \mathsf{Vec}   \;   \ottnt{A}  \;  \ottsym{(}  \ottsym{1}  \ottsym{+}  \ottsym{1}  \ottsym{)}   =   \mathsf{Vec}   \;   \ottnt{A}  \;  \ottsym{2}  $ we write
\begin{code}
conv (join : Vec A 2 = Vec A 2) at (Vec A 2 = Vec A ~(join : 1+1 = 2))
\end{code} 
Not every parallel reduction step can be reached this way, since
substitution is capture-avoiding. For instance, with this choice of
annotations we cannot show an equation like
$\ottsym{(}   \lambda  \ottmv{x}  \ottsym{.}  \ottsym{(}  \lambda  \ottmv{y}  \ottsym{.}  \ottmv{y}  \ottsym{)}  \;  \ottmv{x}   \ottsym{)}  \ottsym{=}  \ottsym{(}  \lambda  \ottmv{x}  \ottsym{.}  \ottmv{x}  \ottsym{)}$.  So far, we have not found this
restriction limiting.

\section{Metatheory}
\label{sec:metatheory}

The main technical contribution of this paper is a proof of type
safety for our language via standard preservation and progress
theorems. The full proof can be found in the appendix. In this
section, we highlight the most interesting parts of it.

\subsection{Annotated and unannotated  type systems}
While the description so far has been in terms of a
type system for annotated terms, we have also developed a type system
for the unannotated language, and it is the unannotated system that is
important for the metatheoretical development.

The unannotated typing judgment is of the form $ \ottnt{H} \vdash \ottnt{m} : \ottnt{M} $,
where the metavariable $\ottnt{H}$ ranges over unannotated typing
environments (i.e., environments of assumptions $ \ottmv{x}  :  \ottnt{M} $). Below
we give an outline of the rules.  The complete definition can be
found in the appendix (section~\ref{sec:unannotated-type-system}).
The two type systems were designed so that there are enough
annotations to make typechecking the annotated language decidable,
and to make erasure into the unannotated system preserve well-typedness:

\begin{lemma}[Decidability of type checking]
There is an algorithm which given $\Gamma$ and $\ottnt{a}$ computes the
unique $\ottnt{A}$ such that $ \Gamma \vdash \ottnt{a} : \ottnt{A} $, or reports that there is no
such $\ottnt{A}$.
\end{lemma}

\begin{lemma}[Annotation soundness]
If $ \Gamma \vdash \ottnt{a} : \ottnt{A} $ then $ \ottsym{\mbox{$\mid$}}  \Gamma  \ottsym{\mbox{$\mid$}} \vdash \ottsym{\mbox{$\mid$}}  \ottnt{a}  \ottsym{\mbox{$\mid$}} : \ottsym{\mbox{$\mid$}}  \ottnt{A}  \ottsym{\mbox{$\mid$}} $ . 
\end{lemma}

In practice, the unannotated rules simply mirror the annotated rules,
except that all the terms in them have gone through erasure. As an example,
compare the annotated and unannotated versions of the
$\ottdrulename{iabs}$ rule:
\[
\begin{array}{lll}
\ottdruletXXiabs{} 
& \qquad & 
\ottdruleetXXiabs{}
\end{array}
\]

Since our operational semantics is defined for unannotated terms, the
preservation and progress theorems will be also stated in terms of
unannotated terms.  One could ask whether it would be possible to
define an operational semantics for the annotated terms and then prove
preservation for the annotated language. The main complication of
doing that is that as terms steps extra type conversions must be
added, which would complicate the step relation.

\subsection{Properties of parallel reduction}
\label{sec:parred-props}

The key intuition in our treatment of equality is that, in an empty
context, propositional equality coincides with joinability under
parallel reduction.  As a result, we will need some basic lemmas about
parallel reduction throughout the proof. These are familiar from,
e.g., the metatheory of PTSs, with the slight difference that the usual
substitution lemma is replaced with two special cases because we work
with CBV reduction.

\begin{lemma}[Substitution of $ \leadsto_{\mathsf{p} } $]
\label{lemma:subst1}
If $\ottnt{N}  \leadsto_{\mathsf{p} }  \ottnt{N'}$, then $\ottsym{[}  \ottnt{N}  \ottsym{/}  \ottmv{x}  \ottsym{]}  \ottnt{M}  \leadsto_{\mathsf{p} }  \ottsym{[}  \ottnt{N'}  \ottsym{/}  \ottmv{x}  \ottsym{]}  \ottnt{M}$.
\end{lemma}

\begin{lemma}[Substitution into $ \leadsto_{\mathsf{p} } $]
\label{lemma:subst2}
If $\ottnt{u}  \leadsto_{\mathsf{p} }  \ottnt{u'}$ and $\ottnt{m}  \leadsto_{\mathsf{p} }  \ottnt{m'}$, then 
$\ottsym{[}  \ottnt{u}  \ottsym{/}  \ottmv{x}  \ottsym{]}  \ottnt{m}  \leadsto_{\mathsf{p} }  \ottsym{[}  \ottnt{u'}  \ottsym{/}  \ottmv{x}  \ottsym{]}  \ottnt{m'}$.
\end{lemma}

\begin{lemma}[Confluence of $ \leadsto_{\mathsf{p} } $]
If $\ottnt{m}  \leadsto^{*}_{\mathsf{p} }  \ottnt{m_{{\mathrm{1}}}}$ and $\ottnt{m}  \leadsto^{*}_{\mathsf{p} }  \ottnt{m_{{\mathrm{2}}}}$, then there exists
some $\ottnt{m'}$ such that $\ottnt{m_{{\mathrm{1}}}}  \leadsto^{*}_{\mathsf{p} }  \ottnt{m'}$ and $\ottnt{m_{{\mathrm{2}}}}  \leadsto^{*}_{\mathsf{p} }  \ottnt{m'}$.
\end{lemma} 

\begin{definition}[Joinability]
We write $ \ottnt{m_{{\mathrm{1}}}} \, \curlyveedownarrow \, \ottnt{m_{{\mathrm{2}}}} $ if there exists some $\ottnt{n}$ such that
$\ottnt{m_{{\mathrm{1}}}}  \leadsto^{*}_{\mathsf{p} }  \ottnt{n}$ and $\ottnt{m_{{\mathrm{2}}}}  \leadsto^{*}_{\mathsf{p} }  \ottnt{n}$.
\end{definition}

Using the above lemmas it is easy to see that $ \curlyveedownarrow $ is an
equivalence relation, and that $ \ottnt{m_{{\mathrm{1}}}} \, \curlyveedownarrow \, \ottnt{m_{{\mathrm{2}}}} $ implies
$ \ottsym{[}  \ottnt{m_{{\mathrm{1}}}}  \ottsym{/}  \ottmv{x}  \ottsym{]}  \ottnt{M} \, \curlyveedownarrow \, \ottsym{[}  \ottnt{m_{{\mathrm{2}}}}  \ottsym{/}  \ottmv{x}  \ottsym{]}  \ottnt{M} $.

\subsection{Preservation}

For the preservation proof we need the usual structural properties:
weakening and substitution. Weakening is standard, but somewhat
unusually substitution is restricted to substituting values $\ottnt{u}$
into the judgment, not arbitrary terms.  This is because our equality
is CBV, so substituting a non-value could block reductions and cause
types to no longer be equal.

\begin{lemma}[Substitution]
If $ \ottnt{H_{{\mathrm{1}}}}  \ottsym{,}   \ottmv{x_{{\mathrm{1}}}}  :  \ottnt{M_{{\mathrm{1}}}}   \ottsym{,}  \ottnt{H_{{\mathrm{2}}}} \vdash \ottnt{m} : \ottnt{M} $ and $ \ottnt{H_{{\mathrm{1}}}} \vdash \ottnt{u_{{\mathrm{1}}}} : \ottnt{M_{{\mathrm{1}}}} $,
then $ \ottnt{H_{{\mathrm{1}}}}  \ottsym{,}  \ottsym{[}  \ottnt{u_{{\mathrm{1}}}}  \ottsym{/}  \ottmv{x_{{\mathrm{1}}}}  \ottsym{]}  \ottnt{H_{{\mathrm{2}}}} \vdash \ottsym{[}  \ottnt{u_{{\mathrm{1}}}}  \ottsym{/}  \ottmv{x_{{\mathrm{1}}}}  \ottsym{]}  \ottnt{m} : \ottsym{[}  \ottnt{u_{{\mathrm{1}}}}  \ottsym{/}  \ottmv{x_{{\mathrm{1}}}}  \ottsym{]}  \ottnt{M} $.
\end{lemma}

Preservation also needs an inversion lemmas for $\lambda$s,
irrelevant $\lambda$s, $\mathsf{rec}$, and data constructors.
They are similar, and we show the one for $\lambda$-abstractions as an
example.

\begin{lemma}[Inversion for $\lambda$s]
If $ \ottnt{H} \vdash \lambda  \ottmv{x}  \ottsym{.}  \ottnt{n} : \ottnt{M} $, then  there exists $\ottnt{m_{{\mathrm{1}}}}$, $\ottnt{M_{{\mathrm{1}}}}$, $\ottnt{N_{{\mathrm{1}}}}$,
such that $ \ottnt{H} \vdash \ottnt{m_{{\mathrm{1}}}} : \ottnt{M}  \ottsym{=}   ( \ottmv{x} \!:\! \ottnt{M_{{\mathrm{1}}}} )  \to \,  \ottnt{N_{{\mathrm{1}}}}  $ and $ \ottnt{H}  \ottsym{,}   \ottmv{x}  :  \ottnt{M_{{\mathrm{1}}}}  \vdash \ottnt{n} : \ottnt{N_{{\mathrm{1}}}} $. 
\end{lemma}

Notice that this is weaker statement than in a language with
computationally relevant conversion. For example, in a PTS we would
have that $\ottnt{M}$ is $\beta$-convertible to the type
$ ( \ottmv{x} \!:\! \ottnt{M_{{\mathrm{1}}}} )  \to \,  \ottnt{N_{{\mathrm{1}}}} $, not just provably equal to it. But in our
language, if the context contained the equality
$\ottsym{(}   \mathsf{Nat}   \to   \mathsf{Nat}   \ottsym{)}  \ottsym{=}   \mathsf{Nat} $, then we could show $ \ottnt{H} \vdash \lambda  \ottmv{x}  \ottsym{.}  \ottmv{x} :  \mathsf{Nat}  $
using a (completely erased) conversion. As we will see, we need to add
extra injectivity rules to the type system to compensate.

Now we are ready to prove the preservation theorem. For type safety we
are only interested in preservation for $ \leadsto_{\mathsf{cbv} } $, but it is
convenient to generalize the theorem to $ \leadsto_{\mathsf{p} } $.

\begin{thm}[Preservation] 
\item If $ \ottnt{H} \vdash \ottnt{m} : \ottnt{M} $ and $\ottnt{m}  \leadsto_{\mathsf{p} }  \ottnt{m'}$, then $ \ottnt{H} \vdash \ottnt{m'} : \ottnt{M} $.
\end{thm}
The proof is mostly straightforward by an induction on the typing
derivation. There are some wrinkles, all of which can be
seen by considering some cases for applications. The typing rule looks
like  
\[
\ottdruleetXXapp{}
\]

First consider the case when $ \ottnt{m}  \;  \ottnt{n} $ steps by congruence, 
$ \ottnt{m}  \;  \ottnt{n}   \leadsto_{\mathsf{p} }   \ottnt{m}  \;  \ottnt{n'} $. Directly by IH we get that $ \ottnt{H} \vdash \ottnt{n'} : \ottnt{M} $,
but because of our CBV-style application rule we also need to
establish $ \ottnt{H} \vdash \ottsym{[}  \ottnt{n'}  \ottsym{/}  \ottmv{x}  \ottsym{]}  \ottnt{N} :  \star  $. But by substitution of $ \leadsto_{\mathsf{p} } $
we know that $\ottsym{[}  \ottnt{n}  \ottsym{/}  \ottmv{x}  \ottsym{]}  \ottnt{N}  \leadsto_{\mathsf{p} }  \ottsym{[}  \ottnt{n'}  \ottsym{/}  \ottmv{x}  \ottsym{]}  \ottnt{N}$, so this also follows by IH
(this is why we generalize the theorem to $ \leadsto_{\mathsf{p} } $).

This showed $ \ottnt{H} \vdash  \ottnt{m}  \;  \ottnt{n'}  : \ottsym{[}  \ottnt{n'}  \ottsym{/}  \ottmv{x}  \ottsym{]}  \ottnt{N} $, but we needed 
$ \ottnt{H} \vdash  \ottnt{m}  \;  \ottnt{n'}  : \ottsym{[}  \ottnt{n}  \ottsym{/}  \ottmv{x}  \ottsym{]}  \ottnt{N} $. Since $\ottsym{[}  \ottnt{n}  \ottsym{/}  \ottmv{x}  \ottsym{]}  \ottnt{N}  \leadsto_{\mathsf{p} }  \ottsym{[}  \ottnt{n'}  \ottsym{/}  \ottmv{x}  \ottsym{]}  \ottnt{N}$ we have
$ \ottnt{H} \vdash \ottkw{join} : \ottsym{[}  \ottnt{n'}  \ottsym{/}  \ottmv{x}  \ottsym{]}  \ottnt{N}  \ottsym{=}  \ottsym{[}  \ottnt{n}  \ottsym{/}  \ottmv{x}  \ottsym{]}  \ottnt{N} $, and we can conclude using the
conv rule. This illustrates how fully erased conversions
generalize the $\beta$-equivalence rule familiar from PTSs.

Second, consider the case when an application steps by
$\beta$-reduction, $ \ottsym{(}  \lambda  \ottmv{x}  \ottsym{.}  \ottnt{m_{{\mathrm{0}}}}  \ottsym{)}  \;  \ottnt{u}   \leadsto_{\mathsf{p} }  \ottsym{[}  \ottnt{u}  \ottsym{/}  \ottmv{x}  \ottsym{]}  \ottnt{m_{{\mathrm{0}}}}$, and we need to show
$ \ottnt{H} \vdash \ottsym{[}  \ottnt{u}  \ottsym{/}  \ottmv{x}  \ottsym{]}  \ottnt{m_{{\mathrm{0}}}} : \ottsym{[}  \ottnt{u}  \ottsym{/}  \ottmv{x}  \ottsym{]}  \ottnt{N} $. The inversion lemma
for $\lambda  \ottmv{x}  \ottsym{.}  \ottnt{m_{{\mathrm{0}}}}$ gives $ \ottnt{H}  \ottsym{,}   \ottmv{x}  :  \ottnt{M_{{\mathrm{1}}}}  \vdash \ottnt{m_{{\mathrm{0}}}} : \ottnt{N_{{\mathrm{1}}}} $ for some
$ \ottnt{H} \vdash \ottkw{join} :  ( \ottmv{x} \!:\! \ottnt{M} )  \to \,  \ottnt{N}  \ottsym{=}   ( \ottmv{x} \!:\! \ottnt{M_{{\mathrm{1}}}} )  \to \,  \ottnt{N_{{\mathrm{1}}}}   $. Now we need to convert the
type of $\ottnt{u}$ to $ \ottnt{H} \vdash \ottnt{u} : \ottnt{M_{{\mathrm{1}}}} $, so that we can apply
substitution and get $ \ottnt{H} \vdash \ottsym{[}  \ottnt{u}  \ottsym{/}  \ottmv{x}  \ottsym{]}  \ottnt{m_{{\mathrm{0}}}} : \ottsym{[}  \ottnt{u}  \ottsym{/}  \ottmv{x}  \ottsym{]}  \ottnt{N_{{\mathrm{1}}}} $, and finally
convert back to $\ottsym{[}  \ottnt{u}  \ottsym{/}  \ottmv{x}  \ottsym{]}  \ottnt{N}$. To do this we need to decompose the
equality proof from the inversion lemma into proofs of $\ottnt{M}  \ottsym{=}  \ottnt{M_{{\mathrm{1}}}}$
and $\ottsym{[}  \ottnt{u}  \ottsym{/}  \ottmv{x}  \ottsym{]}  \ottnt{N_{{\mathrm{1}}}}  \ottsym{=}  \ottsym{[}  \ottnt{u}  \ottsym{/}  \ottmv{x}  \ottsym{]}  \ottnt{N}$. We run into the same issue in the cases
for irrelevant application and pattern matching on datatypes.
So we add a set of injectivity rules to our type system to make these
cases go through (figure~\ref{fig:injectivity}).

\begin{figure}
\ruleline{
\ottdruleetXXinjtcon{} \qquad
\ottdruleetXXinjdom{}
}
\ruleline{
\ottdruleetXXinjrng{}
}
\caption{Injectivity rules (the two rules for $ [  \ottmv{x} \!:\! \ottnt{M_{{\mathrm{1}}}}  ]  \to \,  \ottnt{N_{{\mathrm{1}}}} $ are similar and not shown)}
\label{fig:injectivity}
\end{figure}

\subsection{Progress}

As is common in languages with dependent pattern matching, when
proving progress we have to worry about ``bad''
equations. Specifically, this shows up in the canonical forms
lemma. We want to say that if a closed value has a function type, then
it is actually a function. However, what if we had a proof of
$ \mathsf{Nat}   \ottsym{=}  \ottsym{(}   \mathsf{Nat}   \to   \mathsf{Nat}   \ottsym{)}$?  To rule that out, we start by proving a
lemma characterizing when two expressions can be propositionally
equal. From now on, $H_{D}$ denotes a context which is empty except
that it may contain datatype declarations.

\begin{lemma}[Soundness of equality]
\label{lemma:equality-soundness}
If $ H_{D} \vdash \ottnt{u} : \ottnt{M} $ and $ \ottnt{M} \, \curlyveedownarrow \, \ottsym{(}  \ottnt{m_{{\mathrm{1}}}}  \ottsym{=}  \ottnt{n_{{\mathrm{1}}}}  \ottsym{)} $,
then $ \ottnt{m_{{\mathrm{1}}}} \, \curlyveedownarrow \, \ottnt{n_{{\mathrm{1}}}} $.
\end{lemma}

The proof is by induction on $ H_{D} \vdash \ottnt{u} : \ottnt{M} $. It is not hard, but
it is worth describing briefly. To rule out rules like
$\lambda$-abstraction, we need to know that it is never the case that
$  ( \ottmv{x} \!:\! \ottnt{M} )  \to \,  \ottnt{N}  \, \curlyveedownarrow \, \ottsym{(}  \ottnt{m_{{\mathrm{1}}}}  \ottsym{=}  \ottnt{n_{{\mathrm{1}}}}  \ottsym{)} $, which follows because $ \leadsto_{\mathsf{p} } $
preserves the top-level constructor of a term. To handle the
injectivity rules, we need to know that $ \curlyveedownarrow $ is injective in
the sense that $  ( \ottmv{x} \!:\! \ottnt{M_{{\mathrm{1}}}} )  \to \,  \ottnt{N_{{\mathrm{1}}}}  \, \curlyveedownarrow \,  ( \ottmv{x} \!:\! \ottnt{M_{{\mathrm{2}}}} )  \to \,  \ottnt{N_{{\mathrm{2}}}}  $ implies
$ \ottnt{M_{{\mathrm{1}}}} \, \curlyveedownarrow \, \ottnt{M_{{\mathrm{2}}}} $; again this follows by reasoning about $ \leadsto_{\mathsf{p} } $.
Finally, consider the conversion rule. The case looks like
\[
\ottdrule{
  \ottpremise{ H_{D} \vdash \ottnt{u_{{\mathrm{1}}}} : \ottnt{M_{{\mathrm{1}}}}  \ottsym{=}  \ottnt{N_{{\mathrm{1}}}}  \quad ... \quad  H_{D} \vdash \ottnt{u_{\ottmv{i}}} : \ottnt{M_{\ottmv{i}}}  \ottsym{=}  \ottnt{N_{\ottmv{i}}} }
  \ottpremise{ H_{D} \vdash \ottnt{u} : \ottsym{[}  \ottnt{M_{{\mathrm{1}}}}  \ottsym{/}  \ottmv{x_{{\mathrm{1}}}}  \ottsym{]} \, ... \, \ottsym{[}  \ottnt{M_{\ottmv{i}}}  \ottsym{/}  \ottmv{x_{\ottmv{i}}}  \ottsym{]}  \ottnt{M} }
  \ottpremise{ H_{D} \vdash \ottsym{[}  \ottnt{N_{{\mathrm{1}}}}  \ottsym{/}  \ottmv{x_{{\mathrm{1}}}}  \ottsym{]} \, ... \, \ottsym{[}  \ottnt{N_{\ottmv{i}}}  \ottsym{/}  \ottmv{x_{\ottmv{i}}}  \ottsym{]}  \ottnt{M} :  \star  }
}{ 
   H_{D} \vdash \ottnt{u} : \ottsym{[}  \ottnt{N_{{\mathrm{1}}}}  \ottsym{/}  \ottmv{x_{{\mathrm{1}}}}  \ottsym{]} \, ... \, \ottsym{[}  \ottnt{N_{\ottmv{i}}}  \ottsym{/}  \ottmv{x_{\ottmv{i}}}  \ottsym{]}  \ottnt{M} 
}{
  \ottdrulename{et\_conv}
}
\]
We have as an assumption that $ \ottsym{[}  \ottnt{N_{{\mathrm{1}}}}  \ottsym{/}  \ottmv{x_{{\mathrm{1}}}}  \ottsym{]} \, .. \, \ottsym{[}  \ottnt{N_{\ottmv{i}}}  \ottsym{/}  \ottmv{x_{\ottmv{i}}}  \ottsym{]}  \ottnt{M} \, \curlyveedownarrow \, \ottsym{(}  \ottnt{m_{{\mathrm{1}}}}  \ottsym{=}  \ottnt{n_{{\mathrm{1}}}}  \ottsym{)} $,
and the result would follow from the IH for $\ottnt{u}$ if we knew
that $ \ottsym{[}  \ottnt{M_{{\mathrm{1}}}}  \ottsym{/}  \ottmv{x_{{\mathrm{1}}}}  \ottsym{]} \, ... \, \ottsym{[}  \ottnt{M_{\ottmv{i}}}  \ottsym{/}  \ottmv{x_{\ottmv{i}}}  \ottsym{]}  \ottnt{M} \, \curlyveedownarrow \, \ottsym{(}  \ottnt{m_{{\mathrm{1}}}}  \ottsym{=}  \ottnt{n_{{\mathrm{1}}}}  \ottsym{)} $. But by the IHs for
$\ottnt{u_{\ottmv{i}}}$ we know that $ \ottnt{N_{\ottmv{i}}} \, \curlyveedownarrow \, \ottnt{M_{\ottmv{i}}} $, so this follows by substitution
and transitivity of $ \curlyveedownarrow $.

With the soundness lemma in hand, canonical forms and progress follow
straightforwardly.

\begin{lemma}[Canonical forms]
\label{lemma:canonical-forms}
Suppose $ H_{D} \vdash \ottnt{u} : \ottnt{M} $.
\begin{enumerate}
\item If $ \ottnt{M} \, \curlyveedownarrow \,  ( \ottmv{x} \!:\! \ottnt{M_{{\mathrm{1}}}} )  \to \,  \ottnt{M_{{\mathrm{2}}}}  $, then $u$ is either
  $\lambda  \ottmv{x}  \ottsym{.}  \ottnt{u_{{\mathrm{1}}}}$ or $ \mathsf{rec}\; \ottmv{f} . \ottkw{u} $.
\item If $ \ottnt{M} \, \curlyveedownarrow \,  [  \ottmv{x} \!:\! \ottnt{M_{{\mathrm{1}}}}  ]  \to \,  \ottnt{M_{{\mathrm{2}}}}  $, then $u$ is either
  $\lambda  \ottsym{[}  \ottsym{]}  \ottsym{.}  \ottnt{u_{{\mathrm{1}}}}$ or $ \mathsf{rec}\; \ottmv{f} . \ottkw{u} $.
\item If $ \ottnt{M} \, \curlyveedownarrow \, \ottmv{D} \, \overline{M_i} $ then $u$ is $\ottmv{d} \, \overline{u_i}$,
        where $ \ottkw{data} \, \ottmv{D} \, \Xi^{+} \, \ottkw{where} \, \ottsym{\{} \, \ottcomplu{\ottmv{d_{\ottmv{i}}}  \ottsym{:}  \Xi_{\ottmv{i}}  \to \; D \; \Xi^{+}}{\ottmv{i}}{{\mathrm{1}}}{..}{\ottmv{j}} \, \ottsym{\}} \in H_{D} $
        and $\ottmv{d}$ is one of the $\ottmv{d_{\ottmv{i}}}$.
\end{enumerate}
\end{lemma}

\begin{thm}[Progress]
If $ H_{D} \vdash \ottnt{m} : \ottnt{M} $, then either $\ottnt{m}$ is a value, 
$\ottnt{m}$ is $\ottkw{abort}$, or 
$\ottnt{m}  \leadsto_{\mathsf{cbv} }  \ottnt{m'}$ for some $\ottnt{m'}$.
\end{thm}

\section{Related Work}

\paragraph{Dependent types with nontermination} 
While there are many examples of languages that combine nontermination
with dependent or indexed types, most take care to ensure that
nonterminating expressions can not occur inside types. They do this
either by making the type language completely separate from the
expression language (e.g. DML~\cite{XP98}, ATS~\cite{xi:ats}, $\Omega$mega~\cite{sheard}, Haskell with GADTs~\cite{pj-vytiniotis:wobbly}), or by
restricting dependent application to values or ``pure'' expressions
(e.g. DML~\cite{licata:dml}, F*~\cite{swamy-fstar}, Aura~\cite{jia+08}, and \cite{ou:dependent}). 

In our language, types and expressions are unified and
types can even be computed by general recursive functions. In this
area of the design space, the most comparable languages are
Cayenne~\cite{augustsson98}, Cardelli's Type:Type
language~\cite{Cardelli86apolymorphic}, and
$\Pi\Sigma$~\cite{alti:pisigma-new}. However, none of them have the
particular combination of features that we discuss in this paper,
i.e. irrelevance, CBV, and a built-in propositional equality.

$\lambda^{\cong}$\cite{jia:lambdaeek} is a CBV dependently typed
language with nontermination, which used CBV-respecting parallel
reduction as one possible definitional equivalence.  It proposed an
application rule which is more expressive than just value-dependency,
but not as simple as the one in this paper. $\lambda^{\cong}$ is not
as expressive as our language (no polymorphism, propositional
equality, or Type-in-Type), and has no notion of irrelevance. 

\paragraph{Irrelevance} We already mentioned ICC*~\cite{barras+08},
EPTS~\cite{mishra-linger+08}, and Abel's system~\cite{abel:fossacs11}.
One of the key differences between the systems is whether the variable
$\ottmv{x}$ in an irrelevant arrow type $ [  \ottmv{x} \!:\! \ottnt{A}  ]  \to \,  \ottnt{B} $ is allowed to
occur freely in $\ottnt{B}$ (``Miquel\cite{Miquel01theimplicit}-style
irrelevance'', our choice) or only in irrelevant positions in $\ottnt{B}$
(``Pfenning\cite{pfenning01}-style'', see also~\cite{reed:irrelevance}).  
Agda implements the latter because it interacts better with type-directed equality~\cite{abel:fossacs11},
whereas our equality is not type-directed.

\paragraph{Equality}

The usual equality type in Coq and Agda's standard libraries is
homogenous and has a computationally relevant conversion rule. These
languages also
provide the heterogenous JMeq~\cite{mcbride:motive}, which we
discussed above.

Extensional Type Theory, e.g. Nuprl~\cite{nuprl}, is similar to our
language in that conversion is computationally irrelevant and completely
erased.  ETT terms are similar to our unannotated terms, while our
annotated terms correspond to ETT typing derivations. 
\vs{Say something about ``undecidable type checking''?}
On the other hand, the equational theory of ETT is
different from our language, e.g. it can prove extensionality while our
equality cannot.

Observational Type Theory~\cite{alti:ott-conf} also proves
$\ottsym{(}  \ottkw{conv} \, \ottnt{a} \, \ottkw{at} \, \mathsf{\sim}  \ottnt{b}  \ottsym{)}  \ottsym{=}  \ottnt{a}$, but in a more sophisticated way than by
erasing the conversion. Instead it provides a set of axioms and
ensures that those axioms can never block reduction. It is inherently
type-directed, which means that it validates extensionality but cannot
make use of equations between expressions of genuinely different types.

Guru~\cite{guru09}, like our language, can eliminate equalities where
the two sides have different types, and equalities are
proved by joinability without any type-directed rules. However,
unlike our language the equality formation rule does not require that
the equated expressions are even well-typed. This can be annoying in
practice, because simple programmer errors are not caught by the type
system. Guru does not have our $n$-ary conv rule. 

GHC Core~\cite{sulzmann-fc,vytiniotis-practical11} is similar to our
language in not having a separate notion of definitional and
propositional equality. Instead, all type equivalences---which are
implicit in Haskell source---must be justified by the typechecker
by explicit proof terms. As in our language the presence of
nontermination means that proof terms must by evaluated at runtime, but
there is no notion of irrelevance.

\section{Conclusions and Future Work}

In this paper, we combined {\bf computational
irrelevance} and {\bf nontermination} in a dependently typed
programming language. 

In defining the language, we made concrete choices about
evaluation order and treatment of conversion. Our evaluation order is
CBV, and this is reflected in the equations that the language can prove
(including an inherently CBV rule for error expressions). An effectful
language needs a restriction on the application rule, and we propose a
particularly simple yet expressive one. 

Our conversion rule has a novel combination of features: the equality
proof is computationally irrelevant, conversion can use equalities
where the two sides have different types, and conversion can use
multiple equalities at once. These features are all aimed at making
reasoning about programs easier.

We then proposed typing rules for irrelevant function and constructor
arguments. We gave examples showing that in contrast to previous work
in pure languages, irrelevant application must be restricted, and
described a value-restricted version.

In future work, we plan to integrate this design with
the larger Trellys project.  The Trellys language will be divided into
two fragments: a ``programmatic'' fragment that will resemble the
language presented here, and a ``logical'' fragment that will be
restricted to ensure consistency.  While designing an expressive and
consistent logical fragment will involve substantial additional
challenges, the present work has provided a solid foundation by
identifying and solving many problems that arise from Trellys'
previously unseen combination of features.

\section*{Acknowledgments}
This work was supported by the National Science Foundation (NSF grant
0910500).

\bibliographystyle{eptcs}
\bibliography{refs}

\label{this-is-the-end-of-the-paper}

\newpage
\appendix 

\section{Full Language Specification}
\label{sec:full-specification}

\subsection{Syntax}
\ottgrammartabular{
\otttele\ottinterrule
\ottteleplus\ottinterrule
\ottenv\ottinterrule
\ottdecl\ottinterrule
\ottexp\ottinterrule
\ottP\ottinterrule
\ottval\ottinterrule
\ottexplist\ottinterrule
\ottvallist\ottinterrule
\ottetele\ottinterrule
\otteteleplus\ottinterrule
\otteenv\ottinterrule
\ottedecl\ottinterrule
\otteenvD\ottinterrule
\ottedeclD\ottinterrule
\otteexp\ottinterrule
\otteval\ottinterrule
\otteexplist\ottinterrule
\ottevallist\ottinterrule
\ottevalctx\ottinterrule
}

\subsection{Erasure function}
The erasure function $\ottsym{\mbox{$\mid$}}  \ottnt{a}  \ottsym{\mbox{$\mid$}}$ is defined by:
\[
\begin{array}{ll}
\ottsym{\mbox{$\mid$}}   \star   \ottsym{\mbox{$\mid$}} & =   \star  \\
\ottsym{\mbox{$\mid$}}  \ottmv{x}  \ottsym{\mbox{$\mid$}} & = \ottmv{x} \\
\ottsym{\mbox{$\mid$}}   \ottmv{D}  \;  \overline{A_i}   \ottsym{\mbox{$\mid$}} & 
    = \ottmv{D} \, \ottsym{\mbox{$\mid$}}  \overline{A_i}  \ottsym{\mbox{$\mid$}}  \\
\ottsym{\mbox{$\mid$}}   \ottmv{d}  \; [  \overline{A_i}  ] \;  \overline{a_i}   \ottsym{\mbox{$\mid$}} & 
    = \ottmv{d} \, \ottsym{\mbox{$\mid$}}  \overline{a_i}  \ottsym{\mbox{$\mid$}} \\
\ottsym{\mbox{$\mid$}}   \mathsf{rec}\; \ottmv{f} : \ottnt{A} . \ottnt{v}   \ottsym{\mbox{$\mid$}} & =  \mathsf{rec}\; \ottmv{f} . \ottkw{u}  \\
\ottsym{\mbox{$\mid$}}  \lambda  \ottmv{x}  \ottsym{:}  \ottnt{A}  \ottsym{.}  \ottnt{a}  \ottsym{\mbox{$\mid$}} & = \lambda  \ottmv{x}  \ottsym{.}  \ottsym{\mbox{$\mid$}}  \ottnt{a}  \ottsym{\mbox{$\mid$}} \\
\ottsym{\mbox{$\mid$}}  \lambda  \ottsym{[}  \ottmv{x}  \ottsym{:}  \ottnt{A}  \ottsym{]}  \ottsym{.}  \ottnt{a}  \ottsym{\mbox{$\mid$}} & = \lambda  \ottsym{[}  \ottsym{]}  \ottsym{.}  \ottsym{\mbox{$\mid$}}  \ottnt{a}  \ottsym{\mbox{$\mid$}} \\
\ottsym{\mbox{$\mid$}}   \ottnt{a}  \;  \ottnt{b}   \ottsym{\mbox{$\mid$}} & = \ottsym{\mbox{$\mid$}}  \ottnt{a}  \ottsym{\mbox{$\mid$}}\ \ottsym{\mbox{$\mid$}}  \ottnt{b}  \ottsym{\mbox{$\mid$}} \\
\ottsym{\mbox{$\mid$}}   \ottnt{a}  \; [  \ottnt{b}  ]   \ottsym{\mbox{$\mid$}} & = \ottsym{\mbox{$\mid$}}  \ottnt{a}  \ottsym{\mbox{$\mid$}}  \ottsym{[}  \ottsym{]} \\
\ottsym{\mbox{$\mid$}}   ( \ottmv{x} \!:\! \ottnt{A} )  \to \,  \ottnt{B}   \ottsym{\mbox{$\mid$}} & =  ( \ottmv{x} \!:\! \ottsym{\mbox{$\mid$}}  \ottnt{A}  \ottsym{\mbox{$\mid$}} )  \to \,  \ottsym{\mbox{$\mid$}}  \ottnt{B}  \ottsym{\mbox{$\mid$}}  \\
\ottsym{\mbox{$\mid$}}   [  \ottmv{x} \!:\! \ottnt{A}  ]  \to \,  \ottnt{B}   \ottsym{\mbox{$\mid$}} & =  [  \ottmv{x} \!:\! \ottsym{\mbox{$\mid$}}  \ottnt{A}  \ottsym{\mbox{$\mid$}}  ]  \to \,  \ottsym{\mbox{$\mid$}}  \ottnt{B}  \ottsym{\mbox{$\mid$}}  \\
\ottsym{\mbox{$\mid$}}  \ottnt{a}  \ottsym{=}  \ottnt{b}  \ottsym{\mbox{$\mid$}} & = \ottsym{\mbox{$\mid$}}  \ottnt{a}  \ottsym{\mbox{$\mid$}}  \ottsym{=}  \ottsym{\mbox{$\mid$}}  \ottnt{b}  \ottsym{\mbox{$\mid$}} \\
\ottsym{\mbox{$\mid$}}   \mathsf{join}_{ \ottnt{a} = \ottnt{b} } \;  \ottmv{i}  \;  \ottmv{j}   \ottsym{\mbox{$\mid$}} & =  \ottkw{join}  \\
\ottsym{\mbox{$\mid$}}  \ottkw{injdom} \, \ottnt{a}  \ottsym{\mbox{$\mid$}} &=  \ottkw{join}  \\
\ottsym{\mbox{$\mid$}}  \ottkw{injrng} \, \ottnt{a} \, \ottnt{b}  \ottsym{\mbox{$\mid$}} &=  \ottkw{join}  \\
\ottsym{\mbox{$\mid$}}   \mathsf{injtcon}_{ \ottmv{i} } \;  \ottnt{a}   \ottsym{\mbox{$\mid$}} &=  \ottkw{join}  \\
\ottsym{\mbox{$\mid$}}  \ottkw{case} \, \ottnt{a} \, \ottkw{as} \, \ottsym{[y]} \, \ottkw{of} \, \ottsym{\{} \, \ottcomplu{\ottmv{d_{\ottmv{j}}} \, \Delta_{\ottmv{j}}  \Rightarrow  \ottnt{b_{\ottmv{j}}}}{\ottmv{j}}{{\mathrm{1}}}{..}{\ottmv{k}} \, \ottsym{\}}  \ottsym{\mbox{$\mid$}} & 
  =
  \ottkw{case} \, \ottsym{\mbox{$\mid$}}  \ottnt{a}  \ottsym{\mbox{$\mid$}} \, \ottkw{of} \, \ottsym{\{} \, \ottcomplu{\ottmv{d_{\ottmv{j}}} \, \overline{x_i}_{\ottmv{j}}  \Rightarrow  \ottsym{\mbox{$\mid$}}  \ottnt{b_{\ottmv{j}}}  \ottsym{\mbox{$\mid$}}}{\ottmv{j}}{{\mathrm{1}}}{..}{\ottmv{k}} \, \ottsym{\}} \\
&  \qquad\text{where $\overline{x_i}_{\ottmv{j}}$ are the relevant variables of  $\Delta_{\ottmv{j}}$} \\
\ottsym{\mbox{$\mid$}}  \ottkw{conv} \, \ottnt{a} \, \ottkw{at} \, \ottsym{[}  \mathsf{\sim}  \ottnt{P_{{\mathrm{1}}}}  \ottsym{/}  \ottmv{x_{{\mathrm{1}}}}  \ottsym{]} \, ... \, \ottsym{[}  \mathsf{\sim}  \ottnt{P_{\ottmv{i}}}  \ottsym{/}  \ottmv{x_{\ottmv{i}}}  \ottsym{]}  \ottnt{A}  \ottsym{\mbox{$\mid$}} & = \ottsym{\mbox{$\mid$}}  \ottnt{a}  \ottsym{\mbox{$\mid$}} \\
\ottsym{\mbox{$\mid$}}   \mathsf{abort}_{ \ottnt{A} }   \ottsym{\mbox{$\mid$}} & = \ottkw{abort}\\
& \\
| \cdot |      & = \cdot \\
\ottsym{\mbox{$\mid$}}   \ottnt{a}  \;  \overline{a_i}   \ottsym{\mbox{$\mid$}} &=  \ottsym{\mbox{$\mid$}}  \ottnt{a}  \ottsym{\mbox{$\mid$}}  \;  \ottsym{\mbox{$\mid$}}  \overline{a_i}  \ottsym{\mbox{$\mid$}}  \\
\ottsym{\mbox{$\mid$}}   [  \ottnt{a}  ] \;  \overline{a_i}   \ottsym{\mbox{$\mid$}} &=  [] \;  \ottsym{\mbox{$\mid$}}  \overline{a_i}  \ottsym{\mbox{$\mid$}}  \\
\end{array}
\]

\subsection{CBV evaluation}
\ottdefnsJStepCBV

\subsection{Parallel reduction}
\label{sec:parallel-reduction}
\ottdefnsJStepP
\ottdefnsJJoin

\subsection{Annotated type system}
\label{sec:annotated-type-system}
\ottdefnsJtyp

\subsection{Unannotated type system}
\label{sec:unannotated-type-system}
\ottdefnsJetyp

\section{Proofs}

\subsection{Correctness of annotated system}
\begin{lemma}[Decidability of type checking]
There is an algorithm which given $\Gamma$ and $\ottnt{a}$ computes the
unique $\ottnt{A}$ such that $ \Gamma \vdash \ottnt{a} : \ottnt{A} $, or reports that there is no
such $\ottnt{A}$.
\end{lemma}
\begin{proof}
The algorithm follows the structure of $\ottnt{a}$---for each
syntactic form we see that only one typing rule could apply, and
that the premises of that rule are uniquely determined.
\end{proof}

\begin{lemma}[Correctness of erasure]
If $ \Gamma \vdash \ottnt{a} : \ottnt{A} $, then $ \ottsym{\mbox{$\mid$}}  \Gamma  \ottsym{\mbox{$\mid$}} \vdash \ottsym{\mbox{$\mid$}}  \ottnt{a}  \ottsym{\mbox{$\mid$}} : \ottsym{\mbox{$\mid$}}  \ottnt{A}  \ottsym{\mbox{$\mid$}} $.
\end{lemma}
\begin{proof}
Easy induction---each annotated rule corresponds directly to an
unannotated rule where all terms have gone through erasure.
\end{proof}

\subsection{Facts about parallel reduction}

\begin{definition}
The {\it head constructor} of an expression is defined as follows:
  \begin{itemize}
  \item The head constructor of $ \star $ is $ \star $.
  \item The head constructor of $ \mathsf{Nat} $ is  $ \mathsf{Nat} $.
  \item The head constructor of $ ( \ottmv{x} \!:\! \ottnt{M} )  \to \,  \ottnt{N} $ is
    $\rightarrow$.
  \item The head constructor of $ [  \ottmv{x} \!:\! \ottnt{M}  ]  \to \,  \ottnt{N} $ is
    $[]\rightarrow$.
  \item The head constructor of $\ottmv{D} \, \overline{M_i}$ is $D$.
  \item The head constructor of $\ottmv{d} \, \overline{m_i}$ is $d$.
  \item The head constructor of $\ottnt{a}  \ottsym{=}  \ottnt{b}$ is $=$.
  \item Other expressions do not have a head constructor.
  \end{itemize}
We write $\ottkw{hd} \, \ottsym{(}  \ottnt{M}  \ottsym{)}$ for the partial function mapping $\ottnt{M}$ to
its head constructor.
\end{definition}

\begin{lemma}
\label{lem:hd-step}
If $\ottnt{m}  \leadsto_{\mathsf{p} }  \ottnt{m'}$ and $\ottkw{hd} \, \ottsym{(}  \ottnt{m}  \ottsym{)}$ is defined, then $\ottkw{hd} \, \ottsym{(}  \ottnt{m}  \ottsym{)}
 = \ottkw{hd} \, \ottsym{(}  \ottnt{m'}  \ottsym{)}$.
\end{lemma}
\begin{proof}
By inspecting the definition of $ \leadsto_{\mathsf{p} } $ we see that it always 
preserves the head constructor of a term.
\end{proof}

\begin{lemma}
\label{lem:hd-join}
If $ \ottnt{m} \, \curlyveedownarrow \, \ottnt{m'} $, then $\ottnt{m}$ and $\ottnt{m'}$ do not have 
different head constructors.
\end{lemma}
\begin{proof}
Expanding the definition of $ \curlyveedownarrow $ we know that $\ottnt{m}  \leadsto^{*}_{\mathsf{p} }  \ottnt{n}$
and $\ottnt{m'}  \leadsto^{*}_{\mathsf{p} }  \ottnt{n}$ for some $\ottnt{n}$. If $\ottnt{m}$ and $\ottnt{m'}$
had (defined and) different head constructors, then by repeatedly 
applying Lemma~\ref{lem:hd-step} we would get that $\ottnt{n}$ had two
different head constructors, which is impossible.
\end{proof}

\begin{lemma}[ Injectivity of $ \curlyveedownarrow $]
\label{lem:join-inj}
\hspace*{\fill} \\[-12pt]
\begin{itemize}
\item If $ \ottnt{m_{{\mathrm{1}}}}  \ottsym{=}  \ottnt{n_{{\mathrm{1}}}} \, \curlyveedownarrow \, \ottnt{m_{{\mathrm{2}}}}  \ottsym{=}  \ottnt{n_{{\mathrm{2}}}} $, then $ \ottnt{m_{{\mathrm{1}}}} \, \curlyveedownarrow \, \ottnt{m_{{\mathrm{2}}}} $ and 
$ \ottnt{n_{{\mathrm{1}}}} \, \curlyveedownarrow \, \ottnt{n_{{\mathrm{2}}}} $.
\item If $ \ottmv{D} \, \overline{M_i}_{{\mathrm{1}}} \, \curlyveedownarrow \, \ottmv{D} \, \overline{M_i}_{{\mathrm{2}}} $, then $ \ottnt{M_{\ottmv{i}\,{\mathrm{1}}}} \, \curlyveedownarrow \, \ottnt{M_{\ottmv{i}\,{\mathrm{2}}}} $. 
\item If $  ( \ottmv{x} \!:\! \ottnt{M_{{\mathrm{1}}}} )  \to \,  \ottnt{N_{{\mathrm{1}}}}  \, \curlyveedownarrow \,  ( \ottmv{x} \!:\! \ottnt{M_{{\mathrm{2}}}} )  \to \,  \ottnt{N_{{\mathrm{2}}}}  $ then $ \ottnt{M_{{\mathrm{1}}}} \, \curlyveedownarrow \, \ottnt{M_{{\mathrm{2}}}} $ and
  $ \ottnt{N_{{\mathrm{1}}}} \, \curlyveedownarrow \, \ottnt{N_{{\mathrm{2}}}} $.
\item If $  [  \ottmv{x} \!:\! \ottnt{M_{{\mathrm{1}}}}  ]  \to \,  \ottnt{N_{{\mathrm{1}}}}  \, \curlyveedownarrow \,  [  \ottmv{x} \!:\! \ottnt{M_{{\mathrm{2}}}}  ]  \to \,  \ottnt{N_{{\mathrm{2}}}}  $ then $ \ottnt{M_{{\mathrm{1}}}} \, \curlyveedownarrow \, \ottnt{M_{{\mathrm{2}}}} $ and
  $ \ottnt{N_{{\mathrm{1}}}} \, \curlyveedownarrow \, \ottnt{N_{{\mathrm{2}}}} $.
\end{itemize}
\end{lemma}
\begin{proof}
The lemma is proven in the same way for all the different types of
expressions, so we only show the proof for (1). Expanding the
definition of $ \curlyveedownarrow $, we have that $\ottnt{m_{{\mathrm{1}}}}  \ottsym{=}  \ottnt{n_{{\mathrm{1}}}}  \leadsto^{*}_{\mathsf{p} }  \ottnt{N}$ and
$\ottnt{m_{{\mathrm{2}}}}  \ottsym{=}  \ottnt{n_{{\mathrm{2}}}}  \leadsto_{\mathsf{p} }  \ottnt{N}$ for some $\ottnt{N}$. 

By lemma \ref{lem:hd-join} we know that $\ottnt{N}$ has the shape
$\ottnt{n}  \ottsym{=}  \ottnt{m}$. So it suffices to prove that, for any $\ottnt{n_{{\mathrm{1}}}},
\ottnt{m_{{\mathrm{1}}}}$, if $\ottnt{n_{{\mathrm{1}}}}  \ottsym{=}  \ottnt{m_{{\mathrm{1}}}}  \leadsto^{*}_{\mathsf{p} }  \ottnt{n}  \ottsym{=}  \ottnt{m}$, then $\ottnt{n_{{\mathrm{1}}}}  \leadsto^{*}_{\mathsf{p} }  \ottnt{n}$.
This follows by an easy induction on the chain of reduction, since at
each step the only reduction rule that can apply is congruence.
\end{proof}

\begin{lemma}
\label{lem:stepstep-subst}
If $\ottnt{u}  \leadsto_{\mathsf{p} }  \ottnt{u'}$ and $\ottnt{m}  \leadsto_{\mathsf{p} }  \ottnt{m'}$, then 
$\ottsym{[}  \ottnt{u}  \ottsym{/}  \ottmv{x}  \ottsym{]}  \ottnt{m}  \leadsto_{\mathsf{p} }  \ottsym{[}  \ottnt{u'}  \ottsym{/}  \ottmv{x}  \ottsym{]}  \ottnt{m'}$.
\end{lemma}
\begin{proof}By induction on $\ottnt{m}  \leadsto_{\mathsf{p} }  \ottnt{m'}$ \vs{Perhaps add some more details here.}
\end{proof}

\begin{lemma}
\label{lem:join-valsubst}
If $ \ottnt{M} \, \curlyveedownarrow \, \ottnt{M'} $, then $ \ottsym{[}  \ottnt{u}  \ottsym{/}  \ottmv{x}  \ottsym{]}  \ottnt{M} \, \curlyveedownarrow \, \ottsym{[}  \ottnt{u}  \ottsym{/}  \ottmv{x}  \ottsym{]}  \ottnt{M'} $.
\end{lemma}
\begin{proof}
Expanding the definition of $ \curlyveedownarrow $ we get $\ottnt{M}  \leadsto^{*}_{\mathsf{p} }  \ottnt{M_{{\mathrm{1}}}}$ and 
$\ottnt{M'}  \leadsto_{\mathsf{p} }  \ottnt{M_{{\mathrm{1}}}}$ for some $\ottnt{M_{{\mathrm{1}}}}$. Repeatedly applying
Lemma~\ref{lem:stepstep-subst} we then get $\ottsym{[}  \ottnt{u}  \ottsym{/}  \ottmv{x}  \ottsym{]}  \ottnt{M}  \leadsto^{*}_{\mathsf{p} }  \ottsym{[}  \ottnt{u}  \ottsym{/}  \ottmv{x}  \ottsym{]}  \ottnt{M_{{\mathrm{1}}}}$
and $\ottsym{[}  \ottnt{u}  \ottsym{/}  \ottmv{x}  \ottsym{]}  \ottnt{M'}  \leadsto^{*}_{\mathsf{p} }  \ottsym{[}  \ottnt{u}  \ottsym{/}  \ottmv{x}  \ottsym{]}  \ottnt{M_{{\mathrm{1}}}}$ as required.
\end{proof}

\begin{lemma}[One-step diamond property for $ \leadsto_{\mathsf{p} } $]
\label{lem:step-diamond}
If $\ottnt{m}  \leadsto_{\mathsf{p} }  \ottnt{m_{{\mathrm{1}}}}$ and $\ottnt{m}  \leadsto_{\mathsf{p} }  \ottnt{m_{{\mathrm{2}}}}$, then there exists 
some $\ottnt{m'}$ such that $\ottnt{m_{{\mathrm{1}}}}  \leadsto_{\mathsf{p} }  \ottnt{m'}$ and $\ottnt{m_{{\mathrm{2}}}}  \leadsto_{\mathsf{p} }  \ottnt{m'}$.
\end{lemma}
\begin{proof}
By induction on the structure of $\ottnt{m}$. We only show the case when
$\ottnt{m}$ is an application $ \ottnt{m_{{\mathrm{1}}}}  \;  \ottnt{m_{{\mathrm{2}}}} $, as this case contains all
the ideas of the proof.
\begin{description}
\item[Case $\ottnt{m}$ is $ \ottnt{m_{{\mathrm{1}}}}  \;  \ottnt{m_{{\mathrm{2}}}} $] We consider all possible pairs of
  ways that $ \ottnt{m_{{\mathrm{1}}}}  \;  \ottnt{m_{{\mathrm{2}}}} $ can reduce.
  \begin{itemize}
  \item One reduction is \ottdrulename{sc\_refl}. This case is
    trivial.
  \item Both reductions are \ottdrulename{sc\_app}. That is to say,
    $ \ottnt{m_{{\mathrm{1}}}}  \;  \ottnt{m_{{\mathrm{2}}}}   \leadsto_{\mathsf{p} }   \ottnt{m_{{\mathrm{11}}}}  \;  \ottnt{m_{{\mathrm{21}}}} $ and $ \ottnt{m_{{\mathrm{1}}}}  \;  \ottnt{m_{{\mathrm{2}}}}   \leadsto_{\mathsf{p} }   \ottnt{m_{{\mathrm{12}}}}  \;  \ottnt{m_{{\mathrm{22}}}} $, where
    $\ottnt{m_{{\mathrm{1}}}}  \leadsto_{\mathsf{p} }  \ottnt{m_{{\mathrm{11}}}}$, $\ottnt{m_{{\mathrm{1}}}}  \leadsto_{\mathsf{p} }  \ottnt{m_{{\mathrm{21}}}}$, $\ottnt{m_{{\mathrm{2}}}}  \leadsto_{\mathsf{p} }  \ottnt{m_{{\mathrm{21}}}}$ and
    $\ottnt{m_{{\mathrm{2}}}}  \leadsto_{\mathsf{p} }  \ottnt{m_{{\mathrm{22}}}}$. 

    By the induction hypothesis for $\ottnt{m_{{\mathrm{1}}}}$, there exists $\ottnt{m'_{{\mathrm{1}}}}$,
    such that $\ottnt{m_{{\mathrm{11}}}}  \leadsto_{\mathsf{p} }  \ottnt{m'_{{\mathrm{1}}}}$ and $\ottnt{m_{{\mathrm{21}}}}  \leadsto_{\mathsf{p} }  \ottnt{m'_{{\mathrm{2}}}}$. Similarly
    for $\ottnt{m_{{\mathrm{2}}}}$. So by \ottdrulename{sc\_app} we have
    $ \ottnt{m_{{\mathrm{11}}}}  \;  \ottnt{m_{{\mathrm{21}}}}   \leadsto_{\mathsf{p} }   \ottnt{m'_{{\mathrm{1}}}}  \;  \ottnt{m'_{{\mathrm{2}}}} $ and $ \ottnt{m_{{\mathrm{12}}}}  \;  \ottnt{m_{{\mathrm{22}}}}   \leadsto_{\mathsf{p} }   \ottnt{m'_{{\mathrm{1}}}}  \;  \ottnt{m'_{{\mathrm{2}}}} $ as
    required.
  \item One reduction is \ottdrulename{sc\_appbeta}. So it must be the
    case that $ \ottnt{m_{{\mathrm{1}}}}  \;  \ottnt{m_{{\mathrm{2}}}} $ is $ \ottsym{(}  \lambda  \ottmv{x}  \ottsym{.}  \ottnt{m_{{\mathrm{0}}}}  \ottsym{)}  \;  \ottnt{u} $. By considering cases, we
    see that only only possibilities for the other reduction is
    \ottdrulename{sc\_appbeta} and \ottdrulename{sc\_app}.

    In the case when the other reduction is \ottdrulename{sc\_app}, we
    see that the only way that $\lambda  \ottmv{x}  \ottsym{.}  \ottnt{m_{{\mathrm{0}}}}$ can step is by congruence
    when $\ottnt{m_{{\mathrm{0}}}}  \leadsto_{\mathsf{p} }  \ottnt{m_{{\mathrm{02}}}}$. So we have:
    \begin{align*}
     \ottsym{(}  \lambda  \ottmv{x}  \ottsym{.}  \ottnt{m_{{\mathrm{0}}}}  \ottsym{)}  \;  \ottnt{u}   \leadsto_{\mathsf{p} }  \ottsym{[}  \ottnt{u_{{\mathrm{1}}}}  \ottsym{/}  \ottmv{x}  \ottsym{]}  \ottnt{m_{{\mathrm{01}}}} & \qquad\text{where
      $\ottnt{m_{{\mathrm{0}}}}  \leadsto_{\mathsf{p} }  \ottnt{m_{{\mathrm{01}}}}$ and $\ottnt{u}  \leadsto_{\mathsf{p} }  \ottnt{u_{{\mathrm{1}}}}$.} \\
     \ottsym{(}  \lambda  \ottmv{x}  \ottsym{.}  \ottnt{m_{{\mathrm{0}}}}  \ottsym{)}  \;  \ottnt{u}   \leadsto_{\mathsf{p} }   \ottsym{(}  \lambda  \ottmv{x}  \ottsym{.}  \ottnt{m_{{\mathrm{02}}}}  \ottsym{)}  \;  \ottnt{u_{{\mathrm{2}}}}  & \qquad\text{where
      $\ottnt{m_{{\mathrm{0}}}}  \leadsto_{\mathsf{p} }  \ottnt{m_{{\mathrm{02}}}}$ and $\ottnt{u}  \leadsto_{\mathsf{p} }  \ottnt{u_{{\mathrm{2}}}}$.}\\
    \end{align*}
    Now by IH we get $\ottnt{m'_{{\mathrm{0}}}}$ and $\ottnt{u'}$. By substitution
    (lemma~\ref{lem:stepstep-subst}) we get
    $\ottsym{[}  \ottnt{u_{{\mathrm{1}}}}  \ottsym{/}  \ottmv{x}  \ottsym{]}  \ottnt{m_{{\mathrm{01}}}}  \leadsto_{\mathsf{p} }  \ottsym{[}  \ottnt{u'}  \ottsym{/}  \ottmv{x}  \ottsym{]}  \ottnt{m'_{{\mathrm{0}}}}$, while by
    \ottdrulename{sc\_appbeta} we get
    $ \ottsym{(}  \lambda  \ottmv{x}  \ottsym{.}  \ottnt{m_{{\mathrm{02}}}}  \ottsym{)}  \;  \ottnt{u_{{\mathrm{2}}}}   \leadsto_{\mathsf{p} }  \ottsym{[}  \ottnt{u'}  \ottsym{/}  \ottmv{x}  \ottsym{]}  \ottnt{m'_{{\mathrm{0}}}}$. So the terms are joinable as
    required.

    On the other hand, if both the reductions are by
    \ottdrulename{sc\_apprec}, then we have
    \begin{align*}
     \ottsym{(}  \lambda  \ottmv{x}  \ottsym{.}  \ottnt{m_{{\mathrm{0}}}}  \ottsym{)}  \;  \ottnt{u}   \leadsto_{\mathsf{p} }  \ottsym{[}  \ottnt{u_{{\mathrm{1}}}}  \ottsym{/}  \ottmv{x}  \ottsym{]}  \ottnt{m_{{\mathrm{01}}}} & \qquad\text{where
      $\ottnt{m_{{\mathrm{0}}}}  \leadsto_{\mathsf{p} }  \ottnt{m_{{\mathrm{01}}}}$ and $\ottnt{u}  \leadsto_{\mathsf{p} }  \ottnt{u_{{\mathrm{1}}}}$.} \\
      \ottsym{(}  \lambda  \ottmv{x}  \ottsym{.}  \ottnt{m_{{\mathrm{0}}}}  \ottsym{)}  \;  \ottnt{u}   \leadsto_{\mathsf{p} }  \ottsym{[}  \ottnt{u_{{\mathrm{2}}}}  \ottsym{/}  \ottmv{x}  \ottsym{]}  \ottnt{m_{{\mathrm{02}}}} & \qquad\text{where
       $\ottnt{m_{{\mathrm{0}}}}  \leadsto_{\mathsf{p} }  \ottnt{m_{{\mathrm{02}}}}$ and $\ottnt{u}  \leadsto_{\mathsf{p} }  \ottnt{u_{{\mathrm{2}}}}$.} \\
    \end{align*}
    Then by IH we again get $\ottnt{m'_{{\mathrm{0}}}}$ and $\ottnt{u'}$, and by
    substitution (twice), the two terms are again
    joinable at $\ottsym{[}  \ottnt{u'}  \ottsym{/}  \ottmv{x}  \ottsym{]}  \ottnt{m'_{{\mathrm{0}}}}$.

   \item One reduction is \ottdrulename{sc\_apprec}. So $ \ottnt{m_{{\mathrm{1}}}}  \;  \ottnt{m_{{\mathrm{2}}}} $
     must be $ \ottsym{(}   \mathsf{rec}\; \ottmv{f} . \ottkw{u}   \ottsym{)}  \;  \ottnt{u_{{\mathrm{2}}}} $. By considering cases, we see that
     the other reduction must be either \ottdrulename{sc\_apprec} or \ottdrulename{sc\_app}.

     If the other rule is \ottdrulename{sc\_app} we note that the only
     way $ \mathsf{rec}\; \ottmv{f} . \ottkw{u} $ can step is by congruence to $ \mathsf{rec}\; \ottmv{f} . \ottkw{u}   \leadsto_{\mathsf{p} }   \mathsf{rec}\; \ottmv{f} . \ottkw{u} $,
     so we have 
     \begin{align*}
       \ottsym{(}   \mathsf{rec}\; \ottmv{f} . \ottkw{u}   \ottsym{)}  \;  \ottnt{u_{{\mathrm{2}}}}   \leadsto_{\mathsf{p} }   \ottsym{(}  \ottsym{[}   \mathsf{rec}\; \ottmv{f} . \ottkw{u}   \ottsym{/}  \ottmv{f}  \ottsym{]}  \ottnt{u_{{\mathrm{11}}}}  \ottsym{)}  \;  \ottnt{u_{{\mathrm{21}}}}  &\qquad\text{where
        $\ottnt{u_{{\mathrm{1}}}}  \leadsto_{\mathsf{p} }  \ottnt{u_{{\mathrm{11}}}}$ and $\ottnt{u_{{\mathrm{2}}}}  \leadsto_{\mathsf{p} }  \ottnt{u_{{\mathrm{21}}}}$} \\
       \ottsym{(}   \mathsf{rec}\; \ottmv{f} . \ottkw{u}   \ottsym{)}  \;  \ottnt{u_{{\mathrm{2}}}}   \leadsto_{\mathsf{p} }   \ottsym{(}   \mathsf{rec}\; \ottmv{f} . \ottkw{u}   \ottsym{)}  \;  \ottnt{u_{{\mathrm{22}}}}  & \qquad\text{where
        $\ottnt{u_{{\mathrm{1}}}}  \leadsto_{\mathsf{p} }  \ottnt{u_{{\mathrm{12}}}}$ and $\ottnt{u_{{\mathrm{2}}}}  \leadsto_{\mathsf{p} }  \ottnt{u_{{\mathrm{22}}}}$} \\
     \end{align*}
     
      Now, by IH we have $\ottnt{u'_{{\mathrm{1}}}}$ and $\ottnt{u'_{{\mathrm{2}}}}$. By congruence,
      $ \mathsf{rec}\; \ottmv{f} . \ottkw{u}   \leadsto_{\mathsf{p} }   \mathsf{rec}\; \ottmv{f} . \ottkw{u} $, so by substitution
      (lemma~\ref{lem:stepstep-subst}) we get
      $\ottsym{[}   \mathsf{rec}\; \ottmv{f} . \ottkw{u}   \ottsym{/}  \ottmv{f}  \ottsym{]}  \ottnt{u_{{\mathrm{11}}}}  \leadsto_{\mathsf{p} }  \ottsym{[}   \mathsf{rec}\; \ottmv{f} . \ottkw{u}   \ottsym{/}  \ottmv{f}  \ottsym{]}  \ottnt{u'_{{\mathrm{1}}}}$, and then by
      congruence
      $ \ottsym{(}  \ottsym{[}   \mathsf{rec}\; \ottmv{f} . \ottkw{u}   \ottsym{/}  \ottmv{f}  \ottsym{]}  \ottnt{u_{{\mathrm{11}}}}  \ottsym{)}  \;  \ottnt{u_{{\mathrm{21}}}}   \leadsto_{\mathsf{p} }   \ottsym{(}  \ottsym{[}   \mathsf{rec}\; \ottmv{f} . \ottkw{u}   \ottsym{/}  \ottmv{f}  \ottsym{]}  \ottnt{u'_{{\mathrm{1}}}}  \ottsym{)}  \;  \ottnt{u'_{{\mathrm{2}}}} $. Meanwhile,
      by \ottdrulename{sc\_apprec} we have
      $ \ottsym{(}   \mathsf{rec}\; \ottmv{f} . \ottkw{u}   \ottsym{)}  \;  \ottnt{u_{{\mathrm{22}}}}   \leadsto_{\mathsf{p} }   \ottsym{(}  \ottsym{[}   \mathsf{rec}\; \ottmv{f} . \ottkw{u}   \ottsym{/}  \ottmv{f}  \ottsym{]}  \ottnt{u'_{{\mathrm{1}}}}  \ottsym{)}  \;  \ottnt{u'_{{\mathrm{2}}}} $ as required.

      On the other hand, if both reductions where by
      \ottdrulename{sc\_apprec}, then we proceed in the same way, but
      conclude by using the substitution lemma for both expressions. 

   \item One reduction is \ottdrulename{sc\_abort}. So $ \ottnt{m_{{\mathrm{1}}}}  \;  \ottnt{m_{{\mathrm{2}}}} $
     must be $ \ottkw{abort}  \;  \ottnt{m_{{\mathrm{2}}}} $ or $ \ottnt{u_{{\mathrm{1}}}}  \;  \ottkw{abort} $. Then by considering
     possible cases, we see that the other reduction must be
     \ottdrulename{sc\_abort} or \ottdrulename{sc\_app} (the
     $\beta$-rules cannot match because $\ottkw{abort}$ is not a
     value). If the other rule is \ottdrulename{sc\_avort} we are
     trivially done, if it is \ottdrulename{sc\_app} then the term
     steps to $ \ottnt{u'_{{\mathrm{1}}}}  \;  \ottkw{abort} $, which can step to $\ottkw{abort}$ as required.
  \end{itemize}
\end{description}
\end{proof}

\begin{lemma}[Confluence of $ \leadsto_{\mathsf{p} } $]
\label{lem:step-confluence}
If $\ottnt{m}  \leadsto^{*}_{\mathsf{p} }  \ottnt{m_{{\mathrm{1}}}}$ and $\ottnt{m}  \leadsto^{*}_{\mathsf{p} }  \ottnt{m_{{\mathrm{2}}}}$, then there exists
some $\ottnt{m'}$ such that $\ottnt{m_{{\mathrm{1}}}}  \leadsto^{*}_{\mathsf{p} }  \ottnt{m'}$ and $\ottnt{m_{{\mathrm{2}}}}  \leadsto^{*}_{\mathsf{p} }  \ottnt{m'}$.
\end{lemma} 
\begin{proof}
  This is a simple corollary of the 1-step version (lemma
  \ref{lem:step-diamond}), by ``diagram-chasing to fill in the
  rectangle'' (see e.g. \cite{Bar84}, lemma 3.2.2).
\end{proof}

\begin{lemma}[ $ \curlyveedownarrow $ is an equivalence relation]
\hspace*{\fill} \\[-12pt]
\label{lem:join-eqrel}
\begin{enumerate}
\item For any $\ottnt{m}$, $ \ottnt{m} \, \curlyveedownarrow \, \ottnt{m} $.
\item If $ \ottnt{m} \, \curlyveedownarrow \, \ottnt{n} $ then $ \ottnt{n} \, \curlyveedownarrow \, \ottnt{m} $.
\item If $ \ottnt{m_{{\mathrm{1}}}} \, \curlyveedownarrow \, \ottnt{m_{{\mathrm{2}}}} $ and $ \ottnt{m_{{\mathrm{2}}}} \, \curlyveedownarrow \, \ottnt{m_{{\mathrm{3}}}} $, then $ \ottnt{m_{{\mathrm{1}}}} \, \curlyveedownarrow \, \ottnt{m_{{\mathrm{3}}}} $.
\end{enumerate}
\end{lemma}
\begin{proof}
(1) and (2) are immediate just by expanding the definition of 
$ \ottnt{m} \, \curlyveedownarrow \, \ottnt{n} $ and $\ottnt{m}  \leadsto^{*}_{\mathsf{p} }  \ottnt{n}$.

For (3), by expanding the definition we have some $\ottnt{n_{{\mathrm{1}}}}$ and
$\ottnt{n_{{\mathrm{2}}}}$ such that $\ottnt{m_{{\mathrm{1}}}}  \leadsto^{*}_{\mathsf{p} }  \ottnt{n_{{\mathrm{1}}}}$, $\ottnt{m_{{\mathrm{2}}}}  \leadsto^{*}_{\mathsf{p} }  \ottnt{n_{{\mathrm{1}}}}$,
$\ottnt{m_{{\mathrm{2}}}}  \leadsto^{*}_{\mathsf{p} }  \ottnt{n_{{\mathrm{2}}}}$ and $\ottnt{m_{{\mathrm{3}}}}  \leadsto^{*}_{\mathsf{p} }  \ottnt{n_{{\mathrm{2}}}}$. So by confluence
(lemma~\ref{lem:step-confluence}) applied to the two middle ones, there
exists some $\ottnt{n}$ such that $\ottnt{n_{{\mathrm{1}}}}  \leadsto^{*}_{\mathsf{p} }  \ottnt{n}$ and
$\ottnt{n_{{\mathrm{2}}}}  \leadsto^{*}_{\mathsf{p} }  \ottnt{n}$. Then we have $\ottnt{m_{{\mathrm{1}}}}  \leadsto^{*}_{\mathsf{p} }  \ottnt{n}$ and
$\ottnt{m_{{\mathrm{3}}}}  \leadsto^{*}_{\mathsf{p} }  \ottnt{n}$ as required.
\end{proof}

\begin{lemma}
\label{lem:step-subst}
If $\ottnt{N}  \leadsto_{\mathsf{p} }  \ottnt{N'}$, then $\ottsym{[}  \ottnt{N}  \ottsym{/}  \ottmv{x}  \ottsym{]}  \ottnt{M}  \leadsto_{\mathsf{p} }  \ottsym{[}  \ottnt{N'}  \ottsym{/}  \ottmv{x}  \ottsym{]}  \ottnt{M}$.
\end{lemma}

\begin{lemma}
\label{lem:join-subst}
If $ \ottnt{N} \, \curlyveedownarrow \, \ottnt{N'} $, then $ \ottsym{[}  \ottnt{N}  \ottsym{/}  \ottmv{x}  \ottsym{]}  \ottnt{M} \, \curlyveedownarrow \, \ottsym{[}  \ottnt{N'}  \ottsym{/}  \ottmv{x}  \ottsym{]}  \ottnt{M} $.
\end{lemma}
\begin{proof}
Expanding the definition of $ \curlyveedownarrow $ we have $\ottnt{N}  \leadsto^{*}_{\mathsf{p} }  \ottnt{N_{{\mathrm{1}}}}$
and $\ottnt{N'}  \leadsto_{\mathsf{p} }  \ottnt{N_{{\mathrm{1}}}}$ for some $\ottnt{N_{{\mathrm{1}}}}$. Now repeatedly apply
Lemma~\ref{lem:step-subst}, to get $\ottsym{[}  \ottnt{N}  \ottsym{/}  \ottmv{x}  \ottsym{]}  \ottnt{M}  \leadsto^{*}_{\mathsf{p} }  \ottsym{[}  \ottnt{N_{{\mathrm{1}}}}  \ottsym{/}  \ottmv{x}  \ottsym{]}  \ottnt{M}$
and $\ottsym{[}  \ottnt{N'}  \ottsym{/}  \ottmv{x}  \ottsym{]}  \ottnt{M}  \leadsto^{*}_{\mathsf{p} }  \ottsym{[}  \ottnt{N_{{\mathrm{1}}}}  \ottsym{/}  \ottmv{x}  \ottsym{]}  \ottnt{M}$.
\end{proof}

\begin{lemma}
\label{lem:fv-step}
If $\ottnt{m}  \leadsto_{\mathsf{p} }  \ottnt{m'}$, then $\ottkw{FV} \, \ottsym{(}  \ottnt{m'}  \ottsym{)} \subseteq \ottkw{FV} \, \ottsym{(}  \ottnt{m}  \ottsym{)}$.
\end{lemma}

\subsection{Structural properties}

\begin{lemma}[Free variables in typing judgments]
\label{lem:typing-fv}
If $ \ottnt{H} \vdash \ottnt{m} : \ottnt{M} $, then $\ottkw{FV} \, \ottsym{(}  \ottnt{m}  \ottsym{)} \subseteq \ottkw{dom} \, \ottsym{(}  \ottnt{H}  \ottsym{)}$ and
$\ottkw{FV} \, \ottsym{(}  \ottnt{M}  \ottsym{)} \subseteq \ottkw{dom} \, \ottsym{(}  \ottnt{H}  \ottsym{)}$.
\end{lemma}

\begin{lemma}[Regularity for contexts]
\label{lem:reg-context}
If $ \ottnt{H} \vdash \ottnt{m} : \ottnt{M} $ then $ \vdash   \ottnt{H} $.
\end{lemma}

\begin{lemma}[Regularity for variable lookup]
\label{lem:reg-var}
If $ \ottnt{H_{{\mathrm{1}}}}  \ottsym{,}   \ottmv{x}  :  \ottnt{M}   \ottsym{,}  \ottnt{H_{{\mathrm{2}}}} \vdash \ottnt{n} : \ottnt{N} $, then $ \ottnt{H_{{\mathrm{1}}}} \vdash \ottnt{M} :  \star  $.
\end{lemma}

\begin{lemma}[Context conversion]
\label{lem:context-conversion}
If $ \ottnt{H}  \ottsym{,}   \ottmv{x}  :  \ottnt{M}   \ottsym{,}  \ottnt{H'} \vdash \ottnt{n} : \ottnt{N} $ and $ \ottnt{H} \vdash \ottkw{join} : \ottnt{M}  \ottsym{=}  \ottnt{M'} $
and $ \ottnt{H} \vdash \ottnt{M'} :  \star  $, then $ \ottnt{H}  \ottsym{,}   \ottmv{x}  :  \ottnt{M'}   \ottsym{,}  \ottnt{H'} \vdash \ottnt{n} : \ottnt{N} $.
\end{lemma}

\begin{lemma}[Substitution]
\label{lem:substitution} Suppose $ \ottnt{H_{{\mathrm{1}}}} \vdash \ottnt{u_{{\mathrm{1}}}} : \ottnt{M_{{\mathrm{1}}}} $. Then,
\begin{itemize}
\item If $ \ottnt{H_{{\mathrm{1}}}}  \ottsym{,}   \ottmv{x_{{\mathrm{1}}}}  :  \ottnt{M_{{\mathrm{1}}}}   \ottsym{,}  \ottnt{H_{{\mathrm{2}}}} \vdash \ottnt{m} : \ottnt{M} $,
then $ \ottnt{H_{{\mathrm{1}}}}  \ottsym{,}  \ottsym{[}  \ottnt{u_{{\mathrm{1}}}}  \ottsym{/}  \ottmv{x_{{\mathrm{1}}}}  \ottsym{]}  \ottnt{H_{{\mathrm{2}}}} \vdash \ottsym{[}  \ottnt{u_{{\mathrm{1}}}}  \ottsym{/}  \ottmv{x_{{\mathrm{1}}}}  \ottsym{]}  \ottnt{m} : \ottsym{[}  \ottnt{u_{{\mathrm{1}}}}  \ottsym{/}  \ottmv{x_{{\mathrm{1}}}}  \ottsym{]}  \ottnt{M} $.
\item If $ \vdash   \ottnt{H_{{\mathrm{1}}}}  \ottsym{,}   \ottmv{x_{{\mathrm{1}}}}  :  \ottnt{M_{{\mathrm{1}}}}   \ottsym{,}  \ottnt{H_{{\mathrm{2}}}} $, then $ \vdash   \ottnt{H_{{\mathrm{1}}}}  \ottsym{,}  \ottsym{[}  \ottnt{u_{{\mathrm{1}}}}  \ottsym{/}  \ottmv{x_{{\mathrm{1}}}}  \ottsym{]}  \ottnt{H_{{\mathrm{2}}}} $.
\end{itemize}
\end{lemma}

\begin{lemma}[Regularity]
\label{lem:reg}
If $ \ottnt{H} \vdash \ottnt{m} : \ottnt{M} $, then $ \ottnt{H} \vdash \ottnt{M} :  \star  $.
\end{lemma}

\begin{lemma}[Data constructors are unique in the environment]
\label{lem:dcon-unique}
If $ \vdash   \ottnt{H} $, and
\[
 \ottkw{data} \, \ottmv{D} \, \Xi^{+} \, \ottkw{where} \, \ottsym{\{} \, \ottcomplu{\ottmv{d_{\ottmv{i}}}  \ottsym{:}  \Xi_{\ottmv{i}}  \to \; D \; \Xi^{+}}{\ottmv{i}}{{\mathrm{1}}}{..}{\ottmv{j}} \, \ottsym{\}} \in \ottnt{H} 
\]
and
\[
 \ottkw{data} \, \ottmv{D'} \, {\Xi^{+}}' \, \ottkw{where} \, \ottsym{\{} \,
 \ottcomplu{\ottmv{d'_{\ottmv{i}}}  \ottsym{:}  \Xi'_{\ottmv{i}}  \to
   \; D' \; {\Xi^{+}}'}{\ottmv{i}}{{\mathrm{1}}}{..}{\ottmv{j}} \,
 \ottsym{\}} \in \ottnt{H},
\]
and $\ottmv{d_{\ottmv{k}}} = \ottmv{d_{\ottmv{l}}}'$, then $\ottmv{D} = \ottmv{D'}$ and $\Xi^{+} = {\Xi^{+}}'$
and $\Xi_{\ottmv{k}} = \Xi'_{\ottmv{l}}$.
\end{lemma}

\subsection{Inversion Lemmas}

We need one inversion lemma for each introduction form that has a
computationally irrelevant eliminator.  These proofs are all similar,
so we only show the representative case for $\lambda$. 

We first need some basic facts about equality.

\begin{lemma}[Inversion for equality]
\label{lem:inv-eq}
If $ \ottnt{H} \vdash \ottnt{m}  \ottsym{=}  \ottnt{n} : \ottnt{M_{{\mathrm{0}}}} $ then, $ \ottnt{H} \vdash \ottnt{m} : \ottnt{M} $ and $ \ottnt{H} \vdash \ottnt{n} : \ottnt{N} $.
\end{lemma}
\begin{proof}
  Induction of $ \ottnt{H} \vdash \ottnt{m}  \ottsym{=}  \ottnt{n} : \ottnt{M_{{\mathrm{0}}}} $. The only cases where the
  subject of the conclusion of the rule is an equality are
  \ottdrulename{et\_eq} (where we get the result as a premise to the rule)
  and \ottdrulename{et\_conv} (direct by induction).
\end{proof}

\begin{lemma}[Proof irrelevance for equality proofs]
If $ \ottnt{H} \vdash \ottnt{u} : \ottnt{M}  \ottsym{=}  \ottnt{N} $, then $ \ottnt{H} \vdash \ottkw{join} : \ottnt{M}  \ottsym{=}  \ottnt{N} $
\end{lemma}
\begin{proof}
By regularity (lemma~\ref{lem:reg}) we know
$ \ottnt{H} \vdash \ottnt{M}  \ottsym{=}  \ottnt{N} :  \star  $, so by inversion (lemma~\ref{lem:inv-eq}) we
have $ \ottnt{H} \vdash \ottnt{M} :  \star  $.
So by \ottdrulename{et\_tjoin}, $ \ottnt{H} \vdash \ottkw{join} : \ottnt{M}  \ottsym{=}  \ottnt{M} $.

Now by \ottdrulename{et\_tconv} we get $ \ottnt{H} \vdash \ottkw{join} : \ottnt{M}  \ottsym{=}  \ottnt{N} $ by
using the assumed proof $\ottnt{u}$ to change $\ottnt{M}$ to
$\ottnt{N}$.
\end{proof}

\begin{lemma}[Propositional equality is an equivalence relation]
\label{lem:propeq-eqrel}
\hspace*{\fill} \\[-12pt]
\begin{itemize}
\item If $ \ottnt{H} \vdash \ottnt{m} : \ottnt{M} $, then $ \ottnt{H} \vdash \ottkw{join} : \ottnt{m}  \ottsym{=}  \ottnt{m} $.
\item If $ \ottnt{H} \vdash \ottnt{u} : \ottnt{m_{{\mathrm{1}}}}  \ottsym{=}  \ottnt{m_{{\mathrm{2}}}} $, then $ \ottnt{H} \vdash \ottkw{join} : \ottnt{m_{{\mathrm{2}}}}  \ottsym{=}  \ottnt{m_{{\mathrm{1}}}} $.
\item If $ \ottnt{H} \vdash \ottnt{u} : \ottnt{m_{{\mathrm{1}}}}  \ottsym{=}  \ottnt{m_{{\mathrm{2}}}} $ and $ \ottnt{H} \vdash \ottnt{u'} : \ottnt{m_{{\mathrm{2}}}}  \ottsym{=}  \ottnt{m_{{\mathrm{3}}}} $, then
  $ \ottnt{H} \vdash \ottkw{join} : \ottnt{m_{{\mathrm{1}}}}  \ottsym{=}  \ottnt{m_{{\mathrm{3}}}} $. 
\end{itemize} 
\end{lemma}
\begin{proof}
(1) is just a special case of \ottdrulename{et\_join}.

(2) We have $ \ottnt{H} \vdash \ottkw{join} : \ottnt{m_{{\mathrm{1}}}}  \ottsym{=}  \ottnt{m_{{\mathrm{1}}}} $, so we can use the assumed
proof to change the left $\ottnt{m_{{\mathrm{1}}}}$ to an $\ottnt{m_{{\mathrm{2}}}}$.

(3) Use the assumed proof $\ottnt{u}$ to change the type of $\ottnt{u'}$.
\end{proof}

\begin{lemma}[Inversion for $\lambda$]
\label{lem:inv-lam}
If $ \ottnt{H} \vdash \lambda  \ottmv{x}  \ottsym{.}  \ottnt{n} : \ottnt{M} $, then $ \ottnt{H} \vdash \ottkw{join} :  ( \ottmv{x} \!:\! \ottnt{M_{{\mathrm{1}}}} )  \to \,  \ottnt{N_{{\mathrm{1}}}}  \ottsym{=}  \ottnt{M}  $ for
some $\ottnt{M_{{\mathrm{1}}}}$ and $\ottnt{N_{{\mathrm{1}}}}$, and $ \ottnt{H}  \ottsym{,}   \ottmv{x}  :  \ottnt{M_{{\mathrm{1}}}}  \vdash \ottnt{n} : \ottnt{N_{{\mathrm{1}}}} $. 
\end{lemma}
\begin{proof}
By induction on $ \ottnt{H} \vdash \lambda  \ottmv{x}  \ottsym{.}  \ottnt{n} : \ottnt{M} $. Only two typing rules can have
a $\lambda$ as the subject of the conclusion.
\begin{description}
\item[ Case \ottdrulename{et\_abs}] The rule looks like
\[
\ottdruleetXXabs{}
\]
By \ottdrulename{et\_join} we have
$ \ottnt{H} \vdash \ottkw{join} :  ( \ottmv{x} \!:\! \ottnt{M} )  \to \,  \ottnt{N}  \ottsym{=}   ( \ottmv{x} \!:\! \ottnt{M} )  \to \,  \ottnt{N}   $, and we have
$ \ottnt{H}  \ottsym{,}   \ottmv{x}  :  \ottnt{M}  \vdash \ottnt{n} : \ottnt{N} $ as a premise to the rule.

\item[ Case \ottdrulename{et\_conv}] The rule looks like
\[
\ottdruleetXXconv{}
\]

By IH we get that $\ottsym{[}  \ottnt{M_{{\mathrm{1}}}}  \ottsym{/}  \ottmv{x_{{\mathrm{1}}}}  \ottsym{]} \, ... \, \ottsym{[}  \ottnt{M_{\ottmv{i}}}  \ottsym{/}  \ottmv{x_{\ottmv{i}}}  \ottsym{]}  \ottnt{M}$ is propositionally equal
to an arrow type, with $\ottnt{n}$ being typeable at the ``unwrapping''
of that type. So if we can show that $\ottsym{[}  \ottnt{N_{{\mathrm{1}}}}  \ottsym{/}  \ottmv{x_{{\mathrm{1}}}}  \ottsym{]} \, ... \, \ottsym{[}  \ottnt{N_{\ottmv{i}}}  \ottsym{/}  \ottmv{x_{\ottmv{i}}}  \ottsym{]}  \ottnt{M}$ is
propositionally equal to that same arrow type, then we are done.

But note that by regularity, inversion and reflexivity (lemmas
\ref{lem:reg}, \ref{lem:inv-eq}, \ref{lem:propeq-eqrel}) we have
$ \ottnt{H} \vdash \ottkw{join} : \ottsym{[}  \ottnt{M_{{\mathrm{1}}}}  \ottsym{/}  \ottmv{x_{{\mathrm{1}}}}  \ottsym{]} \, ... \, \ottsym{[}  \ottnt{M_{\ottmv{i}}}  \ottsym{/}  \ottmv{x_{\ottmv{i}}}  \ottsym{]}  \ottnt{M}  \ottsym{=}  \ottsym{[}  \ottnt{M_{{\mathrm{1}}}}  \ottsym{/}  \ottmv{x_{{\mathrm{1}}}}  \ottsym{]} \, ... \, \ottsym{[}  \ottnt{M_{\ottmv{i}}}  \ottsym{/}  \ottmv{x_{\ottmv{i}}}  \ottsym{]}  \ottnt{M} $.
By applying \ottdrulename{et\_conv} using the proof $\ottnt{u_{{\mathrm{1}}}}...\ottnt{u_{\ottmv{i}}}$
we get $ \ottnt{H} \vdash \ottkw{join} : \ottsym{[}  \ottnt{M_{{\mathrm{1}}}}  \ottsym{/}  \ottmv{x_{{\mathrm{1}}}}  \ottsym{]} \, ... \, \ottsym{[}  \ottnt{M_{\ottmv{i}}}  \ottsym{/}  \ottmv{x_{\ottmv{i}}}  \ottsym{]}  \ottnt{M}  \ottsym{=}  \ottsym{[}  \ottnt{N_{{\mathrm{1}}}}  \ottsym{/}  \ottmv{x_{{\mathrm{1}}}}  \ottsym{]} \, ... \, \ottsym{[}  \ottnt{N_{\ottmv{i}}}  \ottsym{/}  \ottmv{x_{\ottmv{i}}}  \ottsym{]}  \ottnt{M} $.
Then by transitivity (lemma~\ref{lem:propeq-eqrel}) we have
that $\ottsym{[}  \ottnt{N_{{\mathrm{1}}}}  \ottsym{/}  \ottmv{x_{{\mathrm{1}}}}  \ottsym{]} \, ... \, \ottsym{[}  \ottnt{N_{\ottmv{i}}}  \ottsym{/}  \ottmv{x_{\ottmv{i}}}  \ottsym{]}  \ottnt{M}$ is propositionally equal to the arrow
type as required.
\end{description}
\end{proof}

The remaining inversion lemmas follow a similar pattern, so we omit
the proofs.

\begin{lemma}[Inversion for irrelevant $\lambda$]
\label{lem:inv-ilam}
If $ \ottnt{H} \vdash \lambda  \ottsym{[}  \ottsym{]}  \ottsym{.}  \ottnt{n} : \ottnt{M} $, then $ \ottnt{H} \vdash \ottkw{join} :  [  \ottmv{x} \!:\! \ottnt{M_{{\mathrm{1}}}}  ]  \to \,  \ottnt{N_{{\mathrm{1}}}}  \ottsym{=}  \ottnt{M}  $ for
some $\ottnt{M_{{\mathrm{1}}}}$ and $\ottnt{N_{{\mathrm{1}}}}$, and $ \ottnt{H}  \ottsym{,}   \ottmv{x}  :  \ottnt{M_{{\mathrm{1}}}}  \vdash \ottnt{n} : \ottnt{N_{{\mathrm{1}}}} $ where  
$ \ottmv{x} \ \notin \ottkw{FV} \, \ottsym{(}  \ottnt{n}  \ottsym{)} \ $.
\end{lemma}

\begin{lemma}[Inversion for rec] 
\label{lem:inv-rec}
If $ \ottnt{H} \vdash  \mathsf{rec}\; \ottmv{f} . \ottkw{u}  : \ottnt{M} $, then $ \ottnt{H} \vdash \ottkw{join} : \ottnt{M}  \ottsym{=}  \ottnt{M_{{\mathrm{1}}}} $
and $ \ottnt{H}  \ottsym{,}   \ottmv{f}  :  \ottnt{M_{{\mathrm{1}}}}  \vdash \ottnt{u} : \ottnt{M_{{\mathrm{1}}}} $ for some $\ottnt{M_{{\mathrm{1}}}}$ such that
$ \ottnt{H} \vdash \ottnt{M_{{\mathrm{1}}}} :  \star  $ and $\ottnt{M_{{\mathrm{1}}}}$ is an relevant or irrelevant arrow type.
\end{lemma}

\begin{lemma}[Inversion for dcon]
\label{lem:inv-dcon}
If $ \ottnt{H} \vdash \ottmv{d} \, \overline{m_i} : \ottnt{M} $, then $ \ottnt{H} \vdash \ottkw{join} : \ottmv{D} \, \overline{N_i}  \ottsym{=}  \ottnt{M} $  for
some $\overline{N_i}$ such that:
\begin{itemize}
\item $ \ottkw{data} \, \ottmv{D} \, \Xi^{+} \, \ottkw{where} \, \ottsym{\{} \, \ottcomplu{\ottmv{d_{\ottmv{i}}}  \ottsym{:}  \Xi_{\ottmv{i}}  \to \; D \; \Xi^{+}}{\ottmv{i}}{{\mathrm{1}}}{..}{\ottmv{j}} \, \ottsym{\}} \in \ottnt{H} $ and
$\ottmv{d}$ is $\ottmv{d_{\ottmv{l}}}$ for one of the constructors in the
declaration.
\item $\ottnt{H}  \vdash  \overline{N_i}  \ottsym{:}  \Xi$
\item $\ottnt{H}  \vdash  \overline{m_i}  \ottsym{:}  \ottsym{[}  \overline{N_i}  \ottsym{/}  \Xi  \ottsym{]}  \Xi_{\ottmv{l}}$
\end{itemize}
\end{lemma}

\subsection{Preservation}

\begin{lemma}[A conversion rule for value lists]
\label{lem:vallist-conv}
If $\ottnt{H}  \vdash  \overline{u_i}  \ottsym{:}  \ottsym{[}  \overline{M_i}  \ottsym{/}  \overline{y_i}  \ottsym{]}  \Xi$ and
$ \forall  \ottmv{i}  .\ \   \ottnt{H} \vdash \ottkw{join} : \ottnt{M_{\ottmv{i}}}  \ottsym{=}  \ottnt{N_{\ottmv{i}}}  $ and
$ \vdash   \ottnt{H}  \ottsym{,}  \ottsym{[}  \overline{N_i}  \ottsym{/}  \overline{y_i}  \ottsym{]}  \Xi $,
then $\ottnt{H}  \vdash  \overline{u_i}  \ottsym{:}  \ottsym{[}  \overline{N_i}  \ottsym{/}  \overline{y_i}  \ottsym{]}  \Xi$.
\end{lemma}
\begin{proof}
We proceed by induction on the structure of $\Xi$.
\begin{description}
\item[Case empty.]  Trivial.
\item[Case $\ottsym{(}  \ottmv{x}  \ottsym{:}  \ottnt{M}  \ottsym{)}  \Xi$.] By inversion on the assumed
  judgments, we know
\[
\ottdrule{
  \ottpremise{ \ottnt{H} \vdash \ottnt{u} : \ottsym{[}  \overline{M_i}  \ottsym{/}  \overline{y_i}  \ottsym{]}  \ottnt{M} }
  \ottpremise{ \ottnt{H} \vdash \ottsym{[}  \overline{M_i}  \ottsym{/}  \overline{y_i}  \ottsym{]}  \ottnt{M} :  \star  }
  \ottpremise{\ottnt{H}  \vdash  \overline{u_i}  \ottsym{:}  \ottsym{[}  \ottnt{u}  \ottsym{/}  \ottmv{x}  \ottsym{]}  \ottsym{[}  \overline{M_i}  \ottsym{/}  \overline{y_i}  \ottsym{]}  \Xi}
}{
 \ottnt{H}  \vdash   \ottnt{u}  \;  \overline{u_i}   \ottsym{:}  \ottsym{(}  \ottmv{x}  \ottsym{:}  \ottsym{[}  \overline{M_i}  \ottsym{/}  \overline{y_i}  \ottsym{]}  \ottnt{M}  \ottsym{)}  \ottsym{[}  \overline{M_i}  \ottsym{/}  \overline{y_i}  \ottsym{]}  \Xi
}{
  \ottdrulename{etl\_cons}
}
\]
and 
\[
 \vdash   \ottnt{H}  \ottsym{,}   \ottmv{x}  :  \ottsym{[}  \overline{N_i}  \ottsym{/}  \overline{y_i}  \ottsym{]}  \ottnt{M}   \ottsym{,}  \ottsym{[}  \overline{N_i}  \ottsym{/}  \overline{y_i}  \ottsym{]}  \Xi .
\]
By inversion on this, we have $ \ottnt{H} \vdash \ottsym{[}  \overline{N_i}  \ottsym{/}  \overline{y_i}  \ottsym{]}  \ottnt{M} :  \star  $.

Now since we know $ \forall  \ottmv{i}  .\ \   \ottnt{H} \vdash \ottkw{join} : \ottnt{M_{\ottmv{i}}}  \ottsym{=}  \ottnt{N_{\ottmv{i}}}  $ and $ \ottnt{H} \vdash \ottsym{[}  \overline{N_i}  \ottsym{/}  \overline{y_i}  \ottsym{]}  \ottnt{M} :  \star  $, 
then by \ottdrulename{et\_conv} we have $ \ottnt{H} \vdash \ottnt{u} : \ottsym{[}  \overline{N_i}  \ottsym{/}  \overline{y_i}  \ottsym{]}  \ottnt{M} $.

By substitution  (lemma~\ref{lem:substitution}) we get $ \vdash   \ottnt{H}  \ottsym{,}  \ottsym{[}  \ottnt{u}  \ottsym{/}  \ottmv{x}  \ottsym{]}  \ottsym{[}  \overline{N_i}  \ottsym{/}  \overline{y_i}  \ottsym{]}  \Xi $.
We know $\ottnt{u}$ is well-typed, so by \ottdrulename{et\_join} we have $ \ottnt{H} \vdash \ottkw{join} : \ottnt{u}  \ottsym{=}  \ottnt{u} $.
Then by IH, taking the multi-substitution to be $[u/x][ \overline{M_i}/\overline{y_i} ]$, we get $\ottnt{H}  \vdash  \overline{u_i}  \ottsym{:}  \ottsym{[}  \ottnt{u}  \ottsym{/}  \ottmv{x}  \ottsym{]}  \ottsym{[}  \overline{N_i}  \ottsym{/}  \overline{y_i}  \ottsym{]}  \Xi$.
So re-applying \ottdrulename{etl\_cons} we get
\[
\ottnt{H}  \vdash   \ottnt{u}  \;  \overline{u_i}   \ottsym{:}  \ottsym{(}  \ottmv{x}  \ottsym{:}  \ottsym{[}  \overline{N_i}  \ottsym{/}  \overline{y_i}  \ottsym{]}  \ottnt{M}  \ottsym{)}  \ottsym{[}  \overline{N_i}  \ottsym{/}  \overline{y_i}  \ottsym{]}  \Xi
\]
as required.

\item[Case $\ottsym{[}  \ottmv{x}  \ottsym{:}  \ottnt{M}  \ottsym{]}  \Xi$.] This case is similar. Inversion on the
  first assumed judgement now gives
\[
\ottdrule{
  \ottpremise{ \ottnt{H} \vdash \ottnt{u} : \ottsym{[}  \overline{M_i}  \ottsym{/}  \overline{y_i}  \ottsym{]}  \ottnt{M} }
  \ottpremise{ \ottnt{H} \vdash \ottsym{[}  \overline{M_i}  \ottsym{/}  \overline{y_i}  \ottsym{]}  \ottnt{M} :  \star  }
  \ottpremise{\ottnt{H}  \vdash  \overline{u_i}  \ottsym{:}  \ottsym{[}  \ottnt{u}  \ottsym{/}  \ottmv{x}  \ottsym{]}  \ottsym{[}  \overline{M_i}  \ottsym{/}  \overline{y_i}  \ottsym{]}  \Xi}
}{
 \ottnt{H}  \vdash   [] \;  \overline{u_i}   \ottsym{:}  \ottsym{(}  \ottmv{x}  \ottsym{:}  \ottsym{[}  \overline{M_i}  \ottsym{/}  \overline{y_i}  \ottsym{]}  \ottnt{M}  \ottsym{)}  \ottsym{[}  \overline{M_i}  \ottsym{/}  \overline{y_i}  \ottsym{]}  \Xi
}{
  \ottdrulename{etl\_cons}
}
\]
By reasoning as in the previous case we get
$ \ottnt{H} \vdash \ottnt{u} : \ottsym{[}  \overline{N_i}  \ottsym{/}  \overline{y_i}  \ottsym{]}  \ottnt{M} $  and $ \ottnt{H} \vdash \ottsym{[}  \overline{N_i}  \ottsym{/}  \overline{y_i}  \ottsym{]}  \ottnt{M} :  \star  $
and $\ottnt{H}  \vdash  \overline{u_i}  \ottsym{:}  \ottsym{[}  \ottnt{u}  \ottsym{/}  \ottmv{x}  \ottsym{]}  \ottsym{[}  \overline{N_i}  \ottsym{/}  \overline{y_i}  \ottsym{]}  \Xi$. Then re-apply \ottdrulename{etl\_icons}.
\end{description}
\end{proof}
 
\begin{thm}[Preservation]
\hspace*{\fill} \\[-12pt]
\label{thm:preservation}
\begin{enumerate}
\item If $ \ottnt{H} \vdash \ottnt{m} : \ottnt{M} $ and $\ottnt{m}  \leadsto_{\mathsf{p} }  \ottnt{m'}$, then $ \ottnt{H} \vdash \ottnt{m'} : \ottnt{M} $.
\item If $\ottnt{H}  \vdash  \overline{m_i}  \ottsym{:}  \ottsym{[}  \ottnt{n_{{\mathrm{1}}}}  \ottsym{/}  \ottmv{y_{{\mathrm{1}}}}  \ottsym{]} \, ... \, \ottsym{[}  \ottnt{n_{\ottmv{l}}}  \ottsym{/}  \ottmv{y_{\ottmv{l}}}  \ottsym{]}  \Xi$
       and $ \forall  \ottmv{i}  .\ \  \ottnt{m_{\ottmv{i}}}  \leadsto_{\mathsf{p} }  \ottnt{m'_{\ottmv{i}}} $ and $ \forall  \ottmv{j}  .\ \  \ottnt{n_{\ottmv{j}}}  \leadsto_{\mathsf{p} }  \ottnt{n'_{\ottmv{j}}} $, then
  $\ottnt{H}  \vdash  \overline{m_i}'  \ottsym{:}  \ottsym{[}  \ottnt{n'_{{\mathrm{1}}}}  \ottsym{/}  \ottmv{y_{{\mathrm{1}}}}  \ottsym{]} \, ... \, \ottsym{[}  \ottnt{n'_{\ottmv{l}}}  \ottsym{/}  \ottmv{y_{\ottmv{l}}}  \ottsym{]}  \Xi$.
\end{enumerate}
\end{thm}
\begin{proof}
By mutual induction on the two judgments. The cases for
 $ \ottnt{H} \vdash \ottnt{m} : \ottnt{M} $ are:
\begin{description}
\item[Cases] \ottdrulename{et\_type},
\ottdrulename{et\_var},
\ottdrulename{et\_abort},
\ottdrulename{et\_join},
\ottdrulename{et\_injdom},
\ottdrulename{et\_injrng},
\ottdrulename{et\_iinjdom},
\ottdrulename{et\_iinjrng},
\ottdrulename{et\_injtcon}.

These expressions can not step except by \ottdrulename{sp\_refl}, so the
result is trivial.

\item[Case \ottdrulename{et\_case}.] The rule looks like
\[
\ottdruletXXcase{}
\]
We consider the ways the expression
$\ottkw{case} \, \ottnt{n} \, \ottkw{of} \, \ottsym{\{} \, \ottcomplu{\ottmv{d_{\ottmv{j}}} \, \overline{x_i}_{\ottmv{j}}  \Rightarrow  \ottnt{m_{\ottmv{j}}}}{\ottmv{j}}{{\mathrm{1}}}{..}{\ottmv{k}} \, \ottsym{\}}$ may
  step:
\begin{itemize}
\item To
  $\ottkw{case} \, \ottnt{n'} \, \ottkw{of} \, \ottsym{\{} \, \ottcomplu{\ottmv{d_{\ottmv{j}}} \, \overline{x_i}_{\ottmv{j}}  \Rightarrow  \ottnt{m'_{\ottmv{j}}}}{\ottmv{j}}{{\mathrm{1}}}{..}{\ottmv{k}} \, \ottsym{\}}$
  by \ottdrulename{sp\_case} when $\ottnt{n}  \leadsto_{\mathsf{p} }  \ottnt{n'}$ and
  $ \forall  \ottmv{j}  .\ \  \ottnt{m_{\ottmv{j}}}  \leadsto_{\mathsf{p} }  \ottnt{m'_{\ottmv{j}}} $.

By IH we get $ \ottnt{H} \vdash \ottnt{n} : \ottmv{D} \, \overline{n_i} $. Also by IH, for each $j$ we
have 
\[
 \ottnt{H}  \ottsym{,}  \ottsym{[}  \overline{n_i}  \ottsym{/}  \Xi  \ottsym{]}  \Xi_{\ottmv{j}}  \ottsym{,}   \ottmv{y}  :  \ottnt{n}  \ottsym{=}  \ottmv{d_{\ottmv{j}}} \, \Xi_{\ottmv{j}}  \vdash \ottnt{m'_{\ottmv{j}}} : \ottnt{M} .
\]
Now by regularity (lemma~\ref{lem:reg-var}) and inversion
(lemma~\ref{lem:inv-eq}) we know that $\ottmv{d_{\ottmv{j}}} \, \Xi_{\ottmv{j}}$ is welltyped in
the context $\ottnt{H}  \ottsym{,}  \ottsym{[}  \overline{n_i}  \ottsym{/}  \Xi^{+}  \ottsym{]}  \Xi_{\ottmv{j}}$. And we already observed that
$\ottnt{n}$ and $\ottnt{n'}$ are welltyped. So $\ottnt{n}  \ottsym{=}  \ottmv{d_{\ottmv{j}}} \, \Xi_{\ottmv{j}}$ and
$\ottnt{n'}  \ottsym{=}  \ottmv{d_{\ottmv{j}}} \, \Xi_{\ottmv{j}}$ are wellformed equations. Since
$\ottnt{n}  \ottsym{=}  \ottmv{d_{\ottmv{j}}} \, \Xi_{\ottmv{j}}  \leadsto_{\mathsf{p} }  \ottnt{n'}  \ottsym{=}  \ottmv{d_{\ottmv{j}}} \, \Xi_{\ottmv{j}}$, by \ottdrulename{et\_join} we have
$ \ottnt{H}  \ottsym{,}  \ottsym{[}  \overline{n_i}  \ottsym{/}  \Xi^{+}  \ottsym{]}  \Xi_{\ottmv{j}} \vdash \ottkw{join} : \ottsym{(}  \ottnt{n}  \ottsym{=}  \ottmv{d_{\ottmv{j}}} \, \Xi_{\ottmv{j}}  \ottsym{)}  \ottsym{=}  \ottsym{(}  \ottnt{n'}  \ottsym{=}  \ottmv{d_{\ottmv{j}}} \, \Xi_{\ottmv{j}}  \ottsym{)} $.
So by context conversion (lemma~\ref{lem:context-conversion}) we have
\[
 \ottnt{H}  \ottsym{,}  \ottsym{[}  \overline{n_i}  \ottsym{/}  \Xi^{+}  \ottsym{]}  \Xi_{\ottmv{j}}  \ottsym{,}   \ottmv{y}  :  \ottnt{n'}  \ottsym{=}  \ottmv{d_{\ottmv{j}}} \, \Xi_{\ottmv{j}}  \vdash \ottnt{m'_{\ottmv{j}}} : \ottnt{M} .
\]
Then we can re-apply \ottdrulename{et\_case} to get the required
\[
 \ottnt{H} \vdash \ottkw{case} \, \ottnt{n'} \, \ottkw{of} \, \ottsym{\{} \, \ottcomplu{\ottmv{d_{\ottmv{j}}} \, \overline{x_i}_{\ottmv{j}}  \Rightarrow  \ottnt{m'_{\ottmv{j}}}}{\ottmv{j}}{{\mathrm{1}}}{..}{\ottmv{k}} \, \ottsym{\}} : \ottnt{M} .
\]

\item To $\ottsym{[}  \overline{u_i}'  \ottsym{/}  \overline{x_i}_{\ottmv{l}}  \ottsym{]}  \ottnt{m'_{\ottmv{l}}}$ by \ottdrulename{sp\_casebeta}
  when $\ottnt{n}$ is $\ottmv{d_{\ottmv{l}}} \, \overline{u_i}$, and $ \forall  \ottmv{i}  .\ \  \ottnt{u_{\ottmv{i}}}  \leadsto_{\mathsf{p} }  \ottnt{u'_{\ottmv{i}}} $ and $\ottnt{m_{\ottmv{l}}}  \leadsto_{\mathsf{p} }  \ottnt{m'_{\ottmv{l}}}$.
Notice that the step rule in particular requires that $\ottmv{d_{\ottmv{l}}}$ is
one of the branches of the case expression. 

By inversion (lemma~\ref{lem:inv-dcon}) on the premise
$ \ottnt{H} \vdash \ottmv{d} \, \overline{u_i} : \ottmv{D} \, \overline{n_i} $, we know that $ \ottnt{H} \vdash \ottnt{n} : \ottmv{D} \, \overline{N_i} $
with $ \ottnt{H} \vdash \ottkw{join} : \ottmv{D} \, \overline{N_i}  \ottsym{=}  \ottmv{D} \, \overline{n_i} $, and $\ottnt{H}  \vdash  \overline{N_i}  \ottsym{:}  \Xi^{+}$
and $\ottnt{H}  \vdash  \overline{u_i}  \ottsym{:}  \ottsym{[}  \overline{N_i}  \ottsym{/}  \Xi^{+}  \ottsym{]}  \Xi_{\ottmv{i}}$. (We know that the $\ottmv{D}$,
$\Xi$ and $\Xi_{\ottmv{i}}$ that come out of the lemma are the same as
the ones in the typing rule because data constructors have a unique
definition in the context (lemma~\ref{lem:dcon-unique})).

By the rule \ottdrulename{injtcon} we get $ \ottnt{H} \vdash \ottkw{join} : \ottnt{N_{\ottmv{i}}}  \ottsym{=}  \ottnt{n_{\ottmv{i}}} $ for each $i$. 
So by value-list conversion (lemma~\ref{lem:vallist-conv}) we have $\ottnt{H}  \vdash  \overline{u_i}  \ottsym{:}  \ottsym{[}  \overline{n_i}  \ottsym{/}  \Xi^{+}  \ottsym{]}  \Xi_{\ottmv{l}}$.

We next claim that $ \ottnt{H}  \ottsym{,}   \ottmv{y}  :  \ottmv{d_{\ottmv{l}}} \, \overline{u_i}  \ottsym{=}  \ottmv{d_{\ottmv{l}}} \, \overline{u_i}'  \vdash \ottsym{[}  \overline{u_i}'  \ottsym{/}  \Xi_{\ottmv{l}}  \ottsym{]}  \ottnt{m'_{\ottmv{l}}} : \ottnt{M} $. To show
this we prove a more general claim: for any prefix $\ottnt{u'_{{\mathrm{1}}}} \dots
\ottnt{u'_{\ottmv{k}}}$ of $\overline{u_i}$, and supposing $\Xi_{\ottmv{l}}$ has the form
$(\ottmv{x_{{\mathrm{1}}}}:\ottnt{M_{{\mathrm{1}}}})\dots \ottsym{(}  \ottmv{x_{\ottmv{k}}}  \ottsym{:}  \ottnt{M_{\ottmv{k}}}  \ottsym{)}  \Xi_{{\mathrm{0}}}$, we have 
\[
\begin{array}{l}
 \ottnt{H}  \ottsym{,}  \ottsym{[}  \ottnt{u'_{{\mathrm{1}}}}  \ottsym{/}  \ottmv{x_{{\mathrm{1}}}}  \ottsym{]} \, ... \, \ottsym{[}  \ottnt{u'_{\ottmv{k}}}  \ottsym{/}  \ottmv{x_{\ottmv{k}}}  \ottsym{]}  \ottsym{[}  \overline{n_i}  \ottsym{/}  \Xi^{+}  \ottsym{]}  \Xi_{{\mathrm{0}}}  \ottsym{,}   \ottmv{y}  :  \ottsym{[}  \ottnt{u'_{{\mathrm{1}}}}  \ottsym{/}  \ottmv{x_{{\mathrm{1}}}}  \ottsym{]} \, ... \, \ottsym{[}  \ottnt{u'_{\ottmv{k}}}  \ottsym{/}  \ottmv{x_{\ottmv{k}}}  \ottsym{]}  \ottsym{(}  \ottmv{d_{\ottmv{l}}} \, \overline{u_i}  \ottsym{=}  \ottmv{d_{\ottmv{l}}} \, \Xi_{\ottmv{l}}  \ottsym{)}  \\
\qquad\qquad\vdash \ottsym{[}  \ottnt{u'_{{\mathrm{1}}}}  \ottsym{/}  \ottmv{x_{{\mathrm{1}}}}  \ottsym{]} \, ... \, \ottsym{[}  \ottnt{u'_{\ottmv{k}}}  \ottsym{/}  \ottmv{x_{\ottmv{k}}}  \ottsym{]}  \ottnt{m_{\ottmv{l}}} : \ottsym{[}  \ottnt{u'_{{\mathrm{1}}}}  \ottsym{/}  \ottmv{x_{{\mathrm{1}}}}  \ottsym{]} \, ... \, \ottsym{[}  \ottnt{u'_{\ottmv{k}}}  \ottsym{/}  \ottmv{x_{\ottmv{k}}}  \ottsym{]}  \ottnt{M} 
\end{array}
\]
This follows by induction on $k$ (by applying substitution, lemma~\ref{lem:substitution},
$k$ times). So in particular, we have
\[
 \ottnt{H}  \ottsym{,}   \ottmv{y}  :  \ottsym{[}  \overline{u_i}'  \ottsym{/}  \Xi_{\ottmv{l}}  \ottsym{]}  \ottsym{(}  \ottmv{d_{\ottmv{l}}} \, \overline{u_i}  \ottsym{=}  \ottmv{d_{\ottmv{l}}} \, \Xi_{\ottmv{l}}  \ottsym{)}  \vdash \ottsym{[}  \overline{u_i}'  \ottsym{/}  \Xi_{\ottmv{l}}  \ottsym{]}  \ottnt{m_{\ottmv{l}}} : \ottsym{[}  \overline{u_i}'  \ottsym{/}  \Xi_{\ottmv{l}}  \ottsym{]}  \ottnt{M} 
\]
But by the premises $ \ottnt{H} \vdash \ottnt{M} :  \star  $ and $ \ottnt{H} \vdash \ottmv{d_{\ottmv{l}}} \, \overline{u_i} : \ottmv{D} \, \overline{n_i} $
together with lemma~\ref{lem:typing-fv} we know that $\ottmv{x_{\ottmv{l}}}$ are
not free in $\ottmv{d_{\ottmv{l}}} \, \overline{u_i}$ or $\ottnt{M}$, so this simplifies to 
\[
 \ottnt{H}  \ottsym{,}   \ottmv{y}  :  \ottmv{d_{\ottmv{l}}} \, \overline{u_i}  \ottsym{=}  \ottmv{d_{\ottmv{l}}} \, \overline{u_i}'  \vdash \ottsym{[}  \overline{u_i}'  \ottsym{/}  \Xi_{\ottmv{l}}  \ottsym{]}  \ottnt{m'_{\ottmv{l}}} : \ottnt{M} 
\]
as we claimed.

Next, we know $\ottmv{d_{\ottmv{l}}} \, \overline{u_i}$ is well-typed because that is a premise to the rule.
By the mutual IH we have that $ \ottnt{H} \vdash \ottmv{d_{\ottmv{l}}} \, \overline{u_i}' : \ottmv{D} \, \overline{n_i} $, so
$\ottmv{d_{\ottmv{l}}} \, \overline{u_i}'$ is well-typed too. So by \ottdrulename{et\_join} we 
have $ \ottnt{H} \vdash \ottkw{join} : \ottmv{d_{\ottmv{l}}} \, \overline{u_i}  \ottsym{=}  \ottmv{d_{\ottmv{l}}} \, \overline{u_i}' $. Then by substitution
(lemma~\ref{lem:substitution}) again we have 
\[
 \ottnt{H} \vdash \ottsym{[}  \ottkw{join}  \ottsym{/}  \ottmv{y}  \ottsym{]}  \ottsym{[}  \overline{u_i}'  \ottsym{/}  \Xi_{\ottmv{l}}  \ottsym{]}  \ottnt{m'_{\ottmv{l}}} : \ottsym{[}  \ottkw{join}  \ottsym{/}  \ottmv{y}  \ottsym{]}  \ottnt{M} .
\]
But as a side-condition to the rule (plus lemma~\ref{lem:fv-step}) we know that
$ \ottmv{y} \ \notin \ottkw{FV} \, \ottsym{(}  \ottnt{m'_{\ottmv{l}}}  \ottsym{)} \ $, and $\ottmv{y}$ is a bound variable which we can
pic so that $ \ottmv{y} \ \notin \ottkw{FV} \, \ottsym{(}  \ottnt{M}  \ottsym{)} \ $. So we have in fact show the required
\[
 \ottnt{H} \vdash \ottsym{[}  \overline{u_i}'  \ottsym{/}  \Xi_{\ottmv{l}}  \ottsym{]}  \ottnt{m'_{\ottmv{l}}} : \ottnt{M} .
\]

\item To $\ottkw{abort}$ by \ottdrulename{sp\_abort}. By regularity
  (lemma~\ref{lem:reg}) on the original typing derivation we know that
  $ \ottnt{H} \vdash \ottnt{M} :  \star  $, so by \ottdrulename{et\_abort} we have
  $ \ottnt{H} \vdash \ottkw{abort} : \ottnt{M} $ as required.
\end{itemize}

\item[Case \ottdrulename{et\_pi}.] The rule looks like
\[
\ottdruleetXXpi{}
\]
The only way $ ( \ottmv{x} \!:\! \ottnt{M} )  \to \,  \ottnt{N} $ can step (except trivially by
\ottdrulename{sp\_refl}) is by \ottdrulename{sp\_pi}:
\[
 ( \ottmv{x} \!:\! \ottnt{M} )  \to \,  \ottnt{N}   \leadsto_{\mathsf{p} }   ( \ottmv{x} \!:\! \ottnt{M'} )  \to \,  \ottnt{N'}  \qquad\text{where $\ottnt{M}  \leadsto_{\mathsf{p} }  \ottnt{M'}$ and $\ottnt{N}  \leadsto_{\mathsf{p} }  \ottnt{N'}$}
\]
We must show $ \ottnt{H} \vdash  ( \ottmv{x} \!:\! \ottnt{M'} )  \to \,  \ottnt{N'}  :  \star  $.

By IH we immediately get $ \ottnt{H} \vdash \ottnt{M'} :  \star  $ and $ \ottnt{H}  \ottsym{,}   \ottmv{x}  :  \ottnt{M}  \vdash \ottnt{N'} :  \star  $.
Since $\ottnt{M}  \leadsto_{\mathsf{p} }  \ottnt{M'}$ we also have $ \ottnt{M} \, \curlyveedownarrow \, \ottnt{M'} $, so applying
\ottdrulename{et\_join} we get $ \ottnt{H} \vdash \ottkw{join} : \ottnt{M}  \ottsym{=}  \ottnt{M'} $. Then by context
conversion (lemma~\ref{lem:context-conversion}) we get 
$ \ottnt{H}  \ottsym{,}   \ottmv{x}  :  \ottnt{M'}  \vdash \ottnt{N'} :  \star  $. We conclude by re-applying \ottdrulename{et\_pi}.

\item[Case \ottdrulename{et\_ipi}] Similar to the previous case.

\item[Case \ottdrulename{et\_abs}]. The rule looks like 
\[
\ottdruleetXXabs{}
\]
The only non-trivial way the expression $\lambda  \ottmv{x}  \ottsym{.}  \ottnt{n}$ can step is by
\ottdrulename{sp\_abs} to $\lambda  \ottmv{x}  \ottsym{.}  \ottnt{n'}$ when $\ottnt{n}  \leadsto_{\mathsf{p} }  \ottnt{n'}$.
By IH we get $ \ottnt{H}  \ottsym{,}   \ottmv{x}  :  \ottnt{M}  \vdash \ottnt{n'} : \ottnt{N} $. So re-applying
\ottdrulename{et\_abs} we get $ \ottnt{H} \vdash \lambda  \ottmv{x}  \ottsym{.}  \ottnt{n'} :  ( \ottmv{x} \!:\! \ottnt{M} )  \to \,  \ottnt{N}  $ as required.

\item[Cases] \ottdrulename{et\_iabs}, \ottdrulename{et\_rec}.

These are similar to the previous case. For \ottdrulename{iabs}, note
that the free variable condition is preserved by lemma \ref{lem:fv-step}.

\item[Case \ottdrulename{et\_tcon}.] The rule looks like
\[
\ottdruleetXXtcon{}
\]
The only way the expression can step is by \ottdrulename{sp\_tcon},
so $ \forall  \ottmv{i}  .\ \  \ottnt{M_{\ottmv{i}}}  \leadsto_{\mathsf{p} }  \ottnt{M'_{\ottmv{i}}} $. By
the mutual IH, we get $\ottnt{H}  \vdash  \overline{M_i}'  \ottsym{:}  \Xi^{+}$. So by re-applying 
\ottdrulename{et\_tcon} we have $ \ottnt{H} \vdash \ottmv{D} \, \overline{M_i} :  \star  $ as 
required.

\item[Case \ottdrulename{et\_abstcon}.] Similar to the previous case.

\item[Case \ottdrulename{et\_dcon}.] The rule looks like
\[
\ottdruleetXXdcon{}
\]
By the mutual induction hypothesis (with an empty substitution) we get 
$\ottnt{H}  \vdash  \overline{M_i}  \ottsym{:}  \Xi$ and $\ottnt{H}  \vdash  \overline{m_i}  \ottsym{:}  \ottsym{[}  \overline{M_i}  \ottsym{/}  \Xi  \ottsym{]}  \Xi_{\ottmv{i}}$. Conclude by
re-applying \ottdrulename{et\_dcon}.

\item[Case \ottdrulename{et\_app}.] The rule looks like 
\[
\ottdruleetXXapp{}
\]
We consider how the expression $ \ottnt{m}  \;  \ottnt{n} $ may step.
\begin{itemize}
\item To $ \ottnt{m'}  \;  \ottnt{n'} $ by \ottdrulename{sp\_app} if $\ottnt{m}  \leadsto_{\mathsf{p} }  \ottnt{m'}$
  and $\ottnt{n}  \leadsto_{\mathsf{p} }  \ottnt{n'}$.

By IH we have $ \ottnt{H} \vdash \ottnt{m'} :  ( \ottmv{x} \!:\! \ottnt{M} )  \to \,  \ottnt{N}  $ and $ \ottnt{H} \vdash \ottnt{n'} : \ottnt{M} $. By
lemma~\ref{lem:step-subst} we know $\ottsym{[}  \ottnt{n}  \ottsym{/}  \ottmv{x}  \ottsym{]}  \ottnt{N}  \leadsto_{\mathsf{p} }  \ottsym{[}  \ottnt{n'}  \ottsym{/}  \ottmv{x}  \ottsym{]}  \ottnt{N}$, so also
by IH we have $ \ottnt{H} \vdash \ottsym{[}  \ottnt{n'}  \ottsym{/}  \ottmv{x}  \ottsym{]}  \ottnt{N} :  \star  $. So re-applying
\ottdrulename{et\_app} we get $ \ottnt{H} \vdash  \ottnt{m'}  \;  \ottnt{n'}  : \ottsym{[}  \ottnt{n'}  \ottsym{/}  \ottmv{x}  \ottsym{]}  \ottnt{N} $.

Finally, by \ottdrulename{et\_join} we have $ \ottnt{H} \vdash \ottkw{join} : \ottsym{[}  \ottnt{n}  \ottsym{/}  \ottmv{x}  \ottsym{]}  \ottnt{N}  \ottsym{=}  \ottsym{[}  \ottnt{n'}  \ottsym{/}  \ottmv{x}  \ottsym{]}  \ottnt{N} $, and hence by \ottdrulename{et\_conv} we get 
$ \ottnt{H} \vdash  \ottnt{m'}  \;  \ottnt{n'}  : \ottsym{[}  \ottnt{n}  \ottsym{/}  \ottmv{x}  \ottsym{]}  \ottnt{N} $ as required.

\item To $\ottsym{[}  \ottnt{u'}  \ottsym{/}  \ottmv{x}  \ottsym{]}  \ottnt{m'_{{\mathrm{1}}}}$ by \ottdrulename{sp\_appbeta} if $\ottnt{m}$ is $\lambda  \ottmv{x}  \ottsym{.}  \ottnt{m_{{\mathrm{1}}}}$ and $\ottnt{n}$
  is $\ottnt{u}$, and $\ottnt{m_{{\mathrm{1}}}}  \leadsto_{\mathsf{p} }  \ottnt{m'_{{\mathrm{1}}}}$ and $\ottnt{u}  \leadsto_{\mathsf{p} }  \ottnt{u'}$.

By IH we have $ \ottnt{H} \vdash \ottnt{u'} : \ottnt{M} $.
Also, $\lambda  \ottmv{x}  \ottsym{.}  \ottnt{m_{{\mathrm{1}}}}  \leadsto_{\mathsf{p} }  \lambda  \ottmv{x}  \ottsym{.}  \ottnt{m'_{{\mathrm{1}}}}$ so by IH we have 
$ \ottnt{H} \vdash \lambda  \ottmv{x}  \ottsym{.}  \ottnt{m'_{{\mathrm{1}}}} :  ( \ottmv{x} \!:\! \ottnt{M} )  \to \,  \ottnt{N}  $.
By inversion (lemma~\ref{lem:inv-lam}) we know that 
$ \ottnt{H}  \ottsym{,}   \ottmv{x}  :  \ottnt{M_{{\mathrm{1}}}}  \vdash \ottnt{m_{{\mathrm{1}}}} : \ottnt{N_{{\mathrm{1}}}} $ for some $\ottnt{M_{{\mathrm{1}}}}, \ottnt{N_{{\mathrm{1}}}}$ such that 
$ \ottnt{H} \vdash \ottkw{join} :  ( \ottmv{x} \!:\! \ottnt{M_{{\mathrm{1}}}} )  \to \,  \ottnt{N_{{\mathrm{1}}}}  \ottsym{=}   ( \ottmv{x} \!:\! \ottnt{M} )  \to \,  \ottnt{N}   $. By
\ottdrulename{et\_injdom} we have $ \ottnt{H} \vdash \ottkw{join} : \ottnt{M_{{\mathrm{1}}}}  \ottsym{=}  \ottnt{M} $, 
and byregularity (lemma~\ref{lem:reg-var}) we have $ \ottnt{H} \vdash \ottnt{M_{{\mathrm{1}}}} :  \star  $, 
so by \ottdrulename{et\_conv} we get $ \ottnt{H} \vdash \ottnt{u'} : \ottnt{M_{{\mathrm{1}}}} $.
Now by substitution (lemma~\ref{lem:substitution}) we get
\[
 \ottnt{H} \vdash \ottsym{[}  \ottnt{u'}  \ottsym{/}  \ottmv{x}  \ottsym{]}  \ottnt{m'_{{\mathrm{1}}}} : \ottsym{[}  \ottnt{u'}  \ottsym{/}  \ottmv{x}  \ottsym{]}  \ottnt{N_{{\mathrm{1}}}} .
\]

Now, by \ottdrulename{et\_injrng} we have
$ \ottnt{H} \vdash \ottkw{join} : \ottsym{[}  \ottnt{u'}  \ottsym{/}  \ottmv{x}  \ottsym{]}  \ottnt{N_{{\mathrm{1}}}}  \ottsym{=}  \ottsym{[}  \ottnt{u'}  \ottsym{/}  \ottmv{x}  \ottsym{]}  \ottnt{N} $. Also, by
lemma~\ref{lem:step-subst} we know $\ottsym{[}  \ottnt{u}  \ottsym{/}  \ottmv{x}  \ottsym{]}  \ottnt{N}  \leadsto_{\mathsf{p} }  \ottsym{[}  \ottnt{u'}  \ottsym{/}  \ottmv{x}  \ottsym{]}  \ottnt{N}$, and 
we noted above that $ \ottnt{H} \vdash \ottsym{[}  \ottnt{u'}  \ottsym{/}  \ottmv{x}  \ottsym{]}  \ottnt{N} :  \star  $, so by
\ottdrulename{et\_join} we have $ \ottnt{H} \vdash \ottkw{join} : \ottsym{[}  \ottnt{u}  \ottsym{/}  \ottmv{x}  \ottsym{]}  \ottnt{N}  \ottsym{=}  \ottsym{[}  \ottnt{u'}  \ottsym{/}  \ottmv{x}  \ottsym{]}  \ottnt{N} $.
By symmetry and transitivity (lemma~\ref{lem:propeq-eqrel}) this yields
$ \ottnt{H} \vdash \ottkw{join} : \ottsym{[}  \ottnt{u'}  \ottsym{/}  \ottmv{x}  \ottsym{]}  \ottnt{N_{{\mathrm{1}}}}  \ottsym{=}  \ottsym{[}  \ottnt{u}  \ottsym{/}  \ottmv{x}  \ottsym{]}  \ottnt{N} $.

Finally, we had $ \ottnt{H} \vdash \ottsym{[}  \ottnt{u}  \ottsym{/}  \ottmv{x}  \ottsym{]}  \ottnt{N} :  \star  $ as a premise to the rule. So
by \ottdrulename{et\_conv} we get the required
\[
 \ottnt{H} \vdash \ottsym{[}  \ottnt{u'}  \ottsym{/}  \ottmv{x}  \ottsym{]}  \ottnt{m_{{\mathrm{1}}}} : \ottsym{[}  \ottnt{u}  \ottsym{/}  \ottmv{x}  \ottsym{]}  \ottnt{N} .
\]

\item To $ \ottsym{(}  \ottsym{[}   \mathsf{rec}\; \ottmv{f} . \ottkw{u}   \ottsym{/}  \ottmv{f}  \ottsym{]}  \ottnt{u'_{{\mathrm{1}}}}  \ottsym{)}  \;  \ottnt{u'_{{\mathrm{2}}}} $ by \ottdrulename{sp\_apprec}
  if $\ottnt{m}$ is $ \mathsf{rec}\; \ottmv{f} . \ottkw{u} $, $\ottnt{n}$ is $\ottnt{u_{{\mathrm{2}}}}$, and
  $\ottnt{u_{{\mathrm{1}}}}  \leadsto_{\mathsf{p} }  \ottnt{u'_{{\mathrm{1}}}}$ and $\ottnt{u_{{\mathrm{2}}}}  \leadsto_{\mathsf{p} }  \ottnt{u'_{{\mathrm{2}}}}$.

By IH we have $ \ottnt{H} \vdash \ottnt{u_{{\mathrm{2}}}} : \ottnt{M} $. Also, since
$ \mathsf{rec}\; \ottmv{f} . \ottkw{u}   \leadsto_{\mathsf{p} }   \mathsf{rec}\; \ottmv{f} . \ottkw{u} $, by IH we have
$ \ottnt{H} \vdash  \mathsf{rec}\; \ottmv{f} . \ottkw{u}  :  ( \ottmv{x} \!:\! \ottnt{M} )  \to \,  \ottnt{N}  $. And since by
lemma~\ref{lem:step-subst} $\ottsym{[}  \ottnt{u_{{\mathrm{2}}}}  \ottsym{/}  \ottmv{x}  \ottsym{]}  \ottnt{N}  \leadsto_{\mathsf{p} }  \ottsym{[}  \ottnt{u'_{{\mathrm{2}}}}  \ottsym{/}  \ottmv{x}  \ottsym{]}  \ottnt{N}$, by IH we get
$ \ottnt{H} \vdash \ottsym{[}  \ottnt{u'_{{\mathrm{2}}}}  \ottsym{/}  \ottmv{x}  \ottsym{]}  \ottnt{N} :  \star  $.

By inversion (lemma~\ref{lem:inv-rec} we know
$ \ottnt{H}  \ottsym{,}   \ottmv{f}  :  \ottnt{M_{{\mathrm{1}}}}  \vdash \ottnt{u'_{{\mathrm{1}}}} : \ottnt{M_{{\mathrm{1}}}} $ for some $\ottnt{M_{{\mathrm{1}}}}$ such that
$ \ottnt{H} \vdash \ottkw{join} : \ottnt{M_{{\mathrm{1}}}}  \ottsym{=}   ( \ottmv{x} \!:\! \ottnt{M} )  \to \,  \ottnt{N}  $ and $ \ottnt{H} \vdash \ottnt{M_{{\mathrm{1}}}} :  \star  $
and such that $\ottnt{M_{{\mathrm{1}}}}$ is an arrow type.

So by the \ottdrulename{et\_rec} rule, we get
$ \ottnt{H} \vdash  \mathsf{rec}\; \ottmv{f} . \ottkw{u}  : \ottnt{M_{{\mathrm{1}}}} $. Then by substitution
(lemma~\ref{lem:substitution}) we have
$ \ottnt{H} \vdash \ottsym{[}   \mathsf{rec}\; \ottmv{f} . \ottkw{u}   \ottsym{/}  \ottmv{f}  \ottsym{]}  \ottnt{u'_{{\mathrm{1}}}} : \ottnt{M_{{\mathrm{1}}}} $.

By regularity (lemma~\ref{lem:reg}) on the original premise
of the rule we know $ \ottnt{H} \vdash  ( \ottmv{x} \!:\! \ottnt{M} )  \to \,  \ottnt{N}  :  \star  $, so by
\ottdrulename{et\_conv} we have
$ \ottnt{H} \vdash \ottsym{[}   \mathsf{rec}\; \ottmv{f} . \ottkw{u}   \ottsym{/}  \ottmv{f}  \ottsym{]}  \ottnt{u'_{{\mathrm{1}}}} :  ( \ottmv{x} \!:\! \ottnt{M} )  \to \,  \ottnt{N}  $. Then re-apply
\ottdrulename{et\_app} to get
\[
 \ottnt{H} \vdash  \ottsym{(}  \ottsym{[}   \mathsf{rec}\; \ottmv{f} . \ottkw{u}   \ottsym{/}  \ottmv{f}  \ottsym{]}  \ottnt{u'_{{\mathrm{1}}}}  \ottsym{)}  \;  \ottnt{u'_{{\mathrm{2}}}}  : \ottsym{[}  \ottnt{u'_{{\mathrm{2}}}}  \ottsym{/}  \ottmv{x}  \ottsym{]}  \ottnt{N} .
\]

As we noted above $\ottsym{[}  \ottnt{u_{{\mathrm{2}}}}  \ottsym{/}  \ottmv{x}  \ottsym{]}  \ottnt{N}  \leadsto_{\mathsf{p} }  \ottsym{[}  \ottnt{u'_{{\mathrm{2}}}}  \ottsym{/}  \ottmv{x}  \ottsym{]}  \ottnt{N}$, and both expressions
are well-kinded, so by \ottdrulename{et\_join} we know
$ \ottnt{H} \vdash \ottkw{join} : \ottsym{[}  \ottnt{u'_{{\mathrm{2}}}}  \ottsym{/}  \ottmv{x}  \ottsym{]}  \ottnt{N}  \ottsym{=}  \ottsym{[}  \ottnt{u_{{\mathrm{2}}}}  \ottsym{/}  \ottmv{x}  \ottsym{]}  \ottnt{N} $. So by finally applying
\ottdrulename{et\_conv} we get the required
\[
 \ottnt{H} \vdash  \ottsym{(}  \ottsym{[}   \mathsf{rec}\; \ottmv{f} . \ottkw{u}   \ottsym{/}  \ottmv{f}  \ottsym{]}  \ottnt{u'_{{\mathrm{1}}}}  \ottsym{)}  \;  \ottnt{u'_{{\mathrm{2}}}}  : \ottsym{[}  \ottnt{u_{{\mathrm{2}}}}  \ottsym{/}  \ottmv{x}  \ottsym{]}  \ottnt{N} .
\]

\item To $\ottkw{abort}$ by \ottdrulename{sp\_abort}. By regularity
  (lemma~\ref{lem:reg}) on the original premise we know
  $ \ottnt{H} \vdash \ottsym{[}  \ottnt{n}  \ottsym{/}  \ottmv{x}  \ottsym{]}  \ottnt{N} :  \star  $. So by \ottdrulename{et\_abort} we have
  $ \ottnt{H} \vdash \ottkw{abort} : \ottsym{[}  \ottnt{n}  \ottsym{/}  \ottmv{x}  \ottsym{]}  \ottnt{N} $ as required.
\end{itemize}

\item[Case \ottdrulename{et\_iapp}.] The typing rule looks like 
\[
\ottdruleetXXiapp{}
\]
We consider how the expression $\ottnt{m}  \ottsym{[}  \ottsym{]}$ may step:
\begin{itemize}
\item To $\ottnt{m'}  \ottsym{[}  \ottsym{]}$ by \ottdrulename{sp\_iapp} if $\ottnt{m}  \leadsto_{\mathsf{p} }  \ottnt{m'}$.
  By IH we know $ \ottnt{H} \vdash \ottnt{m'} :  [  \ottmv{x} \!:\! \ottnt{M}  ]  \to \,  \ottnt{N}  $, so by re-applying
  \ottdrulename{et\_iapp} we get $ \ottnt{H} \vdash \ottnt{m'}  \ottsym{[}  \ottsym{]} : \ottsym{[}  \ottnt{u}  \ottsym{/}  \ottmv{x}  \ottsym{]}  \ottnt{N} $ as
  required.

\item To $\ottnt{m'_{{\mathrm{1}}}}$ by \ottdrulename{sp\_iappbeta} if $\ottnt{m}$ is
  $\lambda  \ottsym{[}  \ottsym{]}  \ottsym{.}  \ottnt{m_{{\mathrm{1}}}}$ and $\ottnt{m_{{\mathrm{1}}}}  \leadsto_{\mathsf{p} }  \ottnt{m'_{{\mathrm{1}}}}$. 

Note that $\lambda  \ottsym{[}  \ottsym{]}  \ottsym{.}  \ottnt{m_{{\mathrm{1}}}}  \leadsto_{\mathsf{p} }  \lambda  \ottsym{[}  \ottsym{]}  \ottsym{.}  \ottnt{m'_{{\mathrm{1}}}}$, so by IH we get
$ \ottnt{H} \vdash \lambda  \ottsym{[}  \ottsym{]}  \ottsym{.}  \ottnt{m'} :  [  \ottmv{x} \!:\! \ottnt{M}  ]  \to \,  \ottnt{N}  $. Then by inversion
(lemma~\ref{lem:inv-ilam}) we know $ \ottnt{H}  \ottsym{,}   \ottmv{x}  :  \ottnt{M_{{\mathrm{1}}}}  \vdash \ottnt{n} : \ottnt{N_{{\mathrm{1}}}} $
for some $\ottnt{M_{{\mathrm{1}}}}$ and $\ottnt{N_{{\mathrm{1}}}}$ with
$ \ottnt{H} \vdash \ottkw{join} : \ottsym{(}   [  \ottmv{x} \!:\! \ottnt{M_{{\mathrm{1}}}}  ]  \to \,  \ottnt{N_{{\mathrm{1}}}}   \ottsym{)}  \ottsym{=}  \ottsym{(}   [  \ottmv{x} \!:\! \ottnt{M}  ]  \to \,  \ottnt{N}   \ottsym{)} $, and
$ \ottmv{x} \ \notin \ottkw{FV} \, \ottsym{(}  \ottnt{m'_{{\mathrm{1}}}}  \ottsym{)} \ $.

Now, we have $ \ottnt{H} \vdash \ottnt{u} : \ottnt{M} $ as an assumption to the rule. By
regularity (lemma~\ref{lem:reg}) on that assumption we get
$ \ottnt{H} \vdash \ottnt{M} :  \star  $, and by \ottdrulename{et\_iinjdom} we have
$ \ottnt{H} \vdash \ottkw{join} : \ottnt{M_{{\mathrm{1}}}}  \ottsym{=}  \ottnt{M} $. So by \ottdrulename{et\_conv} we get
$ \ottnt{H} \vdash \ottnt{u} : \ottnt{M_{{\mathrm{1}}}} $. Then by substitution
(lemma~\ref{lem:substitution}) we get 
\[
 \ottnt{H} \vdash \ottsym{[}  \ottnt{u}  \ottsym{/}  \ottmv{x}  \ottsym{]}  \ottnt{n} : \ottsym{[}  \ottnt{u}  \ottsym{/}  \ottmv{x}  \ottsym{]}  \ottnt{N_{{\mathrm{1}}}} .
\]
Since we know $\ottmv{x}$ is not free in $\ottnt{n}$ this is the same as
saying $ \ottnt{H} \vdash \ottnt{n} : \ottsym{[}  \ottnt{u}  \ottsym{/}  \ottmv{x}  \ottsym{]}  \ottnt{N_{{\mathrm{1}}}} $. Furthermore, by
\ottdrulename{et\_iinjdom} we get
$ \ottnt{H} \vdash \ottkw{join} : \ottsym{[}  \ottnt{u}  \ottsym{/}  \ottmv{x}  \ottsym{]}  \ottnt{N_{{\mathrm{1}}}}  \ottsym{=}  \ottsym{[}  \ottnt{u}  \ottsym{/}  \ottmv{x}  \ottsym{]}  \ottnt{N} $, and by regularity on the
original derivation we have $ \ottnt{H} \vdash \ottsym{[}  \ottnt{u}  \ottsym{/}  \ottmv{x}  \ottsym{]}  \ottnt{N} :  \star  $. So by
\ottdrulename{et\_conv} we get the required
\[ 
 \ottnt{H} \vdash \ottnt{m'} : \ottsym{[}  \ottnt{u}  \ottsym{/}  \ottmv{x}  \ottsym{]}  \ottnt{N} .
\]

\item To $\ottsym{(}  \ottsym{[}   \mathsf{rec}\; \ottmv{f} . \ottkw{u}   \ottsym{/}  \ottmv{f}  \ottsym{]}  \ottnt{u'_{{\mathrm{1}}}}  \ottsym{)}  \ottsym{[}  \ottsym{]}$ by \ottdrulename{sp\_iapprec}
  if $\ottnt{m}$ is $ \mathsf{rec}\; \ottmv{f} . \ottkw{u} $ and $\ottnt{u_{{\mathrm{1}}}}  \leadsto_{\mathsf{p} }  \ottnt{u'_{{\mathrm{1}}}}$.

Note that $ \mathsf{rec}\; \ottmv{f} . \ottkw{u}   \leadsto_{\mathsf{p} }   \mathsf{rec}\; \ottmv{f} . \ottkw{u} $, so by IH we know
$ \ottnt{H} \vdash  \mathsf{rec}\; \ottmv{f} . \ottkw{u}  :  [  \ottmv{x} \!:\! \ottnt{M}  ]  \to \,  \ottnt{N}  $. By inversion
(lemma~\ref{lem:inv-rec}) we get that $ \ottnt{H}  \ottsym{,}   \ottmv{f}  :  \ottnt{M_{{\mathrm{1}}}}  \vdash \ottnt{u'_{{\mathrm{1}}}} : \ottnt{M_{{\mathrm{1}}}} $ for
some arrow type $\ottnt{M_{{\mathrm{1}}}}$ such that $ \ottnt{H} \vdash \ottkw{join} : \ottnt{M_{{\mathrm{1}}}}  \ottsym{=}   [  \ottmv{x} \!:\! \ottnt{M}  ]  \to \,  \ottnt{N}  $
and $ \ottnt{H} \vdash \ottnt{M_{{\mathrm{1}}}} :  \star  $. By \ottdrulename{et\_rec} we then have
$ \ottnt{H} \vdash  \mathsf{rec}\; \ottmv{f} . \ottkw{u}  : \ottnt{M_{{\mathrm{1}}}} $, hence by substitution
(lemma~\ref{lem:substitution}) we have 
\[
 \ottnt{H} \vdash \ottsym{[}   \mathsf{rec}\; \ottmv{f} . \ottkw{u}   \ottsym{/}  \ottmv{f}  \ottsym{]}  \ottnt{u'_{{\mathrm{1}}}} : \ottnt{M_{{\mathrm{1}}}} .
\]

By regularity (lemma~\ref{lem:reg}) applied to the original typing
rule we know $ \ottnt{H} \vdash  [  \ottmv{x} \!:\! \ottnt{M}  ]  \to \,  \ottnt{N}  :  \star  $, so by \ottdrulename{et\_conv}
we then have 
\[
 \ottnt{H} \vdash \ottsym{[}   \mathsf{rec}\; \ottmv{f} . \ottkw{u}   \ottsym{/}  \ottmv{f}  \ottsym{]}  \ottnt{u'_{{\mathrm{1}}}} :  [  \ottmv{x} \!:\! \ottnt{M}  ]  \to \,  \ottnt{N}  .
\]
So re-applying \ottdrulename{et\_iapp} we get the required
\[
 \ottnt{H} \vdash \ottsym{(}  \ottsym{[}   \mathsf{rec}\; \ottmv{f} . \ottkw{u}   \ottsym{/}  \ottmv{f}  \ottsym{]}  \ottnt{u'_{{\mathrm{1}}}}  \ottsym{)}  \ottsym{[}  \ottsym{]} : \ottsym{[}  \ottnt{u}  \ottsym{/}  \ottmv{x}  \ottsym{]}  \ottnt{N} .
\]

\item To $\ottkw{abort}$ by \ottdrulename{sp\_abort}. 

By regularity (lemma~\ref{lem:reg}) applied to the original type rule
we know $ \ottnt{H} \vdash \ottsym{[}  \ottnt{u}  \ottsym{/}  \ottmv{x}  \ottsym{]}  \ottnt{N} :  \star  $, so by \ottdrulename{et\_abort} we
have 
\[
 \ottnt{H} \vdash \ottkw{abort} : \ottsym{[}  \ottnt{u}  \ottsym{/}  \ottmv{x}  \ottsym{]}  \ottnt{N} 
\]
as required.
\end{itemize}


\item[Case \ottdrulename{et\_eq}.] The rule looks like 
\[
\ottdruleetXXeq{}
\]
The only non-trivial way the expression $\ottnt{m}  \ottsym{=}  \ottnt{n}$ can step is by \ottdrulename{sp\_eq} to
$\ottnt{m'}  \ottsym{=}  \ottnt{n'}$, when $\ottnt{m}  \leadsto_{\mathsf{p} }  \ottnt{m'}$ and $\ottnt{n}  \leadsto_{\mathsf{p} }  \ottnt{n'}$. By IH we
immediatly get $ \ottnt{H} \vdash \ottnt{m} : \ottnt{M} $ and $ \ottnt{H} \vdash \ottnt{n} : \ottnt{N} $, and we
conclude by re-applying \ottdrulename{sp\_eq}.

\item[Case \ottdrulename{et\_conv}.] The rule looks like
\[
\ottdruleetXXconv{}
\]
and we know that $\ottnt{m}  \leadsto_{\mathsf{p} }  \ottnt{m'}$. Directly by IH we get
$ \ottnt{H} \vdash \ottnt{m'} : \ottsym{[}  \ottnt{M_{{\mathrm{1}}}}  \ottsym{/}  \ottmv{x_{{\mathrm{1}}}}  \ottsym{]} \, ... \, \ottsym{[}  \ottnt{M_{\ottmv{i}}}  \ottsym{/}  \ottmv{x_{\ottmv{i}}}  \ottsym{]}  \ottnt{M} $, and conclude by re-applying \ottdrulename{et\_conv}.

\end{description}
The cases for $\ottnt{H}  \vdash  \overline{m_i}  \ottsym{:}  \Xi$ are:
\begin{description}
\item[Case \ottdrulename{etl\_empty}.] Trivial.
\item[Case \ottdrulename{etl\_cons}.] After pushing in the
  substitution, the rule looks like:
\[
\ottdrule{
  \ottpremise{ \ottnt{H} \vdash \ottnt{m} : \ottsym{[}  \ottnt{n_{{\mathrm{1}}}}  \ottsym{/}  \ottmv{y_{{\mathrm{1}}}}  \ottsym{]} \, ... \, \ottsym{[}  \ottnt{n_{\ottmv{i}}}  \ottsym{/}  \ottmv{y_{\ottmv{i}}}  \ottsym{]}  \ottnt{M} }
  \ottpremise{ \ottnt{H} \vdash \ottsym{[}  \ottnt{n_{{\mathrm{1}}}}  \ottsym{/}  \ottmv{y_{{\mathrm{1}}}}  \ottsym{]} \, ... \, \ottsym{[}  \ottnt{n_{\ottmv{i}}}  \ottsym{/}  \ottmv{y_{\ottmv{i}}}  \ottsym{]}  \ottnt{M} :  \star  }
  \ottpremise{\ottnt{H}  \vdash  \overline{m_i}  \ottsym{:}  \ottsym{[}  \ottnt{m}  \ottsym{/}  \ottmv{x}  \ottsym{]}  \ottsym{[}  \ottnt{n_{{\mathrm{1}}}}  \ottsym{/}  \ottmv{y_{{\mathrm{1}}}}  \ottsym{]} \, ... \, \ottsym{[}  \ottnt{n_{\ottmv{i}}}  \ottsym{/}  \ottmv{y_{\ottmv{i}}}  \ottsym{]}  \Xi}
}{
  \ottnt{H}  \vdash   \ottnt{m}  \;  \overline{m_i}   \ottsym{:}  \ottsym{(}  \ottmv{x}  \ottsym{:}  \ottsym{[}  \ottnt{n_{{\mathrm{1}}}}  \ottsym{/}  \ottmv{y_{{\mathrm{1}}}}  \ottsym{]} \, ... \, \ottsym{[}  \ottnt{n_{\ottmv{i}}}  \ottsym{/}  \ottmv{y_{\ottmv{i}}}  \ottsym{]}  \ottnt{M}  \ottsym{)}  \ottsym{[}  \ottnt{n_{{\mathrm{1}}}}  \ottsym{/}  \ottmv{y_{{\mathrm{1}}}}  \ottsym{]} \, ... \, \ottsym{[}  \ottnt{n_{\ottmv{i}}}  \ottsym{/}  \ottmv{y_{\ottmv{i}}}  \ottsym{]}  \Xi
}{
  \ottdrulename{etl\_cons}
}
\]

By mutual IH we have $ \ottnt{H} \vdash \ottnt{m'} : \ottsym{[}  \ottnt{n_{{\mathrm{1}}}}  \ottsym{/}  \ottmv{y_{{\mathrm{1}}}}  \ottsym{]} \, ... \, \ottsym{[}  \ottnt{n_{\ottmv{i}}}  \ottsym{/}  \ottmv{y_{\ottmv{i}}}  \ottsym{]}  \ottnt{M} $.

By repeatedly applying lemma~\ref{lem:step-subst} we know
$\ottsym{[}  \ottnt{n_{{\mathrm{1}}}}  \ottsym{/}  \ottmv{y_{{\mathrm{1}}}}  \ottsym{]} \, ... \, \ottsym{[}  \ottnt{n_{\ottmv{i}}}  \ottsym{/}  \ottmv{y_{\ottmv{i}}}  \ottsym{]}  \ottnt{M}  \leadsto_{\mathsf{p} }  \ottsym{[}  \ottnt{n'_{{\mathrm{1}}}}  \ottsym{/}  \ottmv{y_{{\mathrm{1}}}}  \ottsym{]} \, ... \, \ottsym{[}  \ottnt{n'_{\ottmv{i}}}  \ottsym{/}  \ottmv{y_{\ottmv{i}}}  \ottsym{]}  \ottnt{M}$, so 
by mutual IH we get $ \ottnt{H} \vdash \ottsym{[}  \ottnt{n'_{{\mathrm{1}}}}  \ottsym{/}  \ottmv{y_{{\mathrm{1}}}}  \ottsym{]} \, ... \, \ottsym{[}  \ottnt{n'_{\ottmv{i}}}  \ottsym{/}  \ottmv{y_{\ottmv{i}}}  \ottsym{]}  \ottnt{M} :  \star  $.

By \ottdrulename{et\_join} we then have 
$ \ottnt{H} \vdash \ottkw{join} : \ottsym{[}  \ottnt{n_{{\mathrm{1}}}}  \ottsym{/}  \ottmv{y_{{\mathrm{1}}}}  \ottsym{]} \, ... \, \ottsym{[}  \ottnt{n_{\ottmv{i}}}  \ottsym{/}  \ottmv{y_{\ottmv{i}}}  \ottsym{]}  \ottnt{M}  \ottsym{=}  \ottsym{[}  \ottnt{n'_{{\mathrm{1}}}}  \ottsym{/}  \ottmv{y_{{\mathrm{1}}}}  \ottsym{]} \, ... \, \ottsym{[}  \ottnt{n'_{\ottmv{i}}}  \ottsym{/}  \ottmv{y_{\ottmv{i}}}  \ottsym{]}  \ottnt{M} $, so
by \ottdrulename{et\_conv} we get
$ \ottnt{H} \vdash \ottnt{m} : \ottsym{[}  \ottnt{n'_{{\mathrm{1}}}}  \ottsym{/}  \ottmv{y_{{\mathrm{1}}}}  \ottsym{]} \, ... \, \ottsym{[}  \ottnt{n'_{\ottmv{i}}}  \ottsym{/}  \ottmv{y_{\ottmv{i}}}  \ottsym{]}  \ottnt{M} $.

Finally, by the IH (using that $\ottnt{m}  \leadsto_{\mathsf{p} }  \ottnt{m'}$)  we have
 $\ottnt{H}  \vdash  \overline{m_i}'  \ottsym{:}  \ottsym{[}  \ottnt{m'}  \ottsym{/}  \ottmv{x}  \ottsym{]}  \ottsym{[}  \ottnt{n'_{{\mathrm{1}}}}  \ottsym{/}  \ottmv{y_{{\mathrm{1}}}}  \ottsym{]} \, ... \, \ottsym{[}  \ottnt{n'_{\ottmv{i}}}  \ottsym{/}  \ottmv{y_{\ottmv{i}}}  \ottsym{]}  \Xi$. So
re-applying \ottdrulename{etl\_cons} we get the required
\[
\ottnt{H}  \vdash   \ottnt{m'}  \;  \overline{m_i}'   \ottsym{:}  \ottsym{[}  \ottnt{n'_{{\mathrm{1}}}}  \ottsym{/}  \ottmv{y_{{\mathrm{1}}}}  \ottsym{]} \, ... \, \ottsym{[}  \ottnt{n'_{\ottmv{i}}}  \ottsym{/}  \ottmv{y_{\ottmv{i}}}  \ottsym{]}  \Xi.
\]

\item[Case \ottdrulename{etl\_icons}.] After pushing in the substitution, the rule looks like:
\[
\ottdrule{
  \ottpremise{ \ottnt{H} \vdash \ottnt{u} : \ottsym{[}  \ottnt{n_{{\mathrm{1}}}}  \ottsym{/}  \ottmv{y_{{\mathrm{1}}}}  \ottsym{]} \, ... \, \ottsym{[}  \ottnt{n_{\ottmv{i}}}  \ottsym{/}  \ottmv{y_{\ottmv{i}}}  \ottsym{]}  \ottnt{M} }
  \ottpremise{ \ottnt{H} \vdash \ottsym{[}  \ottnt{n_{{\mathrm{1}}}}  \ottsym{/}  \ottmv{y_{{\mathrm{1}}}}  \ottsym{]} \, ... \, \ottsym{[}  \ottnt{n_{\ottmv{i}}}  \ottsym{/}  \ottmv{y_{\ottmv{i}}}  \ottsym{]}  \ottnt{M} :  \star  }
  \ottpremise{\ottnt{H}  \vdash  \overline{m_i}  \ottsym{:}  \ottsym{[}  \ottnt{u}  \ottsym{/}  \ottmv{x}  \ottsym{]}  \ottsym{[}  \ottnt{n_{{\mathrm{1}}}}  \ottsym{/}  \ottmv{y_{{\mathrm{1}}}}  \ottsym{]} \, ... \, \ottsym{[}  \ottnt{n_{\ottmv{i}}}  \ottsym{/}  \ottmv{y_{\ottmv{i}}}  \ottsym{]}  \Xi}
}{
  \ottnt{H}  \vdash   [] \;  \overline{m_i}   \ottsym{:}  \ottsym{[}  \ottmv{x}  \ottsym{:}  \ottsym{[}  \ottnt{n_{{\mathrm{1}}}}  \ottsym{/}  \ottmv{y_{{\mathrm{1}}}}  \ottsym{]} \, ... \, \ottsym{[}  \ottnt{n_{\ottmv{i}}}  \ottsym{/}  \ottmv{y_{\ottmv{i}}}  \ottsym{]}  \ottnt{M}  \ottsym{]}  \ottsym{[}  \ottnt{n_{{\mathrm{1}}}}  \ottsym{/}  \ottmv{y_{{\mathrm{1}}}}  \ottsym{]} \, ... \, \ottsym{[}  \ottnt{n_{\ottmv{i}}}  \ottsym{/}  \ottmv{y_{\ottmv{i}}}  \ottsym{]}  \Xi
}{
  \ottdrulename{etl\_icons}
}
\]

By repeatedly applying lemma~\ref{lem:step-subst} we know
$\ottsym{[}  \ottnt{n_{{\mathrm{1}}}}  \ottsym{/}  \ottmv{y_{{\mathrm{1}}}}  \ottsym{]} \, ... \, \ottsym{[}  \ottnt{n_{\ottmv{i}}}  \ottsym{/}  \ottmv{y_{\ottmv{i}}}  \ottsym{]}  \ottnt{M}  \leadsto_{\mathsf{p} }  \ottsym{[}  \ottnt{n'_{{\mathrm{1}}}}  \ottsym{/}  \ottmv{y_{{\mathrm{1}}}}  \ottsym{]} \, ... \, \ottsym{[}  \ottnt{n'_{\ottmv{i}}}  \ottsym{/}  \ottmv{y_{\ottmv{i}}}  \ottsym{]}  \ottnt{M}$, so 
by mutual IH we get $ \ottnt{H} \vdash \ottsym{[}  \ottnt{n'_{{\mathrm{1}}}}  \ottsym{/}  \ottmv{y_{{\mathrm{1}}}}  \ottsym{]} \, ... \, \ottsym{[}  \ottnt{n'_{\ottmv{i}}}  \ottsym{/}  \ottmv{y_{\ottmv{i}}}  \ottsym{]}  \ottnt{M} :  \star  $.

By \ottdrulename{et\_join} we then have 
$ \ottnt{H} \vdash \ottkw{join} : \ottsym{[}  \ottnt{n_{{\mathrm{1}}}}  \ottsym{/}  \ottmv{y_{{\mathrm{1}}}}  \ottsym{]} \, ... \, \ottsym{[}  \ottnt{n_{\ottmv{i}}}  \ottsym{/}  \ottmv{y_{\ottmv{i}}}  \ottsym{]}  \ottnt{M}  \ottsym{=}  \ottsym{[}  \ottnt{n'_{{\mathrm{1}}}}  \ottsym{/}  \ottmv{y_{{\mathrm{1}}}}  \ottsym{]} \, ... \, \ottsym{[}  \ottnt{n'_{\ottmv{i}}}  \ottsym{/}  \ottmv{y_{\ottmv{i}}}  \ottsym{]}  \ottnt{M} $, so
by \ottdrulename{et\_conv} we get
$ \ottnt{H} \vdash \ottnt{u} : \ottsym{[}  \ottnt{n'_{{\mathrm{1}}}}  \ottsym{/}  \ottmv{y_{{\mathrm{1}}}}  \ottsym{]} \, ... \, \ottsym{[}  \ottnt{n'_{\ottmv{i}}}  \ottsym{/}  \ottmv{y_{\ottmv{i}}}  \ottsym{]}  \ottnt{M} $.
 
Finally, by the IH (using that $\ottnt{u}  \leadsto_{\mathsf{p} }  \ottnt{u}$, reflexively)  we have
 $\ottnt{H}  \vdash  \overline{m_i}'  \ottsym{:}  \ottsym{[}  \ottnt{u}  \ottsym{/}  \ottmv{x}  \ottsym{]}  \ottsym{[}  \ottnt{n'_{{\mathrm{1}}}}  \ottsym{/}  \ottmv{y_{{\mathrm{1}}}}  \ottsym{]} \, ... \, \ottsym{[}  \ottnt{n'_{\ottmv{i}}}  \ottsym{/}  \ottmv{y_{\ottmv{i}}}  \ottsym{]}  \Xi$. So
re-applying \ottdrulename{etl\_cons} we get the required
\[
\ottnt{H}  \vdash   [] \;  \overline{m_i}'   \ottsym{:}  \ottsym{[}  \ottnt{n'_{{\mathrm{1}}}}  \ottsym{/}  \ottmv{y_{{\mathrm{1}}}}  \ottsym{]} \, ... \, \ottsym{[}  \ottnt{n'_{\ottmv{i}}}  \ottsym{/}  \ottmv{y_{\ottmv{i}}}  \ottsym{]}  \Xi.
\]
\end{description}
\end{proof}

\subsection{Progress}

\begin{lemma}[Soundness of equality]
\label{lem:propeq-sound}
If $ H_{D} \vdash \ottnt{u} : \ottnt{M} $ and $ \ottnt{M} \, \curlyveedownarrow \, \ottsym{(}  \ottnt{m_{{\mathrm{1}}}}  \ottsym{=}  \ottnt{n_{{\mathrm{1}}}}  \ottsym{)} $,
then $ \ottnt{m_{{\mathrm{1}}}} \, \curlyveedownarrow \, \ottnt{n_{{\mathrm{1}}}} $.
\end{lemma}
\begin{proof}
\begin{description}
\item[Cases] \ottdrulename{et\_type}, \ottdrulename{et\_pi},
  \ottdrulename{et\_ipi}, \ottdrulename{et\_tcon},
  \ottdrulename{et\_abstcon}, \ottdrulename{et\_dcon}, 
  \ottdrulename{et\_abs},\ottdrulename{et\_iabs}, \ottdrulename{et\_rec},
  \ottdrulename{et\_eq}.

  The $\ottnt{M}$ in the conclusion of these rules have a defined
  head constructor, which is not $=$.
  So by lemma~\ref{lem:hd-join}, $\ottnt{M}$ cannot be joinable with $\ottnt{m_{{\mathrm{1}}}}  \ottsym{=}  \ottnt{n_{{\mathrm{1}}}}$.
\item[Cases \ottdrulename{et\_case},
  \ottdrulename{et\_app},
  \ottdrulename{et\_iapp}, 
  \ottdrulename{et\_abort}.]
These expressions are not values.

\item[Case \ottdrulename{et\_var}.]
This is impossible in an $H_{D}$ context, since it doesn't contain
any variable declarations. 

\item[Case \ottdrulename{et\_join}.] The rule looks like
\[
\ottdruleetXXjoin{}
\]
By injectivity (lemma~\ref{lem:join-inj}) we have have 
$ \ottnt{m} \, \curlyveedownarrow \, \ottnt{m_{{\mathrm{1}}}} $ and $ \ottnt{n} \, \curlyveedownarrow \, \ottnt{n_{{\mathrm{1}}}} $. And by assumption we have
$ \ottnt{m} \, \curlyveedownarrow \, \ottnt{n} $. So by symmetry and transitivity
(lemma~\ref{lem:join-eqrel}) we have $ \ottnt{m_{{\mathrm{1}}}} \, \curlyveedownarrow \, \ottnt{n_{{\mathrm{1}}}} $ as required.

\item[Case \ottdrulename{et\_conv}.] The rule looks like
\[
\ottdruleetXXconv{}
\]
and we are given that that $ \ottsym{[}  \ottnt{N_{{\mathrm{1}}}}  \ottsym{/}  \ottmv{x_{{\mathrm{1}}}}  \ottsym{]} \, ... \, \ottsym{[}  \ottnt{N_{\ottmv{i}}}  \ottsym{/}  \ottmv{x_{\ottmv{i}}}  \ottsym{]}  \ottnt{M} \, \curlyveedownarrow \, \ottsym{(}  \ottnt{m_{{\mathrm{1}}}}  \ottsym{=}  \ottnt{n_{{\mathrm{1}}}}  \ottsym{)} $.
By the IH for $\ottnt{m}$ it
suffices to show that $ \ottsym{[}  \ottnt{M_{{\mathrm{1}}}}  \ottsym{/}  \ottmv{x_{{\mathrm{1}}}}  \ottsym{]} \, ... \, \ottsym{[}  \ottnt{M_{\ottmv{i}}}  \ottsym{/}  \ottmv{x_{\ottmv{i}}}  \ottsym{]}  \ottnt{M} \, \curlyveedownarrow \, \ottsym{(}  \ottnt{m_{{\mathrm{1}}}}  \ottsym{=}  \ottnt{n_{{\mathrm{1}}}}  \ottsym{)} $.

 But by the IH for $\ottnt{u_{\ottmv{i}}}$ we know $ \ottnt{M_{\ottmv{i}}} \, \curlyveedownarrow \, \ottnt{N_{\ottmv{i}}} $, so we get this by repeatedly 
applying lemma~\ref{lem:join-subst}.

\item[Case \ottdrulename{et\_injrng}.] The rule looks like
\[
\ottdruleetXXinjrng{}
\]
and we are given that $ \ottsym{(}  \ottsym{[}  \ottnt{u}  \ottsym{/}  \ottmv{x}  \ottsym{]}  \ottnt{N_{{\mathrm{1}}}}  \ottsym{=}  \ottsym{[}  \ottnt{u}  \ottsym{/}  \ottmv{x}  \ottsym{]}  \ottnt{N_{{\mathrm{2}}}}  \ottsym{)} \, \curlyveedownarrow \, \ottsym{(}  \ottnt{m_{{\mathrm{1}}}}  \ottsym{=}  \ottnt{n_{{\mathrm{1}}}}  \ottsym{)} $.

By IH we get $  ( \ottmv{x} \!:\! \ottnt{M} )  \to \,  \ottnt{N_{{\mathrm{1}}}}  \, \curlyveedownarrow \,  ( \ottmv{x} \!:\! \ottnt{M} )  \to \,  \ottnt{N_{{\mathrm{2}}}}  $, so by injectivity
(lemma~\ref{lem:join-inj}) we know $ \ottnt{N_{{\mathrm{1}}}} \, \curlyveedownarrow \, \ottnt{N_{{\mathrm{2}}}} $. Then by
lemma~\ref{lem:join-valsubst} we get $ \ottsym{[}  \ottnt{u}  \ottsym{/}  \ottmv{x}  \ottsym{]}  \ottnt{N_{{\mathrm{1}}}} \, \curlyveedownarrow \, \ottsym{[}  \ottnt{u}  \ottsym{/}  \ottmv{x}  \ottsym{]}  \ottnt{N_{{\mathrm{2}}}} $
as required.
\item[Case \ottdrulename{et\_injdom}, \ottdrulename{et\_iinjdom},
  \ottdrulename{et\_iinjrng}, \ottdrulename{et\_injtcon}.] Similar to
  the previous case.
\end{description}

\end{proof}

\begin{lemma}[Canonical forms]
\label{lem:canonical}
Suppose $ H_{D} \vdash \ottnt{u} : \ottnt{M} $. Then: 
\begin{enumerate}
\item If $ \ottnt{M} \, \curlyveedownarrow \,  ( \ottmv{x} \!:\! \ottnt{M_{{\mathrm{1}}}} )  \to \,  \ottnt{M_{{\mathrm{2}}}}  $, then $u$ is either
  $\lambda  \ottmv{x}  \ottsym{.}  \ottnt{u_{{\mathrm{1}}}}$ or $ \mathsf{rec}\; \ottmv{f} . \ottkw{u} $.
\item If $ \ottnt{M} \, \curlyveedownarrow \,  [  \ottmv{x} \!:\! \ottnt{M_{{\mathrm{1}}}}  ]  \to \,  \ottnt{M_{{\mathrm{2}}}}  $, then $u$ is either
  $\lambda  \ottsym{[}  \ottsym{]}  \ottsym{.}  \ottnt{u_{{\mathrm{1}}}}$ or $ \mathsf{rec}\; \ottmv{f} . \ottkw{u} $.
\item If $ \ottnt{M} \, \curlyveedownarrow \, \ottmv{D} \, \overline{M_i} $ then $u$ is $\ottmv{d} \, \overline{u_i}$,
        where $ \ottkw{data} \, \ottmv{D} \, \Xi^{+} \, \ottkw{where} \, \ottsym{\{} \, \ottcomplu{\ottmv{d_{\ottmv{i}}}  \ottsym{:}  \Xi_{\ottmv{i}}  \to \; D \; \Xi^{+}}{\ottmv{i}}{{\mathrm{1}}}{..}{\ottmv{j}} \, \ottsym{\}} \in H_{D} $
        and $\ottmv{d}$ is one of the $\ottmv{d_{\ottmv{i}}}$.
\end{enumerate}
\end{lemma}
\begin{proof}
By induction on $ H_{D} \vdash \ottnt{u} : \ottnt{M} $. The cases are:
\begin{description}
\item[Cases] \ottdrulename{et\_type}, \ottdrulename{et\_pi},
  \ottdrulename{et\_ipi}, \ottdrulename{et\_tcon},
  \ottdrulename{et\_abstcon}, \ottdrulename{et\_eq},
  \ottdrulename{et\_join},  \ottdrulename{et\_injrng},
  \ottdrulename{et\_injdom}, \ottdrulename{et\_iinjdom},
  \ottdrulename{et\_iinjrng}, \ottdrulename{et\_injtcon}.

  The $\ottnt{M}$ in the conclusion of these rules have a defined
  head constructor, which is not one of the interesting ones.
  So by lemma~\ref{lem:hd-join}, $\ottnt{M}$ cannot be joinable with one
  of the interesting types.

\item[Cases \ottdrulename{et\_case},
  \ottdrulename{et\_app},
  \ottdrulename{et\_iapp}, 
  \ottdrulename{et\_abort}.]
These expressions are not values.

\item[Case \ottdrulename{et\_var}.]
This is impossible in an $H_{D}$ context, since it doesn't contain
any variable declarations. 

\item[Cases
  \ottdrulename{et\_dcon},\ottdrulename{et\_abs},\ottdrulename{et\_iabs}.]
The type in these expressions is joinable with one of the interesting
ones, and by lemma~\ref{lem:hd-join} it can be joinable with at most
one of them. The expression in the rule does indeed have the required form.

\item[Case \ottdrulename{et\_rec}.] The rule looks like
\[
\ottdruleetXXrec{}
\]
We know from the side condition to the rule that $\ottnt{M}$ is a
relevant or irrelevant arrow. Then the expression does indeed have the
required form.

\item[Case \ottdrulename{et\_conv}.] The rule looks like
\[
\ottdruleetXXconv{}
\]
Suppose, for example, that $ \ottsym{[}  \ottnt{N_{{\mathrm{1}}}}  \ottsym{/}  \ottmv{x_{{\mathrm{1}}}}  \ottsym{]} \, ... \, \ottsym{[}  \ottnt{N_{\ottmv{i}}}  \ottsym{/}  \ottmv{x_{\ottmv{i}}}  \ottsym{]}  \ottnt{M} \, \curlyveedownarrow \,  ( \ottmv{x} \!:\! \ottnt{M_{{\mathrm{1}}}} )  \to \,  \ottnt{N_{{\mathrm{1}}}}  $.
By the IH for $\ottnt{m}$ ut suffices to show that $ \ottsym{[}  \ottnt{M_{{\mathrm{1}}}}  \ottsym{/}  \ottmv{x_{{\mathrm{1}}}}  \ottsym{]} \, ... \, \ottsym{[}  \ottnt{M_{\ottmv{i}}}  \ottsym{/}  \ottmv{x_{\ottmv{i}}}  \ottsym{]}  \ottnt{M} \, \curlyveedownarrow \,  ( \ottmv{x} \!:\! \ottnt{M_{{\mathrm{1}}}} )  \to \,  \ottnt{N_{{\mathrm{1}}}}  $.

But by soundness of equality (lemma~\ref{lem:propeq-sound}), for each
$\ottnt{u_{\ottmv{i}}}$ we know $ \ottnt{M_{\ottmv{i}}} \, \curlyveedownarrow \, \ottnt{N_{\ottmv{i}}} $, so we get this by repeatedly
applying lemma~\ref{lem:join-subst}.
\end{description}
\end{proof}

\begin{thm}[Progress]
If $ H_{D} \vdash \ottnt{m} : \ottnt{M} $, then either $\ottnt{m}$ is a value, 
$\ottnt{m}$ is $\ottkw{abort}$, or 
$\ottnt{m}  \leadsto_{\mathsf{cbv} }  \ottnt{m'}$ for some $\ottnt{m'}$.
\end{thm}
\begin{proof}
By induction on $ H_{D} \vdash \ottnt{m} : \ottnt{M} $. The cases are:
\begin{description}
\item[Cases] \ottdrulename{et\_type}, 
 \ottdrulename{et\_var},
 \ottdrulename{et\_pi},
 \ottdrulename{et\_ipi}
 \ottdrulename{et\_tcon}
 \ottdrulename{et\_abstcon},
 \ottdrulename{et\_eq},
 \ottdrulename{et\_join},
 \ottdrulename{et\_injdom},
 \ottdrulename{et\_injrng},
 \ottdrulename{et\_iinjdom},
 \ottdrulename{et\_iinjrng},
 \ottdrulename{et\_injtcon},
 \ottdrulename{et\_abs},
 \ottdrulename{et\_iabs},
 \ottdrulename{et\_rec}.

These rules have a value as a subject.

\item[Case \ottdrulename{et\_case}.] The typing rule looks like
\[
\ottdruleetXXcase{}
\]
By IH, we have that $\ottnt{n}$ is either a value, is $\ottkw{abort}$, or
steps. If it steps, the entire expression steps by
\ottdrulename{sc\_ctx}. If it is abort, the entire expression steps to
$\ottkw{abort}$ by \ottdrulename{sc\_abort}.  

Finally, suppose $\ottnt{n}$ is a value. By canonical forms
(lemma~\ref{lem:canonical}) we know that it must be of the form
$\ottmv{d} \, \overline{u_i}$, and defined by a datatype declaration for $\ottmv{D}$
in the context. Since datatype declarations are unique
(lemma~\ref{lem:dcon-unique}), it must be the same datatype
declaration that is mentioned in the typing rule above. So the case
expression has a branch for $\ottmv{d}$, and can step by
\ottdrulename{sc\_casebeta}.


\item[Case \ottdrulename{et\_dcon}.] The expression is
  $\ottmv{d} \, \overline{m_i}$. By IH, each of the $\ottnt{m_{\ottmv{i}}}$ is a values, is
  $\ottkw{abort}$, or steps. If they are all values, the entire
  expression is a value. Otherwise, if the first non-value is
  $\ottkw{abort}$ the entire expression steps by
  \ottdrulename{sc\_abort},  and if it steps the expression steps by
  \ottdrulename{sc\_ctx}. 

\item[Case \ottdrulename{et\_app}.] The rule looks like 
\[
\ottdruleetXXapp{}
\]
By IH, $\ottnt{m}$ and $\ottnt{n}$ either, step, are $\ottkw{abort}$ is are
values. If $\ottnt{m}$ steps, the entire expression steps by
\ottdrulename{sc\_ctx}. If it is abort the entire expression steps by
\ottdrulename{sc\_abort}. So in the following we can assume it is a value.

By similar reasoning, we can assume $\ottnt{n}$ is a value.

Now, by canonical forms (lemma~\ref{lem:canonical}) we know that
$\ottnt{m}$ is either $\lambda  \ottmv{x}  \ottsym{.}  \ottnt{m_{{\mathrm{1}}}}$ or $ \mathsf{rec}\; \ottmv{f} . \ottkw{u} $. If it is 
$\lambda  \ottmv{x}  \ottsym{.}  \ottnt{m_{{\mathrm{1}}}}$ the entire expression steps to $\ottsym{[}  \ottnt{n}  \ottsym{/}  \ottmv{x}  \ottsym{]}  \ottnt{m_{{\mathrm{1}}}}$ by
\ottdrulename{sc\_appbeta}, while if it is $ \mathsf{rec}\; \ottmv{f} . \ottkw{u} $
the entire expression steps to $ \ottsym{(}  \ottsym{[}   \mathsf{rec}\; \ottmv{f} . \ottkw{u}   \ottsym{/}  \ottmv{f}  \ottsym{]}  \ottnt{u_{{\mathrm{1}}}}  \ottsym{)}  \;  \ottnt{n} $ by
\ottdrulename{sc\_apprec}.

\item[Case \ottdrulename{et\_iapp}.] The rule looks like
\[
\ottdruleetXXiapp{}
\]
By the IH, $\ottnt{m}$ either steps, is $\ottkw{abort}$ or is a value. If
$\ottnt{m}$ steps, then the entire expresions steps by
\ottdrulename{sc\_ctx} and the context $\bullet  \ottsym{[]}$. If it is
$\ottkw{abort}$, the entire expression steps to abort by
\ottdrulename{sc\_abort} and the same context.

Finally, $\ottnt{m}$ may be a value. In that case, by canonical forms
(lemma~\ref{lem:canonical}), $\ottnt{m}$ is either $\lambda  \ottsym{[}  \ottsym{]}  \ottsym{.}  \ottnt{m_{{\mathrm{1}}}}$ or 
$ \mathsf{rec}\; \ottmv{f} . \ottkw{u} $. If it is $\lambda  \ottsym{[}  \ottsym{]}  \ottsym{.}  \ottnt{m_{{\mathrm{1}}}}$, the entire expression steps
to $\ottnt{m_{{\mathrm{1}}}}$ by \ottdrulename{sc\_iappbeta}. If it is $ \mathsf{rec}\; \ottmv{f} . \ottkw{u} $, then the entire expression steps to $\ottsym{(}  \ottsym{[}   \mathsf{rec}\; \ottmv{f} . \ottkw{u}   \ottsym{/}  \ottmv{f}  \ottsym{]}  \ottnt{u_{{\mathrm{1}}}}  \ottsym{)}  \ottsym{[}  \ottsym{]}$
by \ottdrulename{sc\_iapprec}.

\item[Case \ottdrulename{et\_abort}.] The subject of the typing rule
  is $\ottkw{abort}$.
\item[Case \ottdrulename{et\_conv}.] Follows directly by IH.
\end{description}
\end{proof}

\subsection{Regularity and substitution for the annotated language}
\label{sec:ann-theory}

While not needed for the type safety proof, in this section we supply
proofs of regularity and substitution for the \emph{annotated}
language. This is of interest because it proves that the
``value-dependent'' application rule is admissible in our system.

\begin{lemma}
\label{lem:ann-typing-fv}
If $ \Gamma \vdash \ottnt{a} : \ottnt{A} $, then $\ottkw{FV} \, \ottsym{(}  \ottnt{a}  \ottsym{)} \subseteq \ottkw{dom} \, \ottsym{(}  \Gamma  \ottsym{)}$ and
$\ottkw{FV} \, \ottsym{(}  \ottnt{A}  \ottsym{)} \subseteq \ottkw{dom} \, \ottsym{(}  \Gamma  \ottsym{)}$.
\end{lemma}

\begin{lemma}[Weakening for $ \vdash   \Gamma $.]
\label{lem:ann-weaken-envwf}
If $ \vdash   \Gamma  \ottsym{,}  \Gamma' $ then $ \vdash   \Gamma $.
\end{lemma}

\begin{lemma}[Weakening for the annotated language]
\label{lem:ann-weakening}
If $ \Gamma \vdash \ottnt{a} : \ottnt{A} $ and $ \vdash   \Gamma  \ottsym{,}  \Gamma' $ then $ \Gamma  \ottsym{,}  \Gamma' \vdash \ottnt{a} : \ottnt{A} $.
\end{lemma}

\begin{lemma}[Substitution commutes with erasure]
\label{lem:erase-subst}
We always have $\ottsym{\mbox{$\mid$}}  \ottsym{[}  \ottnt{a}  \ottsym{/}  \ottmv{x}  \ottsym{]}  \ottnt{b}  \ottsym{\mbox{$\mid$}} = \ottsym{[}  \ottsym{\mbox{$\mid$}}  \ottnt{a}  \ottsym{\mbox{$\mid$}}  \ottsym{/}  \ottmv{x}  \ottsym{]}  \ottsym{\mbox{$\mid$}}  \ottnt{b}  \ottsym{\mbox{$\mid$}}$.
\end{lemma}
\begin{proof}
By induction on $\ottnt{b}$.
\end{proof}

\begin{lemma}
\label{lem:cstep-subst}
If $\ottnt{m}  \leadsto_{\mathsf{cbv} }  \ottnt{m'}$, then $\ottsym{[}  \ottnt{u_{{\mathrm{0}}}}  \ottsym{/}  \ottmv{x_{{\mathrm{0}}}}  \ottsym{]}  \ottnt{m}  \leadsto_{\mathsf{cbv} }  \ottsym{[}  \ottnt{u_{{\mathrm{0}}}}  \ottsym{/}  \ottmv{x_{{\mathrm{0}}}}  \ottsym{]}  \ottnt{m'}$.
\end{lemma}
\begin{proof}
By induction on $\ottnt{m}  \leadsto_{\mathsf{cbv} }  \ottnt{m'}$. The cases are:
\begin{description}
\item [\ottdrulename{sc\_appbeta}.] The assumed step is
  $ \ottsym{(}  \lambda  \ottmv{x}  \ottsym{.}  \ottnt{m}  \ottsym{)}  \;  \ottnt{u}   \leadsto_{\mathsf{cbv} }  \ottsym{[}  \ottnt{u}  \ottsym{/}  \ottmv{x}  \ottsym{]}  \ottnt{m}$, and we must show
  $\ottsym{[}  \ottnt{u_{{\mathrm{0}}}}  \ottsym{/}  \ottmv{x_{{\mathrm{0}}}}  \ottsym{]}  \ottsym{(}   \ottsym{(}  \lambda  \ottmv{x}  \ottsym{.}  \ottnt{m}  \ottsym{)}  \;  \ottnt{u}   \ottsym{)}  \leadsto_{\mathsf{cbv} }  \ottsym{[}  \ottnt{u_{{\mathrm{0}}}}  \ottsym{/}  \ottmv{x_{{\mathrm{0}}}}  \ottsym{]}  \ottsym{[}  \ottnt{u}  \ottsym{/}  \ottmv{x}  \ottsym{]}  \ottnt{m}$.

Pushing the substitution down we know $\ottsym{[}  \ottnt{u_{{\mathrm{0}}}}  \ottsym{/}  \ottmv{x_{{\mathrm{0}}}}  \ottsym{]}  \ottsym{(}   \ottsym{(}  \lambda  \ottmv{x}  \ottsym{.}  \ottnt{m}  \ottsym{)}  \;  \ottnt{u}   \ottsym{)} =
 \ottsym{(}  \lambda  \ottmv{x}  \ottsym{.}  \ottsym{[}  \ottnt{u_{{\mathrm{0}}}}  \ottsym{/}  \ottmv{x_{{\mathrm{0}}}}  \ottsym{]}  \ottnt{m}  \ottsym{)}  \;  \ottsym{(}  \ottsym{[}  \ottnt{u_{{\mathrm{0}}}}  \ottsym{/}  \ottmv{x_{{\mathrm{0}}}}  \ottsym{]}  \ottnt{u}  \ottsym{)} $, which steps to
$\ottsym{[}  \ottsym{[}  \ottnt{u_{{\mathrm{0}}}}  \ottsym{/}  \ottmv{x_{{\mathrm{0}}}}  \ottsym{]}  \ottnt{u}  \ottsym{/}  \ottmv{x}  \ottsym{]}  \ottsym{[}  \ottnt{u_{{\mathrm{0}}}}  \ottsym{/}  \ottmv{x_{{\mathrm{0}}}}  \ottsym{]}  \ottnt{m}$. Since $x$ is a bound variable we can
pick it so that $ \ottmv{x} \ \notin \ottkw{FV} \, \ottsym{(}  \ottnt{u_{{\mathrm{0}}}}  \ottsym{)} \ $. Then
$\ottsym{[}  \ottsym{[}  \ottnt{u_{{\mathrm{0}}}}  \ottsym{/}  \ottmv{x_{{\mathrm{0}}}}  \ottsym{]}  \ottnt{u}  \ottsym{/}  \ottmv{x}  \ottsym{]}  \ottsym{[}  \ottnt{u_{{\mathrm{0}}}}  \ottsym{/}  \ottmv{x_{{\mathrm{0}}}}  \ottsym{]}  \ottnt{m} = \ottsym{[}  \ottnt{u_{{\mathrm{0}}}}  \ottsym{/}  \ottmv{x_{{\mathrm{0}}}}  \ottsym{]}  \ottsym{[}  \ottnt{u}  \ottsym{/}  \ottmv{x}  \ottsym{]}  \ottnt{m}$ as required.

\item [\ottdrulename{sc\_casebeta}.] Similar to the previous case.

\item [\ottdrulename{sc\_apprec}.] The assumed step is
  $ \ottsym{(}   \mathsf{rec}\; \ottmv{f} . \ottkw{u}   \ottsym{)}  \;  \ottnt{u_{{\mathrm{2}}}}   \leadsto_{\mathsf{cbv} }   \ottsym{(}  \ottsym{[}   \mathsf{rec}\; \ottmv{f} . \ottkw{u}   \ottsym{/}  \ottmv{f}  \ottsym{]}  \ottnt{u_{{\mathrm{1}}}}  \ottsym{)}  \;  \ottnt{u_{{\mathrm{2}}}} $, and we must show
\[
\ottsym{[}  \ottnt{u_{{\mathrm{0}}}}  \ottsym{/}  \ottmv{x_{{\mathrm{0}}}}  \ottsym{]}  \ottsym{(}   \ottsym{(}   \mathsf{rec}\; \ottmv{f} . \ottkw{u}   \ottsym{)}  \;  \ottnt{u_{{\mathrm{2}}}}   \ottsym{)}  \leadsto_{\mathsf{cbv} }  \ottsym{[}  \ottnt{u_{{\mathrm{0}}}}  \ottsym{/}  \ottmv{x_{{\mathrm{0}}}}  \ottsym{]}  \ottsym{(}   \ottsym{(}  \ottsym{[}   \mathsf{rec}\; \ottmv{f} . \ottkw{u}   \ottsym{/}  \ottmv{f}  \ottsym{]}  \ottnt{u_{{\mathrm{1}}}}  \ottsym{)}  \;  \ottnt{u_{{\mathrm{2}}}}   \ottsym{)}.
\]

Pushing down the subsitution we know $\ottsym{[}  \ottnt{u_{{\mathrm{0}}}}  \ottsym{/}  \ottmv{x_{{\mathrm{0}}}}  \ottsym{]}  \ottsym{(}   \ottsym{(}   \mathsf{rec}\; \ottmv{f} . \ottkw{u}   \ottsym{)}  \;  \ottnt{u_{{\mathrm{2}}}}   \ottsym{)} =
 \ottsym{(}   \mathsf{rec}\; \ottmv{f} . \ottkw{u}   \ottsym{)}  \;  \ottsym{(}  \ottsym{[}  \ottnt{u_{{\mathrm{0}}}}  \ottsym{/}  \ottmv{x_{{\mathrm{0}}}}  \ottsym{]}  \ottnt{u_{{\mathrm{2}}}}  \ottsym{)} $, which steps to
$ \ottsym{(}  \ottsym{[}  \ottsym{(}   \mathsf{rec}\; \ottmv{f} . \ottkw{u}   \ottsym{)}  \ottsym{/}  \ottmv{f}  \ottsym{]}  \ottsym{[}  \ottnt{u_{{\mathrm{0}}}}  \ottsym{/}  \ottmv{x_{{\mathrm{0}}}}  \ottsym{]}  \ottnt{u_{{\mathrm{1}}}}  \ottsym{)}  \;  \ottsym{(}  \ottsym{[}  \ottnt{u_{{\mathrm{0}}}}  \ottsym{/}  \ottmv{x_{{\mathrm{0}}}}  \ottsym{]}  \ottnt{u_{{\mathrm{2}}}}  \ottsym{)} $. 

By picking the bound variable $f$ so that $ \ottmv{f} \ \notin \ottkw{FV} \, \ottsym{(}  \ottnt{u_{{\mathrm{0}}}}  \ottsym{)} \ $ we
have $\ottsym{[}   \mathsf{rec}\; \ottmv{f} . \ottkw{u}   \ottsym{/}  \ottmv{f}  \ottsym{]}  \ottsym{[}  \ottnt{u_{{\mathrm{0}}}}  \ottsym{/}  \ottmv{x_{{\mathrm{0}}}}  \ottsym{]}  \ottnt{u_{{\mathrm{1}}}} = \ottsym{[}  \ottnt{u_{{\mathrm{0}}}}  \ottsym{/}  \ottmv{x_{{\mathrm{0}}}}  \ottsym{]}  \ottsym{[}   \mathsf{rec}\; \ottmv{f} . \ottkw{u}   \ottsym{/}  \ottmv{f}  \ottsym{]}  \ottnt{u_{{\mathrm{1}}}}$
as required.

\item [\ottdrulename{sc\_iapprec}.] Similar to the previous case.

\item [\ottdrulename{sc\_iappbeta}, \ottdrulename{sc\_abort}, \ottdrulename{sc\_ctx}.] Immediate by just pushing in the substitution.
\end{description}
\end{proof}

\begin{lemma}
\label{lem:ann-subst-step}
If $ \ottsym{\mbox{$\mid$}}  \ottnt{a}  \ottsym{\mbox{$\mid$}}   \leadsto_{\mathsf{cbv} }  ^{ \ottmv{i} }  \ottnt{n} $, then $ \ottsym{\mbox{$\mid$}}  \ottsym{[}  \ottnt{v}  \ottsym{/}  \ottmv{x}  \ottsym{]}  \ottnt{a}  \ottsym{\mbox{$\mid$}}   \leadsto_{\mathsf{cbv} }  ^{ \ottmv{i} }  \ottsym{[}  \ottsym{\mbox{$\mid$}}  \ottnt{v}  \ottsym{\mbox{$\mid$}}  \ottsym{/}  \ottmv{x}  \ottsym{]}  \ottnt{n} $.  
\end{lemma}
\begin{proof}
By commuting the substitution (lemma \ref{lem:erase-subst}) we know
$\ottsym{\mbox{$\mid$}}  \ottsym{[}  \ottnt{v}  \ottsym{/}  \ottmv{x}  \ottsym{]}  \ottnt{a}  \ottsym{\mbox{$\mid$}} = \ottsym{[}  \ottsym{\mbox{$\mid$}}  \ottnt{v}  \ottsym{\mbox{$\mid$}}  \ottsym{/}  \ottmv{x}  \ottsym{]}  \ottsym{\mbox{$\mid$}}  \ottnt{a}  \ottsym{\mbox{$\mid$}}$. Then apply lemma
\ref{lem:cstep-subst} to each step of the reduction sequence
$ \ottsym{\mbox{$\mid$}}  \ottnt{a}  \ottsym{\mbox{$\mid$}}   \leadsto_{\mathsf{cbv} }  ^{ \ottmv{i} }  \ottnt{n} $.
\end{proof}

\begin{lemma}[Substitution for the annotated language]
\label{lem:ann-subst}
Suppose $ \Gamma_{{\mathrm{1}}} \vdash \ottnt{v_{{\mathrm{1}}}} : \ottnt{A_{{\mathrm{1}}}} $. Then,
\begin{enumerate}
\item If $ \Gamma_{{\mathrm{1}}}  \ottsym{,}   \ottmv{x_{{\mathrm{1}}}}  :  \ottnt{A_{{\mathrm{1}}}}   \ottsym{,}  \Gamma_{{\mathrm{2}}} \vdash \ottnt{a} : \ottnt{A} $, then
 $ \Gamma_{{\mathrm{1}}}  \ottsym{,}  \ottsym{[}  \ottnt{v_{{\mathrm{1}}}}  \ottsym{/}  \ottmv{x_{{\mathrm{1}}}}  \ottsym{]}  \Gamma_{{\mathrm{2}}} \vdash \ottsym{[}  \ottnt{v_{{\mathrm{1}}}}  \ottsym{/}  \ottmv{x_{{\mathrm{1}}}}  \ottsym{]}  \ottnt{a} : \ottsym{[}  \ottnt{v_{{\mathrm{1}}}}  \ottsym{/}  \ottmv{x_{{\mathrm{1}}}}  \ottsym{]}  \ottnt{A} $.
\item If $ \vdash   \Gamma_{{\mathrm{1}}}  \ottsym{,}   \ottmv{x_{{\mathrm{1}}}}  :  \ottnt{A_{{\mathrm{1}}}}   \ottsym{,}  \Gamma_{{\mathrm{2}}} $, then $ \vdash   \Gamma_{{\mathrm{1}}}  \ottsym{,}  \ottsym{[}  \ottnt{v_{{\mathrm{1}}}}  \ottsym{/}  \ottmv{x_{{\mathrm{1}}}}  \ottsym{]}  \Gamma_{{\mathrm{2}}} $.
\end{enumerate}
\end{lemma}
\begin{proof}
By mutual induction on $ \Gamma_{{\mathrm{1}}}  \ottsym{,}   \ottmv{x_{{\mathrm{1}}}}  :  \ottnt{A_{{\mathrm{1}}}}   \ottsym{,}  \Gamma_{{\mathrm{2}}} \vdash \ottnt{a} : \ottnt{A} $ and
$ \vdash   \Gamma_{{\mathrm{1}}}  \ottsym{,}   \ottmv{x_{{\mathrm{1}}}}  :  \ottnt{A_{{\mathrm{1}}}}   \ottsym{,}  \Gamma_{{\mathrm{2}}} $. Most cases follow
directly by IH. Two interesting cases are:
\begin{description}
\item[Case \ottdrulename{t\_var}.] We get $ \vdash   \Gamma_{{\mathrm{1}}}  \ottsym{,}  \ottsym{[}  \ottnt{v_{{\mathrm{1}}}}  \ottsym{/}  \ottmv{x_{{\mathrm{1}}}}  \ottsym{]}  \Gamma_{{\mathrm{2}}} $ by
  the mutual IH. Then do a case-split on where in the context $\ottmv{x}$
  occurs:

\begin{itemize}
\item If $  \ottmv{x}  :  \ottnt{A}   \in  \Gamma_{{\mathrm{1}}} $, then by \ottdrulename{t\_var} we have
$ \Gamma_{{\mathrm{1}}}  \ottsym{,}  \ottsym{[}  \ottnt{v_{{\mathrm{1}}}}  \ottsym{/}  \ottmv{x_{{\mathrm{1}}}}  \ottsym{]}  \Gamma_{{\mathrm{2}}} \vdash \ottmv{x} : \ottnt{A} $. 

By  $ \vdash   \Gamma_{{\mathrm{1}}}  \ottsym{,}   \ottmv{x_{{\mathrm{1}}}}  :  \ottnt{A_{{\mathrm{1}}}}   \ottsym{,}  \Gamma_{{\mathrm{2}}} $ we
know $ \Gamma_{{\mathrm{0}}} \vdash \ottnt{A} :  \star  $ for some prefix $\Gamma_{{\mathrm{0}}}$ of $\Gamma_{{\mathrm{1}}}$, so
in particular by lemma~\ref{lem:ann-typing-fv} we know $\ottkw{FV} \, \ottsym{(}  \ottnt{A}  \ottsym{)}
\subseteq \ottkw{dom} \, \ottsym{(}  \Gamma_{{\mathrm{0}}}  \ottsym{)}$, so $\ottmv{x_{{\mathrm{1}}}} \not\in \ottkw{FV} \, \ottsym{(}  \ottnt{A}  \ottsym{)}$. 

Also, by $ \vdash   \Gamma_{{\mathrm{1}}}  \ottsym{,}   \ottmv{x_{{\mathrm{1}}}}  :  \ottnt{A_{{\mathrm{1}}}}   \ottsym{,}  \Gamma_{{\mathrm{2}}} $ we know $x \neq x_1$. So
$\ottsym{[}  \ottnt{v_{{\mathrm{1}}}}  \ottsym{/}  \ottmv{x_{{\mathrm{1}}}}  \ottsym{]}  \ottmv{x} = x$ and $\ottsym{[}  \ottnt{v_{{\mathrm{1}}}}  \ottsym{/}  \ottmv{x_{{\mathrm{1}}}}  \ottsym{]}  \ottnt{A} = \ottnt{A}$, so we have showed
$ \Gamma  \ottsym{,}  \ottsym{[}  \ottnt{v_{{\mathrm{1}}}}  \ottsym{/}  \ottmv{x_{{\mathrm{1}}}}  \ottsym{]}  \Gamma_{{\mathrm{2}}} \vdash \ottsym{[}  \ottnt{v_{{\mathrm{1}}}}  \ottsym{/}  \ottmv{x_{{\mathrm{1}}}}  \ottsym{]}  \ottmv{x} : \ottsym{[}  \ottnt{v_{{\mathrm{1}}}}  \ottsym{/}  \ottmv{x_{{\mathrm{1}}}}  \ottsym{]}  \ottnt{A} $ as required.

\item If $\ottmv{x} = \ottmv{x_{{\mathrm{1}}}}$, then $\ottsym{[}  \ottnt{v_{{\mathrm{1}}}}  \ottsym{/}  \ottmv{x_{{\mathrm{1}}}}  \ottsym{]}  \ottmv{x} = \ottnt{v_{{\mathrm{1}}}}$, so by
  assumption we have  $ \Gamma_{{\mathrm{1}}} \vdash \ottsym{[}  \ottnt{v_{{\mathrm{1}}}}  \ottsym{/}  \ottmv{x_{{\mathrm{1}}}}  \ottsym{]}  \ottmv{x} : \ottnt{A_{{\mathrm{1}}}} $. By the assumption
  $ \vdash   \Gamma_{{\mathrm{1}}}  \ottsym{,}   \ottmv{x_{{\mathrm{1}}}}  :  \ottnt{A_{{\mathrm{1}}}}   \ottsym{,}  \Gamma_{{\mathrm{2}}} $ we know that $\ottmv{x_{{\mathrm{1}}}}$ is not free in
  $\ottnt{A_{{\mathrm{1}}}}$, so $\ottsym{[}  \ottnt{v_{{\mathrm{1}}}}  \ottsym{/}  \ottmv{x_{{\mathrm{1}}}}  \ottsym{]}  \ottnt{A} = \ottnt{A_{{\mathrm{1}}}}$ and so we have shown
  $ \Gamma_{{\mathrm{1}}} \vdash \ottsym{[}  \ottnt{v_{{\mathrm{1}}}}  \ottsym{/}  \ottmv{x_{{\mathrm{1}}}}  \ottsym{]}  \ottmv{x} : \ottsym{[}  \ottnt{v_{{\mathrm{1}}}}  \ottsym{/}  \ottmv{x_{{\mathrm{1}}}}  \ottsym{]}  \ottnt{A} $. Finally by weakening (lemma
  \ref{lem:ann-weakening}) we have
  $ \Gamma_{{\mathrm{1}}}  \ottsym{,}  \ottsym{[}  \ottnt{v_{{\mathrm{1}}}}  \ottsym{/}  \ottmv{x_{{\mathrm{1}}}}  \ottsym{]}  \Gamma_{{\mathrm{2}}} \vdash \ottsym{[}  \ottnt{v_{{\mathrm{1}}}}  \ottsym{/}  \ottmv{x_{{\mathrm{1}}}}  \ottsym{]}  \ottmv{x} : \ottsym{[}  \ottnt{v_{{\mathrm{1}}}}  \ottsym{/}  \ottmv{x_{{\mathrm{1}}}}  \ottsym{]}  \ottnt{A} $ as required.

\item If $  \ottmv{x}  :  \ottnt{A}   \in  \Gamma_{{\mathrm{2}}} $, then $  \ottmv{x}  :  \ottsym{[}  \ottnt{v_{{\mathrm{1}}}}  \ottsym{/}  \ottmv{x_{{\mathrm{1}}}}  \ottsym{]}  \ottnt{A}   \in  \ottsym{[}  \ottnt{v_{{\mathrm{1}}}}  \ottsym{/}  \ottmv{x_{{\mathrm{1}}}}  \ottsym{]}  \Gamma_{{\mathrm{2}}} $,
  so we have $ \Gamma_{{\mathrm{1}}}  \ottsym{,}  \ottsym{[}  \ottnt{v_{{\mathrm{1}}}}  \ottsym{/}  \ottmv{x_{{\mathrm{1}}}}  \ottsym{]}  \Gamma_{{\mathrm{2}}} \vdash \ottmv{x} : \ottsym{[}  \ottnt{v_{{\mathrm{1}}}}  \ottsym{/}  \ottmv{x_{{\mathrm{1}}}}  \ottsym{]}  \ottnt{A} $ by
  \ottdrulename{t\_var}. By the same reasoning as above we know $\ottmv{x_{{\mathrm{1}}}}
  \neq \ottmv{x}$, so this shows
  $ \Gamma_{{\mathrm{1}}}  \ottsym{,}  \ottsym{[}  \ottnt{v_{{\mathrm{1}}}}  \ottsym{/}  \ottmv{x_{{\mathrm{1}}}}  \ottsym{]}  \Gamma_{{\mathrm{2}}} \vdash \ottsym{[}  \ottnt{v_{{\mathrm{1}}}}  \ottsym{/}  \ottmv{x_{{\mathrm{1}}}}  \ottsym{]}  \ottmv{x} : \ottsym{[}  \ottnt{v_{{\mathrm{1}}}}  \ottsym{/}  \ottmv{x_{{\mathrm{1}}}}  \ottsym{]}  \ottnt{A} $ as required.
\end{itemize}

\item[Case \ottdrulename{t\_join}.] The typing rule looks like
\[
\ottdruletXXjoin{}
\]
By lemma \ref{lem:ann-subst-step} we get $ \ottsym{\mbox{$\mid$}}  \ottsym{[}  \ottnt{v_{{\mathrm{1}}}}  \ottsym{/}  \ottmv{x_{{\mathrm{1}}}}  \ottsym{]}  \ottnt{a}  \ottsym{\mbox{$\mid$}}   \leadsto_{\mathsf{cbv} }  ^{ \ottmv{i} }  \ottnt{n} $
and $ \ottsym{\mbox{$\mid$}}  \ottsym{[}  \ottnt{v_{{\mathrm{1}}}}  \ottsym{/}  \ottmv{x}  \ottsym{]}  \ottnt{b}  \ottsym{\mbox{$\mid$}}   \leadsto_{\mathsf{cbv} }  ^{ \ottmv{i} }  \ottnt{n} $. By IH we have
$ \Gamma_{{\mathrm{1}}}  \ottsym{,}  \ottsym{[}  \ottnt{v_{{\mathrm{1}}}}  \ottsym{/}  \ottmv{x_{{\mathrm{1}}}}  \ottsym{]}  \Gamma_{{\mathrm{2}}} \vdash \ottsym{[}  \ottnt{v_{{\mathrm{1}}}}  \ottsym{/}  \ottmv{x_{{\mathrm{1}}}}  \ottsym{]}  \ottsym{(}  \ottnt{a}  \ottsym{=}  \ottnt{b}  \ottsym{)} :  \star  $. Then re-apply \ottdrulename{t\_join}.
\end{description}
\end{proof}

\begin{lemma}[Regularity inversion for the annotated language]
\hspace*{\fill} \\[-12pt]
\label{lem:ann-inv-reg}
\begin{enumerate}
\item If $ \Gamma \vdash  ( \ottmv{x} \!:\! \ottnt{A} )  \to \,  \ottnt{B}  : \ottnt{A_{{\mathrm{0}}}} $ for some $\ottnt{A_{{\mathrm{0}}}}$, then
  $ \Gamma \vdash \ottnt{A} :  \star  $ and $ \Gamma  \ottsym{,}   \ottmv{x}  :  \ottnt{A}  \vdash \ottnt{B} :  \star  $.
\item If $ \Gamma \vdash  [  \ottmv{x} \!:\! \ottnt{A}  ]  \to \,  \ottnt{B}  : \ottnt{A_{{\mathrm{0}}}} $ for some $\ottnt{A_{{\mathrm{0}}}}$, then
  $ \Gamma \vdash \ottnt{A} :  \star  $ and $ \Gamma  \ottsym{,}   \ottmv{x}  :  \ottnt{A}  \vdash \ottnt{B} :  \star  $.
\item If $ \Gamma \vdash \ottnt{a}  \ottsym{=}  \ottnt{b} : \ottnt{A_{{\mathrm{0}}}} $ for some $\ottnt{A_{{\mathrm{0}}}}$, then $ \Gamma \vdash \ottnt{a} : \ottnt{A} $ and
  $ \Gamma \vdash \ottnt{b} : \ottnt{B} $ for some types $\ottnt{A}$ and $\ottnt{B}$.
\end{enumerate}
\end{lemma}
\begin{proof}
By induction on the assumed typing derivation. The only rules that can
apply are the intro rule, which has the required statements as
assumptions, and conversion, which goes directly by IH.
\end{proof}

\begin{lemma}[Regularity for the annotated language]
\label{lem:ann-regularity}
If $ \Gamma \vdash \ottnt{a} : \ottnt{A} $, then $ \Gamma \vdash \ottnt{A} :  \star  $ and $ \vdash   \Gamma $.
\end{lemma}
\begin{proof}
Induction on $ \Gamma \vdash \ottnt{a} : \ottnt{A} $. The cases are:
\begin{description}
\item[Cases] \ottdrulename{t\_type}, \ottdrulename{t\_pi},
  \ottdrulename{t\_ipi}, \ottdrulename{t\_tcon},
  \ottdrulename{t\_abstcon}, \ottdrulename{t\_eq}.

By \ottdrulename{t\_type} we have $ \Gamma \vdash  \star  :  \star  $ as required. We
get $ \vdash   \Gamma $ by IH (or assumption in the \ottdrulename{type} case). 

\item[Cases] \ottdrulename{t\_case},
  \ottdrulename{t\_rec}, \ottdrulename{t\_app},
  \ottdrulename{t\_abort}, \ottdrulename{t\_join}, \ottdrulename{t\_conv}.

We have $ \Gamma \vdash \ottnt{A} :  \star  $ as a
  premise to the rule, and $ \vdash   \Gamma $ by IH.

\item[Case \ottdrulename{t\_var}.] By inversion on $ \vdash   \Gamma $ plus
  weakening (lemma \ref{lem:ann-weakening}).

\item[Case \ottdrulename{t\_dcon}.] By \ottdrulename{t\_tcon}, using
  the premise $\Gamma  \vdash  \overline{A_i}  \ottsym{:}  \Delta^{+}$.

\item[Case \ottdrulename{t\_abs}.] By IH we get $ \Gamma  \ottsym{,}   \ottmv{x}  :  \ottnt{A}  \vdash \ottnt{B} :  \star  $ and $ \vdash   \Gamma  \ottsym{,}   \ottmv{x}  :  \ottnt{A}  $.  Inversion on the latter gives 
$ \Gamma \vdash \ottnt{A} :  \star  $ so by \ottdrulename{t\_pi} we get
  $ \Gamma \vdash  ( \ottmv{x} \!:\! \ottnt{A} )  \to \,  \ottnt{B}  :  \star  $ as required.

Meanwhile, weakening (lemma \ref{lem:ann-weaken-envwf}) on
$ \vdash   \Gamma  \ottsym{,}   \ottmv{x}  :  \ottnt{A}  $ gives $ \vdash   \Gamma $ as required.

\item[Case \ottdrulename{t\_iabs}.] Similar to the previous case.

\item[Case \ottdrulename{t\_iapp}.] By the IH we have
  $ \Gamma \vdash  [  \ottmv{x} \!:\! \ottnt{A}  ]  \to \,  \ottnt{B}  :  \star  $ and $ \vdash   \Gamma $.

Now by inversion on $ \Gamma \vdash  [  \ottmv{x} \!:\! \ottnt{A}  ]  \to \,  \ottnt{B}  :  \star  $ (lemma
\ref{lem:ann-inv-reg}) we get $ \Gamma  \ottsym{,}   \ottmv{x}  :  \ottnt{A}  \vdash \ottnt{B} :  \star  $.
Then by substitution (lemma \ref{lem:ann-subst}) we have
$ \Gamma \vdash \ottsym{[}  \ottnt{v}  \ottsym{/}  \ottmv{x}  \ottsym{]}  \ottnt{B} :  \star  $ as required.

\item[Case \ottdrulename{t\_injdom}.] By IH we have $ \vdash   \Gamma $. Also
  by IH we have $ \Gamma \vdash \ottsym{(}   ( \ottmv{x} \!:\! \ottnt{A_{{\mathrm{1}}}} )  \to \,  \ottnt{B_{{\mathrm{1}}}}   \ottsym{)}  \ottsym{=}  \ottsym{(}   ( \ottmv{x} \!:\! \ottnt{A_{{\mathrm{2}}}} )  \to \,  \ottnt{B_{{\mathrm{2}}}}   \ottsym{)} :  \star  $, so by
  applying inversion (lemma \ref{lem:ann-inv-reg}) twice we get
  $ \Gamma \vdash \ottnt{A_{{\mathrm{1}}}} :  \star  $ and $ \Gamma \vdash \ottnt{A_{{\mathrm{2}}}} :  \star  $. Then by
  \ottdrulename{t\_eq} we have $ \Gamma \vdash \ottnt{A_{{\mathrm{1}}}}  \ottsym{=}  \ottnt{A_{{\mathrm{2}}}} :  \star  $ as required.

\item[Case \ottdrulename{t\_injrng}.] By similar reasoning to the
  previous case we get $ \Gamma  \ottsym{,}   \ottmv{x}  :  \ottnt{A}  \vdash \ottnt{B_{{\mathrm{1}}}} :  \star  $ and
  $ \Gamma  \ottsym{,}   \ottmv{x}  :  \ottnt{A}  \vdash \ottnt{B_{{\mathrm{1}}}} :  \star  $. Then by substitution (lemma
  \ref{lem:ann-subst}) we have $ \Gamma \vdash \ottsym{[}  \ottnt{v}  \ottsym{/}  \ottmv{x}  \ottsym{]}  \ottnt{B_{{\mathrm{1}}}} :  \star  $ and
  $ \Gamma \vdash \ottsym{[}  \ottnt{v}  \ottsym{/}  \ottmv{x}  \ottsym{]}  \ottnt{B_{{\mathrm{2}}}} :  \star  $, so by \ottdrulename{t\_eq} we have
  $ \Gamma \vdash \ottsym{[}  \ottnt{v}  \ottsym{/}  \ottmv{x}  \ottsym{]}  \ottnt{B_{{\mathrm{1}}}}  \ottsym{=}  \ottsym{[}  \ottnt{v}  \ottsym{/}  \ottmv{x}  \ottsym{]}  \ottnt{B_{{\mathrm{2}}}} :  \star  $ as required.

\item[Case \ottdrulename{t\_iinjdom}, \ottdrulename{t\_iinjrng}.]
  Similar to the previous two cases.

\item[Case \ottdrulename{t\_injtcon}.] By IH we have
  $ \Gamma \vdash  \ottmv{D}  \;  \overline{A_i}   \ottsym{=}   \ottmv{D}  \;  \overline{A_i}'  :  \star  $, so by applying inversion (lemma
  \ref{lem:ann-inv-reg}) twice we have $\Gamma  \vdash  \overline{A_i}  \ottsym{:}  \Delta$ for
  some $\Delta$. By inversion on that judgment we get
  $ \Gamma \vdash \ottnt{A_{\ottmv{k}}} :  \star  $, and similarly  $ \Gamma \vdash \ottnt{A'_{\ottmv{k}}} :  \star  $. So by
  \ottdrulename{t\_eq} we have $ \Gamma \vdash \ottnt{A_{\ottmv{k}}}  \ottsym{=}  \ottnt{A'_{\ottmv{k}}} :  \star  $ as required.

\end{description}
\end{proof}

\begin{lemma}[Strengthening for the annotated language]
\label{lem:ann-strengthening}
If $ \Gamma_{{\mathrm{1}}}  \ottsym{,}   \ottmv{x_{{\mathrm{1}}}}  :  \ottnt{A_{{\mathrm{1}}}}   \ottsym{,}  \Gamma_{{\mathrm{2}}} \vdash \ottnt{a} : \ottnt{A} $ and $\ottmv{x_{{\mathrm{1}}}}$ is not free in
$\Gamma_{{\mathrm{2}}}$, $\ottnt{a}$ or $\ottnt{A}$, then $ \Gamma_{{\mathrm{1}}}  \ottsym{,}  \Gamma_{{\mathrm{2}}} \vdash \ottnt{a} : \ottnt{A} $.
\end{lemma}
\begin{proof}
By induction on $ \Gamma_{{\mathrm{1}}}  \ottsym{,}   \ottmv{x_{{\mathrm{1}}}}  :  \ottnt{A_{{\mathrm{1}}}}   \ottsym{,}  \Gamma_{{\mathrm{2}}} \vdash \ottnt{a} : \ottnt{A} $.
\end{proof}

\begin{lemma}[Value application] The following rule is admissible.
\[
\ottdruleappXXval{}
\]
\end{lemma}
\begin{proof}
By regularity (lemma \ref{lem:ann-regularity}) we have
$ \Gamma \vdash  ( \ottmv{x} \!:\! \ottnt{A} )  \to \,  \ottnt{B}  :  \star  $. So by inversion (lemma
\ref{lem:ann-inv-reg}) we know $ \Gamma  \ottsym{,}   \ottmv{x}  :  \ottnt{A}  \vdash \ottnt{B} :  \star  $.
Then by substitution (lemma \ref{lem:ann-subst}) we 
have $ \Gamma \vdash \ottsym{[}  \ottnt{v}  \ottsym{/}  \ottmv{x}  \ottsym{]}  \ottnt{B} :  \star  $, so we can apply \ottdrulename{t\_app}.
\end{proof}

\begin{lemma}[Nondependent application] The following rule is
  admissible.
\[
\ottdruleappXXnondep{}
\]
\end{lemma}
\begin{proof}
By regularity (lemma \ref{lem:ann-regularity}) we have
$ \Gamma \vdash \ottnt{A}  \to  \ottnt{B} :  \star  $, so by inversion (lemma \ref{lem:ann-inv-reg})
we have $ \Gamma  \ottsym{,}   \ottmv{x}  :  \ottnt{A}  \vdash \ottnt{B} :  \star  $. By strengthening (lemma
\ref{lem:ann-strengthening}) we have $ \Gamma \vdash \ottnt{B} :  \star  $. Since
$\ottmv{x}$ is not free we know $\ottsym{[}  \ottnt{b}  \ottsym{/}  \ottmv{x}  \ottsym{]}  \ottnt{B} = \ottnt{B}$, so this also
shows $ \Gamma \vdash \ottsym{[}  \ottnt{b}  \ottsym{/}  \ottmv{x}  \ottsym{]}  \ottnt{B} :  \star  $, and we can apply \ottdrulename{t\_app}. 
\end{proof}

\end{document}